\documentclass[12pt]{article}
\pdfoutput=1
\usepackage{commands}
\usepackage{putex} 
\usepackage[all,cmtip]{xy}
\usepackage{fancybox}
\usepackage{subcaption}
\usepackage{graphicx}
\usepackage{cite}
\usepackage{dsfont}
\usepackage{physics}
\usepackage{mciteplus}
\usepackage{comment}
\usepackage{skak}
\usepackage{esint}
\usepackage{amsmath}
 \usepackage[compat=1.1.0]{tikz-feynman}\tikzfeynmanset{warn luatex=false}
\usepackage{multirow}
\usepackage{calc}
\usepackage{mathtools,amssymb}
\usepackage{tikz}
\usepackage{pgfplots}
\pgfplotsset{compat=1.18}
\usetikzlibrary{decorations.markings,decorations.pathmorphing}
\usetikzlibrary{arrows.meta}
\DeclareFontFamily{OT1}{pzc}{}
\DeclareFontShape{OT1}{pzc}{m}{it}{<-> s * [1.10] pzcmi7t}{}
\DeclareMathAlphabet{\mathpzc}{OT1}{pzc}{m}{it}

\def\({\left(}
\def\){\right)}

\usepackage{tikz-cd}
\makeatletter
\DeclareFontFamily{OMX}{MnSymbolE}{}
\DeclareSymbolFont{MnLargeSymbols}{OMX}{MnSymbolE}{m}{n}
\SetSymbolFont{MnLargeSymbols}{bold}{OMX}{MnSymbolE}{b}{n}
\DeclareFontShape{OMX}{MnSymbolE}{m}{n}{
    <-6>  MnSymbolE5
   <6-7>  MnSymbolE6
   <7-8>  MnSymbolE7
   <8-9>  MnSymbolE8
   <9-10> MnSymbolE9
  <10-12> MnSymbolE10
  <12->   MnSymbolE12
}{}
\DeclareFontShape{OMX}{MnSymbolE}{b}{n}{
    <-6>  MnSymbolE-Bold5
   <6-7>  MnSymbolE-Bold6
   <7-8>  MnSymbolE-Bold7
   <8-9>  MnSymbolE-Bold8
   <9-10> MnSymbolE-Bold9
  <10-12> MnSymbolE-Bold10
  <12->   MnSymbolE-Bold12
}{}

\let\llangle\@undefined
\let\rrangle\@undefined
\DeclareMathDelimiter{\llangle}{\mathopen}%
                     {MnLargeSymbols}{'164}{MnLargeSymbols}{'164}
\DeclareMathDelimiter{\rrangle}{\mathclose}%
                     {MnLargeSymbols}{'171}{MnLargeSymbols}{'171}
\makeatother

\usepackage{latexsym}
\usepackage{stmaryrd}

\usepackage{cite}
\usepackage{framed}
\definecolor{shadecolor}{rgb}{0.95,0.95,0.97}
\definecolor{refkey}{rgb}{0.5,0.5,0}
\definecolor{labelkey}{rgb}{0.5,0.5,0}
\definecolor{citekey}{rgb}{0.5,0.5,0}
\definecolor{darkgreen}{rgb}{0,0.5,0}
\definecolor{darkblue}{cmyk}{0.9,0.9,0,0}
\definecolor{darkred}{rgb}{0.6,0,0.3}
\colorlet{mydarkblue}{blue!50!black}
\colorlet{myred}{red!65!black}
\usepackage{ifpdf}
\ifpdf
\usepackage{graphicx} 
\usepackage[setpagesize=false,pagebackref=false, linktocpage, bookmarksopen=true, colorlinks=true, linkcolor=blue,citecolor=blue,urlcolor=blue]{hyperref}
\usepackage{hyperref}

\else
\usepackage{graphicx}

\fi

\renewcommand{\Im}{{\rm Im}}
\renewcommand{\Re}{{\rm Re}}

\def\eqref#1{(\ref{#1})}
\def\comma{\,,}
\def\period{\,.}
\def\figref#1{Figure \ref{#1}}

\def\ket#1{|#1\rangle}
\def\bra#1{\langle #1 |}

\def\XXint#1#2#3{{\setbox0=\hbox{$#1{#2#3}{\int}$}
		\vcenter{\hbox{$#2#3$}}\kern-.5\wd0}}

\newcommand{\beq}{\begin{equation}}
\newcommand{\eeq}{\end{equation}}
\def\red#1{\textcolor[rgb]{1, 0, 0}{#1}}

\def\nullify#1{}
\definecolor{darkgreen}{cmyk}{0.9,0,0.9,0}

\makeatletter
\def\section{\@startsection {section}{1}{\z@}{-3.5ex plus -1ex minus 
		-.2ex}{2.3ex plus .2ex}{\large\bf}}
\makeatother
\makeatletter
\def\subsection{\@startsection {subsection}{1}{\z@}{-3.5ex plus -1ex minus 
		-.2ex}{2.3ex plus .2ex}{\normalsize\bf}}
\makeatother

\begin{document}
	
	\preprint{DESY-23-155\\
 CERN-TH-2023-197}
	\institution{DESY}{Deutsches Elektronen-Synchrotron DESY, Notkestr. 85, 22607 Hamburg, Germany}
 	\institution{CERN}{Department of Theoretical Physics, CERN, 1211 Meyrin, Switzerland}
	
	\title{
	2d QCD and Integrability\\
      \LARGE Part I: 't Hooft model
 	}
		
	\authors{Federico Ambrosino\worksat{\DESY} and Shota Komatsu\worksat{\CERN} }

		\abstract
	{We study analytic properties and integrable structures of the meson spectrum in large $N_c$ QCD$_2$. We show that the integral equation that determines the masses of the mesons, often called the 't Hooft equation, is equivalent to finding solutions to a TQ-Baxter equation.  Our analysis extends some of previous results  by Fateev et al.\ to general quark masses $m=m_1=m_2$, as a perturbative series of the mass parameter. This reformulation, together with its relation to an inhomogeneous Fredholm equation, makes accessible the analytic structure of the spectrum in the complex plane of the quark masses. We also comment on applications of our techniques to non-perturbative topological string partition functions.
}
	\date{\today}
	\maketitle	
	\tableofcontents
 	\section{Introduction} \label{intro}
Understanding the mechanism of confinement in gauge theories has proven to be one of the most challenging and intriguing unresolved questions in quantum field theory. The enduring objective of analytically describing confinement in Quantum Chromodynamics (QCD) in 3+1 dimensions, as well as more generally, in strongly coupled gauge theories across various dimensions, remains beyond our current capabilities. Consequently, in order to advance our comprehension of the fundamental mechanisms underlying confining dynamics, it is often advantageous to thoroughly analyse toy models that exhibit similar characteristics within more controlled settings, enabling a more effective analytic approach to the problems.

Following this spirit, in this paper we focus on the 1+1-dimensional version of QCD, coupled with quarks in the fundamental representation of the gauge group. This is one of the most celebrated toy model for confining gauge theories, and significant effort has been devoted, both analytically and numerically, to exploring its various aspects over the years (See \cite{Abdalla_1996} and reference therein for a review of the topic).
Upon considering the double scaling limit in which the number of colors $N_c$ is sent to infinity while taking the 't Hooft coupling $g = g^2_{\rm YM} N_c$ fixed, the theory,  widely known as \textit{'t Hooft model} in this limit and firstly introduced in \cite{THOOFT1974461}, becomes particularly simple and tractable. By taking the large $N_c$ limit, all but the Feynman diagrams that can be drawn on a two-dimensional sphere are suppressed, and many emergent features of QCD$_2$, that would not be evident from the Lagrangian, are made accessible by this simplification\cite{HOOFT1974461}. Some of those characteristics, such as the absence of deconfined quark states  and the Regge-like behavior of meson spectra, are shared with the four-dimensional version, making the 't Hooft model one of the most intriguing low-dimensional models for mesons in QCD$_4$.  

Remarkably the problem of determining the mass spectrum of mesons and their wavefunctions in the 't Hooft model  can be reformulated as an eigenvalue problem, which we present here for reference, deferring a comprehensive explanation of its significance to the main text:
\be \label{thooftequation}
2\pi^2 \lambda_n \,\phi_n(x) = \bigg[ \frac{\alpha_1}{x} + \frac{\alpha_2}{1-x} \bigg] \phi_n(x) - \fint_0^1 \dd{y} \cP \frac{1}{(x-y)^2} \, \phi_n(y)
\ee 
Here $\alpha_{1,2}$ and $\lambda_n$ are related to the quark masses $m_{1,2}$ and the meson mass $M_n$ by 
\beq
\alpha_{1,2}=\frac{\pi m_{1,2}^2}{g}-1 \comma \qquad \lambda_n =\frac{M_n^2}{\pi g}\comma
\eeq
with $g$ being the 't Hooft coupling $g=g_{\rm YM}^2N_c$.   
This eigenproblem, often referred to as \textit{'t Hooft Equation}, can be solved numerically with arbitrary precision and accuracy. 
The main benchmark characteristics of this theory that we have listed above  can be directly inferred by direct numerical solution of \eqref{thooftequation} \cite{THOOFT1974461, PhysRevD.13.1649, Anand:2021qnd}. However, we believe 
that the eigenproblem \eqref{thooftequation} deserves a more thorough analytical treatment for reasons that we describe below. 

Firstly, in the seminal work \cite{Fateev_2009}, Fateev, Lukyanov and Zamolodchikov found that the 't Hooft equation, when quark masses are set to a special value ($\alpha_1=\alpha_2=0$), can be reformulated as a Baxter TQ-system --- the equations that describe the spectrum of general $(1+1)$-dimensional integrable systems. While this observation was intriguing, it remained unclear whether it was just a mathematical coincidence or indicative of deeper underlying structures, as their analysis was limited to the specific mass value\footnote{In \cite{Fateev_2009}, they reported preliminary results for general $\alpha$'s without derivations, announcing an upcoming publication. However, the paper announced never appeared to our knowledge.}. In this paper, we extend their findings by demonstrating that these structures persist for general masses, providing compelling evidence for the existence of hidden structures in the 't Hooft model. Secondly, there has been a recent resurgence of interest in the study of {\it adjoint QCD} in two dimensions, i.e.~two-dimensional Yang-Mills theory coupled to quarks in the adjoint representation of the gauge group \cite{Dalley:1992yy,Kutasov:1993gq,Boorstein:1993nd,Bhanot:1993xp,Demeterfi:1993rs,Smilga:1994hc,Lenz:1994du,Katz:2013qua,Katz:2014uoa,Cherman:2019hbq,Komargodski:2020mxz,Dempsey:2021xpf,Popov:2022vud,Dempsey_2023,Dempsey:2023fvm}. This serves as an interesting toy model of confining fluxtubes since the quarks in the adjoint representation can introduce nontrivial dynamics without making the fluxtubes to break\footnote{This is true only for massive quarks. As discussed in \cite{Gross:1995bp,Antonuccio:1998uz,Komargodski:2020mxz}, the theory is not confining when the quark masses are zero.}. In particular, there are indications that the dynamics on the confining fluxtube may become integrable in certain limits \cite{Dubovsky:2018dlk,Donahue:2019adv,Donahue:2019fgn,Donahue:2022jxu,Asrat:2022aov}. To address this question more concretely, it is valuable to thoroughly study an integrable structure of a simpler, solvable theory like the 't Hooft model. This is precisely what we aim to achieve in this paper.

To be more specific, we illustrate that, in the Fourier space relative to the rapidities $\theta=\frac{1}{2}\log(\frac{x}{1-x})$, 
 't Hooft equation can be can be transformed into a finite-difference operator that has the same structure as a Baxter TQ-relation:
\be \label{tqrelation}
Q(\nu + 2i) + Q(\nu - 2i) - 2Q(\nu)  = \frac{-4\pi \lambda}{\frac{2\alpha}{\pi} + \nu \coth(\frac \pi 2 \nu)} \, Q(\nu), \qquad \alpha = \frac{\alpha_1 + \alpha_2}{2} 
\ee
On top of unveiling a very interesting structure typical of integrable systems, this reformulation of the problem is particularly suitable for analytical methods, especially for the case $\alpha_1 = \alpha_2 = \alpha$. Starting from the TQ system, we can construct  expressions for the following spectral sums:
\be
G_+^{(s)}(\alpha) = \sum_{n=0}^{\infty} \frac{1}{\lambda^s_{2n}}\, , \qquad \qquad G_-^{(s)}(\alpha) = \sum_{n=0}^{\infty} \frac{1}{\lambda^s_{2n+1}} \, ,
\ee
Such results for $\alpha=0$ have been obtained by Fateev, Lukyanov and Zamolodchikov \cite{Fateev_2009}. We will extend those results by developing a systematic perturbation in the mass parameter $\alpha$.

Another noteworthy result that we can extract from the TQ-system is an asymptotic approximation for the meson wavefunctions, which, although being constructed to be only asymptotic, turn out to approximate remarkably well the numerical solutions. These results are potentially useful for the evaluation of scattering amplitudes of mesons, which are known to be given by overlaps of meson wavefunctions \cite{PhysRevD.13.1649,BROWER1977131,Abdalla_1996}.

In addition to this, we deduce important properties of the model that we anticipate here: as a function of the complex parameter $\alpha$, $\lambda(\alpha)$ exhibits a square-root branch point at $\alpha_1 = \alpha_2 = -1$ where the chiral symmetry becomes exact and a massless meson emerges. One of the pivotal advantages of recasting 't Hooft equation in $\nu$-space is that it makes straightforward to extend the analysis to the complex plane of the masses $\alpha\in\mathbb{C}$. Indeed, we illustrate that as we analytically continue onto the second sheet beyond the chiral-symmetry branch cut, there are infinitely many branch-point singularities accumulating at $\lambda = \infty$ at  special values of the (complex) masses corresponding to which one of the even eigenvalues turns zero.

Notably, this emergent structure is not specific to the 't Hooft model. We can consider an extension of the 't Hooft model known as the generalized Yang-Mills (gYM$_2$), obtained by replacing the kinetic term for the field strength $F$ with a $BF$ coupling involving an adjoint scalar field $B$ and introducing an arbitrary potential $V(B)$ for it:
\be    
\mathcal{L} = \frac{N}{8\pi} \Tr B \wedge F -  \frac{N}{4\pi}V(B) + \overline{\psi}_a \left(i \gamma^\mu D_\mu -m_a\right)\psi_a
\ee
For this very large class of models, we demonstrate that, employing the same spectral representation, the associated bound state 't Hooft equation can be casted into the exact same form of \eqref{tqrelation}, with a transfer matrix $T(\nu)$ determined in a closed form for any given $V(B)$. Many of the key features of the 't Hooft model that we demonstrate in this paper persist in this generalization. Presenting  the properties of these models is the subject of the second part of this project \cite{Ambrosino:2024prz}.  This, together with other results, suggest that the analytical structures unveiled in this problem, and in particular the singularities in the complex plane of the couplings, might be a universal feature of confining two-dimensional theories and are therefore worth investigation.

Furthermore, similar spectral problems of integral operators and TQ-systems appear in the context of topological string \cite{Marino:2011eh,kallen2014instanton,grassi2015topological,Marino:2016rsq,marino2018wavefunctions}; we comment on possible applications of our techniques  in Section  \ref{topstring}.

{\bf The outline of the paper} is the following: In {\bf Section} \ref{t Hooft model}, we review the derivation of the 't Hooft equation, which determines the spectrum of mesons. In {\bf Section} \ref{sectq}, we explain how to rewrite the 't Hooft equation into a TQ-Baxter equation and extend it to arbitrary values of the meson mass $\lambda$ by introducing an inhomogeneous term to the 't Hooft equation. This can be achieved systematically using the framework of the inhomogeneous Fredholm equation. In {\bf Section} \ref{criticality}, we discuss the properties of the meson spectrum as we analytically continue the quark masses to complex values. In particular, we show that the complex quark mass plane contains infinitely many branch cuts, each of which starts from a point where one of the mesons becomes massless. We suggest that each of those points describes a ``critical point"; namely their infrared dynamics is described by a nontrivial CFT. Further results, both numerical and analytical, as well as the results for wavefunctions are provided in {\bf Section} \ref{sec:further}. Quite surprisingly, the asymptotic expansion of the wavefunction approximates them quite well even for the low-lying spectrum. In {\bf Section} \ref{spectraldata}, we explain how to extract the spectral data from the TQ-equation, generalizing the discussion in \cite{Fateev_2009} for a special quark mass. The techniques employed here are likely to be useful in other setups in which the inhomogeneous Fredholm equation and TQ-equations show up, such as non-perturbative topological string discussed in {\bf Section} \ref{topstring}. Finally in {\bf Section} \ref{conclusions}, we give a summary and discuss various future directions. Appendices are included to explain technical details.

\paragraph{Notes Added:} After the first version of this paper, a beautiful paper \cite{Litvinov:2024riz} appeared in arXiv, which determined the spectral sums and the asymptotic expansion of the $Q$-functions and the meson masses, fully non-perturbatively in $\alpha$. The methods employed here are different since we performed the perturbative expansion in $\alpha$ which leads to different analytic structures of the relevant function. Nevertheless the resulting expressions are in agreement with those in \cite{Litvinov:2024riz}, once expanded in $\alpha$. Also, the first version of this paper contained some analysis on the asymptotic expansion of the meson spectrum. However, the argument employed there relied on the use of conjectural identities (generalizing (2.17) of \cite{Fateev_2009}), which, upon further inspection, we concluded is not trustworthy. We thus removed a relevant subsection. 
\section{Review: The 't Hooft model for mesons}
\label{t Hooft model}
\subsection{Conventions}
Before starting our discussion, let us summarize notations and conventions used in this paper. We work with two-dimensional Minkwoski metric  $g^{00} = - g^{11} = 1$, and we take the $\gamma$-matrices $\gamma^0 = \begin{psmallmatrix}
    0 & 1\\ 1& 0
\end{psmallmatrix},  \, \gamma^1 = \begin{psmallmatrix}
    0 & -1\\ 1& 0
\end{psmallmatrix}$. We define lightcone coordinates as: $x^\pm = \frac{1}{\sqrt{2}}(x^0 \pm x^1)$, $p_\pm =  \frac{1}{\sqrt{2}}(p_0 \pm p_1)$. In this coordinates, the $\gamma$-matrix algebra takes the form $\{\gamma_+, \gamma_-\} = 2$, $\gamma^2_+ = \gamma^2_-=0$.
\subsection{Derivation of the 't Hooft equation}
We now review the key features of the 't Hooft model that will be useful for our discussion. For more details, see the original paper \cite{THOOFT1974461}, which is quite pedagogical. 't Hooft considered an $SU(N)$ YM theory in two dimensions coupled  to fundamental massive fermions. In the large number of colors limit $N\to \infty$ the theory admits a double scaling limit  taken for the 't Hooft coupling $g= g_{\YM}^2 N$ being fixed. 
The Lagrangian is the standard YM Lagrangian with $a= 1,\cdots, N_f$ being the flavour index:
\be
\mathcal{L} = -\frac{1}{4}\Tr F \wedge \star F + \overline{\Psi}_a (i\slashed{D}\, - m_a) \Psi_a
\ee
As customary in 2-dimensional theories, we take lightcone coordinates, where the theory greatly simplifies upon fixing lightcone gauge $A_- = 0$ as the field strength is linear in the fields, there are no ghosts
and non-linear interactions between gauge fields are set to zero. Denoting $A_+=A$, in this gauge the Lagrangian is:
\be \label{lightconelagrangian}
\mathcal{L} = -\frac{1}{2} (\partial_- A)^2 + \overline{\Psi}_a (i \gamma_\mu \partial^\mu\, - m_a - g_{\YM} \gamma_- A) \Psi_a
\ee 

\paragraph{Exact quark propagator.} The lightcone gauge leads to a drastic simplification of Feynman rules. Since the quark-antiquark-gluon vertex only involves $\gamma_-$ satisfying $\gamma_{+}^2=\gamma_{-}^2=0$ and $\gamma_{+}\gamma_{-}\gamma_{+}=2\gamma_{+}$, the terms proportional to $ip_{+}\gamma_{-}+m_a$ in the quark propagator (highlighted in red below) do not contribute when they appear in internal legs\footnote{These terms do give non-vanishing contributions if the fermion propagators show up in the external legs. We can however project them out by multiplying $\gamma_{-}$'s from outside. In the subsequent analysis, we implicitly perform such a projection.}:
\be
\la \overline{\Psi}_a (p)\, \Psi_a(-p)\ra=\frac{ip_{-}\gamma_{+}\red{+ip_{+}\gamma_{-}+m_a}}{2p_{+}p_{-}-m_a^2+i\epsilon}\comma
\ee
 As a result, we can focus on the component proportional to $\gamma_{+}$ and the Feynman rules simplify to
\begin{alignat}{3}
&\begin{tikzpicture}[baseline={(current bounding box.center)}]
\begin{feynman}
\vertex[label = {[left ]$i$}] (a1) at (0,0);
\vertex[right=2cm of a1, label={right}:$i$] (a2);
\vertex[below=0.50em of a1, label = {[xshift=-0.5em, yshift= -1.2em] $j$}] (b1);
\vertex[below=0.50em of a2,label = {[xshift=  0.5em, yshift= -1.2em] $j$}] (b2);
\diagram* [edges=fermion]{
(b2)--(b1),
(a1)--(a2),
};
\end{feynman}
\end{tikzpicture} \qquad && \la A(p)\, A(-p)\ra &&= -i \frac{1}{p^2_-}  \; , \\
&\begin{tikzpicture}[baseline={(current bounding box.center)}]
\begin{feynman}
\vertex   [label = {[left ] $i,a$}]                 (a1) ;
\vertex[right=2cm of a1,label = {[right]  $i,a$}]   (a2);
\diagram* [edges=fermion]{
(a1)--(a2),
};\end{feynman}
\end{tikzpicture}\qquad &&  \la \overline{\Psi}_a (p)\, \Psi_a(-p)\ra  &&= \frac{ip_-}{2 p_- p_+ -m^2_a + i\epsilon}  \; ,\\
&\begin{tikzpicture}[baseline={(current bounding box.center)}]
\begin{feynman}
\vertex [label= {[left ] $i,a$}   ]         ;
\vertex[below right=1 cm and 1cm of a1]   (a2) ;
\vertex[below=0.5 em of a2]   (a3);
\vertex[below left= 1cm and 1cm of a3, label = {[left] $j,a$ }]   (a4);
\vertex[right= 1cm of a2, label={[xshift = 0.2em, yshift = -0.1em] $i$}]   (a5);
\vertex[right= 1cm of a3, label={[xshift = 0.2em, yshift = -1.2em] $j$}]   (a6);
\diagram* [edges=fermion]{
(a1)--(a2)--(a5),
(a6)--(a3)--(a4),
};\end{feynman}
\end{tikzpicture}  && \la \overline{\Psi}_a \, \Psi_a \,  A\ra  &&= - 2i g_{\YM} \; ,
\end{alignat}
while fermionic loops are subleading in a $1/N$ expansion and are absent in the planar limit. These simplifications make it possible to obtain an explicit expression for the full (non-perturbative) fermion propagator $S(p)$ obtained  by a resummation of the 1PI contribution $\Sigma(m_a, p)$:
\begin{align} \nonumber
\begin{tikzpicture}[baseline={(a.base)}]
\begin{feynman}[every blob={/tikz/fill=gray!30,/tikz/inner sep=2pt}]
\vertex[blob] (m) {S};
\vertex[left= 1.5cm of m] (a);
 \vertex[right= 1.5cm of m] (b);
 \diagram* [edges=fermion]{
(a)--[ edge label=$p$](m)--[ edge label=$p$](b),
};
\end{feynman}
\end{tikzpicture} 
&{=} \begin{tikzpicture}[baseline={(a.base)}]
\begin{feynman}[every blob={/tikz/fill=gray!30,/tikz/inner sep=2pt}]
\vertex (a);
 \vertex[right= 1cm of a] (b);
 \diagram* [edges=fermion]{
(a)--[edge label = $p$](b),
};
\end{feynman}
\end{tikzpicture} + \begin{tikzpicture}[baseline={(a.base)}]
\begin{feynman}[every blob={/tikz/fill=gray!30,/tikz/inner sep=2pt}]
\vertex[blob] (m) {1PI};
\vertex[left= 1.5cm of m] (a);
 \vertex[right= 1.5cm of m] (b);
 \diagram* [edges=fermion]{
(a)--[edge label=$p$](m)--[edge label=$p$](b),
};
\end{feynman}
\end{tikzpicture} + \begin{tikzpicture}[baseline={(a.base)}]
\begin{feynman}[every blob={/tikz/fill=gray!30,/tikz/inner sep=2pt}]
\vertex[blob] (m) {1PI};
\vertex[left= 1.5cm of m] (a);
 \vertex[blob, right = 1.2cm of m] (n) {1PI};
 \vertex[right= 1.5cm of n] (b);
 \diagram* [edges=fermion]{
(a)--[edge label=$p$](m)--(n)--[edge label=$p$](b),
};
\end{feynman}
\end{tikzpicture} + \cdots \\
&= \frac{ip_-}{2p_+p_- - m^2_a - p_- \Sigma(m_a, p) + i\epsilon}
\end{align}
In the planar limit, the 1PI contribution can be bootstrapped through the Dyson equation:
\begin{align}\label{dyson} \Sigma(m_a, p) = 
\begin{tikzpicture}[baseline={(a.base)}]
\begin{feynman}[every blob={/tikz/fill=gray!30,/tikz/inner sep=2pt}]
\vertex[blob] (m) {1PI};
\vertex[left= 1.5cm of m] (a);
 \vertex[right= 1.5cm of m] (b);
 \diagram* [edges=fermion]{
(a)--[edge label=$p$](m)--[edge label=$p$](b),
};
\end{feynman}
\end{tikzpicture} = \begin{tikzpicture}[baseline={(a.base)}]
\begin{feynman}[every blob={/tikz/fill=gray!30,/tikz/inner sep=2pt}]
\vertex (a);
 \vertex[right= 1cm of a] (b);
  \vertex[right= 0.5em of b] (c);
   \vertex[right= 2.8cm of b] (d);
   \vertex[left= 0.5em of d] (e);
   \vertex[blob, right = 1cm of b] (f) {$S$} ;
   \vertex[right = 1cm of d] (g);
 \diagram* [edges=fermion]{
(a)--[ edge label=$p$](b),
(b)--[ edge label=$k$,half left] (d),
(e)--[half right] (c),
(c) -- (f),
(f) -- (e),
(d)--[edge label=$p$](g),
};
\end{feynman}
\node at (2.4,-0.6) {$p-k$};
\end{tikzpicture}
\end{align}
that can be remarkably solved explicitly in this case (see \cite{THOOFT1974461} for a derivation) leading to
\be\begin{split}\label{propagator}
\Sigma(m_a, p) &= \frac{g}{\pi} \left( \frac{\sgn(p_-)}{\rho} - \frac{1}{p_-} \right)\\ S(p) &= \frac{ip_-}{2p_+p_- - m^2 + g /\pi - g \abs{p_-}/(\rho\pi)}
\end{split}\ee
where $\rho$ is an IR regulator that must be introduced to cancel the infrared divergence.  The introduction of this sharp cutoff should be regarded as an intermediate step since, in any gauge-invariant observables, factors containing $\rho$ cancel in the limit $\rho\to 0$. On the other hand, any UV divergence is absent (upon symmetric regularisation) as a consequence of the super-renomalizability of QCD$_2$.

The form of the propagator illustrates the confining nature of the interaction: upon sending $\rho \to 0$, the pole of the full propagator, i.e.\ the effective mass that re-summes all the quantum corrections, is moved to $\infty$, signalling the absence of on-shell free quark states in this model. One can also understand it directly from the Lagrangian \eqref{lightconelagrangian} by integrating out the gauge field and obtaining a confining linear potential between quarks: 
\be
\begin{split}
A(x_+,x_-) &= - \sum_a \frac g 2 \int \dd{y_-} |y_- -x_-| \, j_a(x_+,y_-), \qquad j_a = \overline{\Psi}_a \Psi_a \\
\mathcal{L} &\supset \mathcal \int dx_1\,dx_2\,j^0(x_0,x_1)|x_1-y_1|\,j^0(x_0,y_1) 
\end{split} 
\ee
These arguments already capture the qualitative behavior of the quark-antiquark states and lead to the expectation that, in the quark-antiquark vertex, only bound state are formed and the spectrum is purely discrete. 

\paragraph{Bethe-Salpeter equation.}The remarkable feature of this model is that we can go beyond this rudimentary analysis and construct explicitly the meson states.
Generically, bound states arise as a result of summing up infinitely many interaction diagrams between their constituents. Such a resummation leads to the so-called {\it Bethe-Salpeter equation}, which can be viewed as a Schrödinger equation for bound states. Its derivation is standard textbook-material by now and we briefly review it here. In quantum field theory, the spectrum of particles can be read off from poles of the Green function. This statement, well-known for the one-particle spectrum, holds true also for bound states. Namely, two-particle bound states appear as poles of the two-particle Green function $\cG^{(4)} \equiv \cG^{(4)}(x_1,x_2; x_3,x_4)$. The Green function satisfies an {\it inhomogeneous} Dyson equation analogous to \eqref{dyson},
 \be \begin{split}
 \label{Dyson4pt}
\cG^{(4)}(x_1,x_2;& x_3,x_4) = S(x_1-x_3) \, S(x_2-x_4) \\ &+ \int\dd[2]{y_1}\cdots \dd[2]{y_4} S(x_1 - y_1) \, S(x_2 - y_2) \, K(y_1,y_2; y_3,y_4) \, \cG^{(4)}(y_3,y_4; x_3,x_4)
\end{split}\ee
where  $K(y_1,y_2; y_3,y_4)$ is the full irreducible interaction kernel and $S$ in the full two-point function of the constituents of the meson (the $S$ we computed above).  As stated above, $\cG^{(4)}(x_1,x_2; x_3,x_4)$ will have poles in momentum space corresponding to the mass of the bound state $M^2$, so that, in the vicinity of a pole, the Green function 
can be approximated as 
 $\cG^{(4)}(p,r,P) \sim \frac{\overline{\Phi}(r,p) \Phi(p,r)}{P^2 - M^2}$ where $\Phi$ is bound state wavefunction\footnote{More precisely, it is a transition amplitude between the constituents and the bound state: $\bra{\Omega} \overline{\psi}\, \psi \ket{\Phi}$.}. Then, the 4-point Dyson equation near the pole reduces to the following {\it homogeneous} Dyson equation,
\be 
\Phi(p,r) = S(p)\, S(p-r) \int\dd[2]{k} K(k,p,r)\,\Phi(p+k,r) 
\ee
which is known as the Bethe-Salpeter equation. In general, $K(k,p,r) $ is a complicated object, containing all the interactions among the constituents. Normally, one then approximates the full interaction kernel by considering only a subset of diagrams. The most common approximation is the so-called rainbow-ladder approximation in which one only resums rainbow diagrams for each constituents and ladder diagrams between constituents. Normally, this is just a (non-systematic) approximation scheme. However, in the planar limit, it gives an exact answer since fermionic loops are suppressed by $1/N_c$
\begin{align}
\label{bssinglequark}
\Phi^A_{B  }(p,r) = \begin{tikzpicture}[baseline={([yshift=-.5ex]current bounding box.center)}]
\begin{feynman}[every blob={/tikz/fill=blue!30,/tikz/inner sep=2pt}]
\vertex[blob] (m) {$\Phi^A_B$};
\vertex[above left=0.35cm  and 0.27cm of m] (m1);
\vertex[below left = 0.35cm and 0.27cm of m] (m2);
\vertex[ left= 0.9cm of m1] (a);
\vertex[small,draw,circle,inner sep=0pt,minimum size=0.5cm,fill=gray!30, left = 0.9cm of m1] {$S$};
\vertex[ left= 0.5cm of a] (a0);
\vertex[ left= 0.9cm of a0] (a01);
\vertex[ left= 0.9cm of m2] (b);
\vertex[small,draw,circle,inner sep=0pt,minimum size=0.5cm,fill=gray!30, left = 0.9cm of m2] {$S$};
\vertex[ left= 0.5cm of b] (b0);
\vertex[ left= 0.9cm of b0] (b01);
\diagram* [edges=fermion]{
(a)--[edge label= $p-r$](m1),
(a01)--[edge label= $p-r$](a0),
(m2)--[edge label= $p$](b),
(b0)--[edge label= $p$](b01),
};
\end{feynman}
\end{tikzpicture}  = \begin{tikzpicture}[baseline={([yshift=-.5ex]current bounding box.center)}]
\begin{feynman}[every blob={/tikz/fill=blue!30,/tikz/inner sep=2pt}]
\vertex (a0);
\vertex[above =0.35cm of a0] (a1);
\vertex[below = 0.35cm of a0](b1);
\vertex[right = 0.9cm of a1] (a11);
\vertex[small,draw,circle,inner sep=0pt,minimum size=0.5cm, fill=gray!30,right = 0.9cm of a1] {$S$};
\vertex[right = 0.5cm of a11] (a12);
\vertex[right = 0.9cm of a12] (a2);
\vertex[right = 0.9cm of b1] (b11);
\vertex[small,draw,circle,inner sep=0pt,minimum size=0.5cm, fill=gray!30,right = 0.9cm of b1] {$S$};
\vertex[right = 0.5cm of b11] (b12);
\vertex[right = 0.9cm of b12] (b2);
\vertex[right = 0.5em of a2] (a3);
\vertex[right = 0.5em of b2] (b3);
\vertex[right = 0.8cm of a3] (a4);
\vertex[right = 0.8cm of b3] (b4);
\vertex[small,draw,circle,inner sep=0pt,minimum size=0.5cm, fill=gray!30,right = 0.8cm of a3] {$S$};
\vertex[small,draw,circle,inner sep=0pt,minimum size=0.5cm,fill=gray!30, right = 0.8cm of b3] {$S$};
\vertex[right = 0.5 cm of a4] (a5);
\vertex[right = 0.5 cm of b4] (b5);
\vertex[blob, right = 5.3cm of a0] (m) {$\Phi^A_B$};
\vertex[above left=0.35cm  and 0.27cm of m] (a6);
\vertex[below left = 0.35cm and 0.27cm of m] (b6);
\diagram* [edges=fermion]{
(a1)--[edge label= $p-r$](a11),
(a12)--[edge label= $p-r$](a2),
(a3)--(a4),
(b2)--[edge label= $p$](b12),
(b11)--[edge label= $p$](b1),
(b4)--(b3),
(b3)--[edge label'= $k$](a3),
(a2)--(b2),
(a5)--[edge label= $p+k-r$](a6),
(b6)--[edge label= $p+k$](b5),
};
\end{feynman} \end{tikzpicture}
\end{align}
and contains the full quantum dynamics of the quark-antiquark bound states.  Explicitly, it takes the form: 
\be
\Phi^A_{B  }(p,r) = - \frac{4 g}{i}\, S(p) S(p-r) \int\frac{\dd{k_+}\dd{k_-}}{(2\pi)^2} \, \frac{\Phi^A_{B  }(p+k,r)}{k_-^2}\, 
\ee
with $S(p)$ being the full propagator in \eqref{propagator}. 
Although it is not possible to solve analytically for this equation, it can be recast into a one dimensional eigenvalues problem. Namely, upon employing the variables:
\be
\alpha_{a,b} = \frac{\pi\, m_{a,b}^2}{g} -1 , \qquad 
2 r_+\,r_- =\pi g\lambda, \qquad x = \frac{p_-}{r_-}, \quad x \in [0,1], 
\ee the ``average'' function $\phi^a_b(x) = \int\dd{p_+} \Phi^a_b(p,r)$  is readily shown to satisfy the following equation\cite{THOOFT1974461}:
\be\label{thoofteq}
2\pi^2\lambda\,\phi^a_b(x) = \left(\frac{\alpha_a}{x}+ \frac{\alpha_b}{1-x}\right)\phi^a_b(x)- \fint_0^1\dd{y} \frac{\phi^a_b(y)}{(y-x)^2} =: \cH\,\phi^a_b(x)
\ee
widely referred to as the 't Hooft equation\footnote{Since in the planar limit fermion loops are suppressed, 't Hooft equation depends on the flavour symmetry only through the masses of the quarks pairwise. In addition, since this equation describes gauge-invariant bound states, any dependence on the regulator $\rho$ has disappeared.}.

\paragraph{Boundary condition and hermiticity.}
The eigenproblem \eqref{thoofteq} is the main focus of our paper and therefore its characteristics deserve a thorough discussion. The behavior of the wavefunction close to the boundaries at $x=0$ and $1$ cannot be simply deduced from the equation itself; rather it requires careful analysis on the regularity near the boundaries and the hermiticity of the Hamiltonian. 

Assume that the solutions have a power-law behavior near the boundaries of$[0,1]$:
\be 
\label{boundarycond}
\phi^a_b(x\sim 0 ) \simeq x^{\beta_a},  \qquad \phi^a_b(x\sim 1)\simeq(1-x)^{\beta_b} \period
\ee 
As shown in \cite{THOOFT1974461}, the hermiticity of the Hamiltonian requires the wave function to vanish at the boundaries i.e.~$\beta_i \geq 0$. In addition, it turns out that one needs to impose $\beta_i<1$. To see this, let us analyze the behavior of \eqref{thoofteq} near $x=0$. The first term on the RHS, $\alpha_a\frac{\phi(x)}{x}$, behaves as $\propto x^{\beta_1-1}$. To reproduce the LHS, which decays faster as $x^{\beta_1}$, it needs to be cancelled by the integral term, which can be approximated as
\beq
-\fint_0^1\dd{y} \frac{\phi(y)}{(x-y)^2}\quad \overset{x\sim 0}{\sim}\quad \begin{cases} \pi\beta_1 \cot(\pi\beta_1) x^{\beta_1 -1} \qquad & \beta_1< 1\comma\\
\mathcal{O}(1) & \beta_1\geq 1 \period
\end{cases}
\eeq
As shown above, for $\beta_1<1$ the boundary behavior is singular and can be computed simply by  substituting $\phi(x)=x^{\beta_1}$ to the integral. On the other hand, for $\beta_1\geq 1$ the integral is dominated by the behavior of $\phi$ away from $x\sim 0$, and is generically $\mathcal{O}(1)$. Thus, requiring the cancellation forces us to take $\beta_1<1$. A similar analysis at $x\sim 1$ shows that $\beta_2<1$ as well. To be more precise, the cancellation is guaranteed if $\beta_i$ satisfies the following\footnote{For integral equations of the Cauchy kind on a finite interval on the complex plane, which only contains  first-order poles (unlike ours which contains second-order poles), the boundary behavior of solutions is   well-studied \cite{muskhelishvili2013singular}. The most general boundary condition in such cases is $\beta_i = \Re\, \beta_i  + i\Im \,\beta_i $ with $-1\leq \Re\, \beta_i\leq 0$, $\Im\, \beta_i \in \mathbb{R}$. Our results can be retrieved from it by imposing the reality condition and performing a simple differentiation \cite{Kaushal:2023ezo} since the double pole principal value regularization (Hadamard regularization) is nothing but the derivative of the Cauchy principal value, i.e.\ $\fint_a^b \frac{f(t)}{(t-x)^2} = \dv{x} \fint_a^b \frac{f(t)}{t-x}$. Complex boundary conditions will show up also in our analysis, as we extend the problem to imaginary masses ($\alpha_i < -1$).}
:
\be\label{exponents}
    \alpha_i + \pi \beta_i \cot(\pi \beta_i) = 0, \qquad 0\leq \beta_i < 1 \qquad (i = a,b )\period
\ee
See also \S 29 of \cite{muskhelishvili2013singular} or recent discussions in \cite{Kaushal:2023ezo}.
For any fixed $\alpha_i\geq-1$, the solution to these conditions \eqref{exponents} is unique and it unambiguously determines the boundary behavior\footnote{For $\alpha_i>-1$, the exponents are strictly positive, and henceforth \eqref{boundarycond} correspond to Dirichlet boundary conditions in the box $[0,1]$. On the other hand  for $\alpha = -1$, the eigenfunction does not vanish at the boundary. }.

For this space of functions with the boundary conditions, the ``Hamiltonian'' $\cH$ is  Hermitian and positive definite \cite{THOOFT1974461}:
\small
\be\label{scalarproduct}
\left( \phi, \cH\, \psi \right) = \int_0^1 \dd{x}\left[ \frac{\alpha_1 +1 }{x} +\frac{\alpha_2 +1 }{1-x} \right]\phi^*(x)\,\psi(x)  + \frac{1}{2}\int_0^1\dd{x}\int_0^1\dd{y} \frac{\left(\phi^*(x)-\phi^*(y)\right)\left( \psi(x)- \psi(y) \right)}{(x-y)^2}
\ee 
\normalsize
Furthermore, the divergence of the potential at boundary of the box guarantees that the spectrum is purely discrete and non-degenerate.  This confirms our expectations and shows that the spectrum  of QCD$_2$ (at least at the leading order in $1/N$) only contains mesonic bound states and no free quark states, as the latter would be associated with a continuous spectrum. We remark that as a function of $\alpha_i \in \mathbb{C}$, the Hamiltonian is explicitly not-Hermitian. For $\mathbb R \ni \alpha < -1$, i.e.\ for purely imaginary quark masses, the scalar product is not longer positive definite and the Hamiltonian in this case is non-Hermitian as well.

\paragraph{Charge conjugation/parity.}Solving the (discrete) eigenproblem defined by \eqref{thoofteq}, allows us to extract the physical spectrum of mesons in the 't Hooft model: the eigenfunction $(\phi^{a}_b)_n$ correspond to the $n$-th excited bound state of the quark-antiquark meson and the corresponding eigenvalues $\mu^2_n$ is related to the invariant mass of the bound state. The eigenstates can be chosen to be real $\phi^*(x) = \phi(x)$, given that the equation is real and the spectrum is non-degenerate. Solutions of \eqref{thoofteq} are also eigenstates of the charge conjugation  $\cC$, which is an exact symmetry of $\cH$ and acts as a parity in $x$-space\footnote{To see that this is the correct parity in the 
box, one can go to the ``Regge" limit $\mu^2 \to \infty$ where $\phi_n \sim \sin(n\pi x)$.}: 
\be 
(\phi^b_a) _n(x) = (-1)^n ( \phi_b^a)_n(1-x)\period
\ee

\section{'t Hooft equation as a TQ-system}
\label{sectq}
For many years, the eigensystem \eqref{thoofteq} has been investigated either by means of numerical methods or by the WKB approximation (for instance \cite{Brower:1978wm}). Although this approach has been fruitful in illustrating many interesting features of the model, here we pursue the approach initiated in \cite{Fateev_2009} that illustrates rich and emergent analytical structures that can be used to construct explicit analytical solutions in terms of asymptotic series.  The paper \cite{Fateev_2009} focused on the special case $\alpha_1 = \alpha_2 = 0$\footnote{Recall that this does not correspond to any special physical limit of the problem but simplifies greatly the eigenproblem.}. Here, we  extend their results to the more general case  $\alpha_1 = \alpha_2 \neq 0$, following closely the arguments in \cite{Fateev_2009}. Some of our results apply also to the most general case of $\alpha_1\neq \alpha_2$. 
\subsection{Spectral representation}\label{subsec:spectral}
The crucial idea is considering a suitably defined ``Fourier space'' (or the ``Mellin space") defined by the Fourier transform with respect to the ``rapidity" variable\footnote{This representation was considered also in \cite{Narayanan_2005}.} $\theta = \frac 1 2 \log(\frac{x}{1-x})$:
\be \label{tranf}
\begin{split}
\Psi(\nu) = \mathcal{F}[\phi(x)](\nu) &= \int_0^1 \frac{\dd{x}}{2 x(1-x)} \left(\frac{x}{1-x}\right)^{\frac{i\nu}{2}}  \phi(x) = \infint \dd{\theta} e^{i\nu\theta}  \phi(\theta)
\end{split}\ee
Being a Fourier transform, this integral transformation can be inverted as follows:
\be \label{invtranf}
\begin{split}
\phi(x) = \mathcal{F}^{-1}[\Psi(\nu)](x) &= \infint\frac{\dd{\nu}}{2\pi} \left(\frac{x}{1-x}\right)^{-\frac{i\nu}{2}}  \Psi(\nu)
\end{split}\ee
It is worth noting that essentially the same rewriting was used in the analysis of the so-called ladder-limit of Wilson loop in $\mathcal{N}=4$ super Yang-Mills \cite{Correa:2012nk,Kim:2017sju,Cavaglia:2018lxi}, in order to reformulate the Bethe-Salpeter equation governing the spectrum of operators into the Baxter TQ-system.

\paragraph{'t Hooft equation in the spectral space.} Let us express \eqref{thoofteq} in the $\nu$-space. To do that, it is convenient to multiply $x(1-x)$ to  \eqref{thoofteq}. Then upon going to rapidity space, $x = \frac{1}{1+ e^{-2\theta}}$, we can just apply the usual properties of the Fourier transform, together with the identity:
\be
    x(1-x)\,\infint\frac{\dd{\nu}}{2\pi}\,\fint_0^1\dd{y}\,\frac{1}{(x-y)^2} \left(\frac{y}{1-y}\right)^{i \frac{\nu}{2}} \, \Psi(\nu) = \mathcal{F}^{-1}\left[ \frac{\pi \nu}{2} \coth(\frac\pi 2 \nu) \, \Psi(\nu)\right]\, ,
\ee
to obtain the following form of \eqref{thoofteq} in the $\nu$-space:
\be
\begin{split}
\label{thooftnuspacediffalpha}
   \lambda \int_{-\infty}^{+\infty} \dd{\nu'} \frac{\pi (\nu - \nu') }{2 \sinh(\frac{\pi}{2}(\nu - \nu') )}\,\Psi^a_b(\nu') = &\left[ \nu\coth(\frac \pi 2 \nu )  + \frac{\alpha_1 + \alpha_2}{\pi}  \right]\Psi^a_b (\nu)   \\ &+ \int_{-\infty}^{+\infty} \dd{\nu'}  \cP \frac{i(\alpha_2 - \alpha_1)}{2 \sinh(\frac{\pi}{2}(\nu - \nu') )} \Psi^a_b(\nu') \period
\end{split}
\ee
The factor of $i$ multiplying the term proportional to the mass difference is crucial. Charge-conjugation $\mathcal{C}$, that is an exact symmetry of the system for real values of $\alpha$, is readily shown to act in $\nu$-space with complex conjugation of the wavefunctions: \be \left(\Psi^b_a\right)_n^*(\nu) =\int_0^1 \frac{\dd{x}}{2 x(1-x)} \left(\frac{x}{1-x}\right)^{-\frac{i\nu}{2}} \left( \phi^a_b\right)_n(1-x) = (-1)^n \left(\Psi^a_b\right)_n(\nu) \; . \ee
Then, the complex conjugate of \eqref{thooftnuspacediffalpha} is satisfied by $\left(\Psi^a_b\right)(\nu)$ rather than $\left(\Psi^b_a\right)(\nu)$, and the factor of $i$ in the last term takes care of exchanging the masses of the two quark constituents $\alpha_a\leftrightarrow\alpha_b$ under the complex conjugation.
 
Let us now specialize to $\alpha_1 = \alpha_2 = \alpha$ and use the notation $\Psi^a_a(\nu) = \Psi(\nu)$. We then arrive at the following form of the 't Hooft equation in the $\nu$-space
\be \label{thooftnuspace}
   \lambda \int_{-\infty}^{+\infty} \dd{\nu'} \frac{\pi (\nu - \nu') }{2 \sinh(\frac{\pi}{2}(\nu - \nu') )}\,\Psi(\nu') = \left[ \nu\coth(\frac \pi 2 \nu )  + \frac{2\alpha}{\pi}  \right]\Psi (\nu) \period
\ee
The solutions $\Psi(\nu)$ have to decay at $\nu\to \pm \infty$ to guarantee that the norm of the eigenfunctions is finite; we will refine this below by showing that the decay is actually exponential.

A key advantage of recasting the problem in the $\nu$-space \eqref{thooftnuspace} is that this transformation turns a highly singular kernel affected by a (double-pole) principal-value prescription to a regular one. In fact, upon redefining further the function $\Psi$ as:
\be \label{defsqrt}
    \phi(\nu) = \sqrt{f(\nu)}\, \Psi(\nu), \qquad f(\nu) = \nu\coth(\frac \pi 2 \nu )  + \frac{2 \alpha}{\pi},
\ee
the equation \eqref{thooftnuspace} takes a form of a homogeneous Fredholm equation of the second kind, a well-studied subject in mathematics: \cite{IntegralEquations,weisstein}:
\be \label{fred}
   \phi(\nu) =\lambda \int_{-\infty}^{\infty}\dd{\nu'} \kappa(\nu,\nu')\,\phi(\nu'), \qquad  \kappa(\nu,\nu') := \frac{\pi (\nu - \nu')}{2 \sinh(\frac{\pi}{2}(\nu - \nu') )} \frac{1}{\sqrt{f(\nu)\, f(\nu')}}\; .
\ee
For $\alpha > -1$,  the kernel $\kappa(\nu,\nu^{\prime})$ defines a Hilbert–Schmidt integral operator\footnote{Note the similarities with the kernel in (2.1) of \cite{Kallen:2013qla}, also considered in \cite{Zamolodchikov_1994}. See also related works \cite{kallen2014spectral,kallen2014instanton,Bonelli:2016idi,grassi2015topological} and references therein,}\cite{IntegralEquations}. This follows from the following properties of $f(\nu)$,
\be
    \pdv{f(\nu)}{(\frac{2\alpha}{\pi})} =  \nu \coth(\frac\pi 2 \nu) +1  > 0 \quad \forall \nu, \qquad \eval{\min_{\nu\in \mathbb{R}} 
f(\nu)}_{\alpha= -1} = 0\comma
\ee 
which guarantees that 
 $f(\nu)$ never vanishes in the considered ranges of $\alpha_i$ and  $\kappa(\nu,\nu')$ has a finite $L^2(\mathbb{R}^2)$ norm.  This, together with (fast) decaying of the wavefunctions at infinity, is sufficient to show the uniqueness of the solution of \eqref{fred} in $L^2(\mathbb{R})$ \cite{IntegralEquations}. By contrast, the analysis for $\alpha=-1$ is more subtle as we will discuss below.
\paragraph{Parity and norm.}The Fourier-space function also enjoys similar parity properties, as illustrated by the simple calculation:
    \be
    \label{eq:paritynu}
    \begin{split}
        \Psi_n(-\nu) &= \int_{-\infty}^{+\infty} \dd{x}\frac{\phi_n(x)}{x(1-x)}  \left(\frac{x}{1-x}\right)^{-\frac{i\nu}{2}} =  \int_{-\infty}^{+\infty} \dd{x} \frac{\phi_n(1-x)}{x(1-x)}  \left(\frac{x}{1-x}\right)^{\frac{i\nu}{2}} \\  &=(-1)^n\,\Psi_n(\nu);
            \end{split}
    \ee 
As discussed in \cite{THOOFT1974461}, the wave functions $\phi_n(x)$ form an orthonormal and complete set in the considered Hilbert space of functions. This can be rephrased in $\nu$-space as:
\begin{align}
    \lambda_n\,\delta_{nm} &=  \lambda_n \int_0^1\dd{x} \phi_n(x)\,\phi^*_m(x) =\\ &= \lambda_n \int_{-\infty}^{+\infty} \frac{\dd{\nu}}{2\pi} \int_{-\infty}^{+\infty} \frac{\dd{\mu}}{2\pi}\, \Psi_n(\nu)\, \Psi_m^*(\mu)\int_0^1\dd{x}  \left(\frac{x}{1-x}\right)^{\frac{i}{2}(\nu-\mu)} \nonumber\\
    &=  \lambda_n \int_{-\infty}^{+\infty} \frac{\dd{\nu}}{2\pi} \int_{-\infty}^{+\infty} \frac{\dd{\mu}}{2\pi} \, \Psi_n(\nu)\, \Psi_m^*(\mu)\,\frac{\pi (\nu-\mu)}{2 \sinh(\frac\pi 2 (\nu-\mu))}\nonumber\\
    &\overset{\eqref{thooftnuspace}}{=} \int_{-\infty}^{+\infty} \frac{\dd{\mu}}{2\pi} \frac{f(\mu)}{2\pi}\, \Psi_n(\mu)\, \Psi_m^*(\mu)
\end{align}
The measure $f(\mu)$ is the same as the one defined in \eqref{defsqrt}. 
Another important fact is the asymptotic behavior of the $\Psi(\nu)$: in order to guarantee finiteness of the $\mathbb{L}^2$-norm,  $\Psi(\nu)$ needs to decay at least polynomially fast on the real axis. 
 
\paragraph{Massless quarks.}At $\alpha= -1$, the quarks are exactly massless and the theory possesses exact chiral symmetry. $f(\nu)$ has a (double) zero in $\nu = 0$ and the conditions above are not satisfied. Indeed we can see that, for $\alpha=-1$ and $\lambda = 0$, \eqref{thooftnuspace} reduces to:
\be
    \left[ \nu\coth(\frac \pi 2 \nu )  - \frac{2}{\pi}  \right] \Psi (\nu) = 0
\ee
and both  $\Psi (\nu) = \delta(\nu)$ and $\Psi(\nu) = \delta'(\nu)$ solve the equation above. While the first solution in $x$-space is just the constant $\phi(x) = 1$, the second correspond to a logarithmic solution $\mathcal{F}[\delta'(\nu)] = i\log(\frac{x}{1-x})$, with finite $\mathbb{L}^2[0,1]$ norm:
\be
    \int_0^1 \left(\log(\frac{x}{1-x})\right)^2 = \frac{\pi^2}{3} \; .
\ee
Although the solution diverges logarithmically at the boundaries, it can be verified explicitly that it solves the 't Hooft equation in $x$-space with $\lambda = 0$:
\begin{align}
    & \fint_0^1\dd{x}\frac{\log(\frac{x}{1-x})}{(x-y)^2} = \lim_{\epsilon\to 0^+}\frac{1}{2}\left[\int_0^1\dd{x} \frac{\log(\frac{x}{1-x})}{(x-y + i \epsilon)^2} +  \int_0^1\dd{x} \frac{\log(\frac{x}{1-x})}{(x-y - i \epsilon)^2} \right] \\
    &= -\lim_{\epsilon\to 0^+}\frac{1}{2}\left[\frac{\log (-y +i \epsilon )-\log (1-y +i \epsilon)}{(1-y + i \epsilon) (y-i \epsilon )}  + \frac{\log (y +i \epsilon )-\log (-1+y +i \epsilon)}{(1-y - i \epsilon) (y+i \epsilon )}\right]
\end{align}
Now, given that $y\in (0,1)$, we have to be careful while performing the limit $\epsilon\to 0^+$ to take the correct logarithmic branch:
\begin{equation}
    \lim_{\epsilon\to 0^+}\log (-\zeta +i \epsilon ) = \log(\zeta\, e^{+i\pi}) = \log(\zeta)  + i \pi; \qquad \zeta =  y, (1-y) >0
\end{equation}
Thus, we have
\begin{equation}
    \fint_0^1\dd{x}\frac{\log(\frac{x}{1-x})}{(x-y)^2} = -\frac{\log (\frac{y}{1-y})}{y(1-y)}\comma
\end{equation}
from which we deduce that the 't Hooft equation \eqref{thoofteq} with $\alpha=-1$ is satisfied by $\log(\frac{x}{1-x})$. Nevertheless this solution does not correspond to a physical mesonic state.  This degeneracy between physical and unphysical solutions leads to  interesting analytic structure of the spectrum (as a function of mass squared) as we will discuss thoroughly in section \ref{criticality}.

\paragraph{Bethe-Salpeter approximation in Ising Field Theory.}
\label{sec:bsift}
Before proceeding further, let us point out a similarity with the so-called {\it Ising field theory} (IFT), i.e.~Ising CFT perturbed by both magnetic field and temperature. In the ``rapidity" variable, the 't Hooft equation \eqref{thoofteq} takes the form
\be
\label{thetaspace}
\left(\frac{\pi^2\lambda}{2\cosh^2\theta} -  \frac{\alpha_a + \alpha_b}{2} + \frac{\alpha_a - \alpha_b}{2} \tanh\theta  \right) \phi_a^b(\theta) =  -2\fint_{-\infty}^{+\infty} \dd{\varphi} \frac{1}{\sinh^2(\theta-\varphi)} \phi_a^b(\varphi)\period
\ee
In the equal mass limit, this equation is similar to the Bethe-Salpeter approximation of IFT \cite{fonseca2006ising}. In a small external magnetic field  approximation of IFT, the stable particles in the spectrum share similarities with bound states in QCD$_2$ behaving exactly as mesons (see \ \cite{fonseca2006ising}).  In this limit, one can expand around the free theory of Majorana fermions with mass $m$ and obtains the Bethe-Salpeter equation  by looking at the eigenstates of the perturbed Hamiltonian $H_{\text{free}} + h \int \sigma$ that are only of the form of two-particle state, and ignoring all the higher-particle contributions:
\be
\left(m^2 - \frac{M^2}{\cosh^2 \theta} \right) \Psi(\theta) = f_0\, \fint_{-\infty}^{+\infty} \frac{\dd{\theta'}}{2\pi}\left(2\frac{\cosh(\theta-\theta')}{\sinh^2(\theta-\theta')} + \frac{1}{4} \frac{\sinh\theta\,\sinh\theta'}{\cosh^2\theta\, \cosh^2 \theta'}\right) \Psi(\theta')
\ee
(Here $f_0$ is the  ``string tension" scale and is proportional to the strength of the magnetic field $h$.) This  equation determines the spectrum of the \textit{Ising mesons} that share many interesting features with the ones of the 't Hooft model, which we will discuss in Section \ref{IFTmesons}. Higher-particle states will contribute at higher order in $h$, with the two particle sector being the dominant in $h\to 0$. In this sense, $h$ plays the same role of $1/N$ in the 't Hooft model. 
\subsection{TQ-relation}
\label{TQsec}
Let us consider again the 't Hooft equation with equal masses \eqref{thooftnuspace}, that we display here for convenience:
\be
\begin{split}
   \lambda \int_{-\infty}^{+\infty} \dd{\nu'} \frac{\pi (\nu - \nu') }{2 \sinh(\frac{\pi}{2}(\nu - \nu') )}\,\Psi(\nu') = & \underbrace{\left[\nu\coth(\frac \pi 2 \nu )  + \frac{2\alpha}{\pi} \right]}_{=:f(\nu)} \Psi (\nu) \; .
\end{split}
\ee
The function$\sqrt{f(\nu)} \Psi(\nu)$ is a solution to a Fredholm equation of the second kind with a Hilbert-Schmidt kernel. In particular the right hand side $f(\nu)\Psi(\nu)$ does not have any singularities in the strip $(-2i,2i)$ where we can analytically continue both the l.h.s.\ and the r.h.s.\ of the equation\footnote{If we analytically continue outside of the strip, the integral kernel gets modified because of poles crossing the integration contour. We will discuss this in detail in Section \ref{criticality}.}. Then, the {\it Q-function} defined by
\begin{equation}
\label{qdefinition}
    Q(\nu) = \sinh(\frac\pi 2 \nu)\, f(\nu)\,\Psi(\nu)
\end{equation}
does not have any singularity in the strip $[-2i,2i]$, grows slower than an exponential at $\nu \to \pm \infty$ on the real line, and satisfies the related integral equation:
\be
\label{inteqq}
       Q(\nu) = \lambda\sinh(\frac\pi 2 \nu) \int_{-\infty}^{+\infty} \dd{\nu'} \frac{\pi (\nu - \nu') }{2 \sinh(\frac{\pi}{2}(\nu - \nu') )} \frac{Q(\nu')}{\sinh(\frac \pi 2 \nu')f(\nu')} \comma
\ee
Furthermore, it follows from the absence of poles of $f(\nu)\,\Psi(\nu)$ as well as \eqref{inteqq} that 
\beq\label{eq:quanti}
Q(0) = Q(\pm 2i)=0 \comma
\eeq
where $\sinh(\frac{\pi}{2}\nu)$ vanishes. The equation \eqref{eq:quanti} can be regarded as the ``quantization" condition as we will discuss in Section \ref{inhomo}. 

We can extend the asymptotic behavior of the $Q$-functions to the entire strip $[-2i,2i]$. To do so, we use \eqref{inteqq} to verify that the Fourier transform of $\cQ(\nu) := \frac{Q(\nu)}{\sinh(\frac{\pi}{2}\nu)}$ ,
\be\label{defcq}
\widehat\cQ (y):=\cF\left[\cQ(\nu)\right](y) = \infint\dd{\nu} \frac{Q(\nu)}{\sinh(\frac{\pi}{2}\nu)} e^{i \nu y}\comma \qquad  \widehat\cQ_{\xi} (y):=\mathcal{F}[\cQ (\nu+i\xi)](y)\comma
\ee
satisfies the following relation:
\be
\label{shiftQ}
\widehat{\cQ}_{\xi} (y) = \lambda \frac{\pi e^{- y \xi}}{\cosh^2(y)} \cF\left[{\cQ}/{f}\right](y) = \lambda \frac{\pi e^{- y \xi}}{\cosh^2(y)} \cF\left[\Psi\right](y)\period 
\ee
Since $\Psi(\nu)$  is $\mathbb{L}^2$ and decays at least polynomially on the real axis, \eqref{shiftQ} implies automatically that also $\widehat{\cQ}_{\xi} (y)$ and consequently $\cQ(\nu + i \xi)$ are $\mathbb{L}^2$ for any $\abs{\xi}<2$. We thus deduce that $\cQ({\nu})$ decays at least polynomially for any $\Im\,\nu\in [-2i,2i]$ at $\abs{\Re\, \nu} \to \infty$. Upon evaluating \eqref{shiftQ} at $\xi = \pm 2$, one can easily show 
\be
\widehat{\cQ}_{2}(y) + \widehat{\cQ}_{-2}(y) =  -2\widehat{\cQ}(y) + (4\pi \lambda) \, \cF\left[\cQ/f \right](y) \period
\ee
This is the Fourier transform of the following difference equation\footnote{We also present an alternative and more direct proof of \eqref{tqtilde} and \eqref{tq} in Appendix \ref{appderivation}.},
\be\label{tqtilde}
\cQ(\nu + 2i) + \cQ(\nu - 2i) + 2\cQ(\nu) = \frac{4\pi \lambda}{f(\nu)} \cQ(\nu)\comma
\ee
which can be further rewritten in terms of the original $Q$-function as follows\footnote{The reason for which we prefer this latter formulation in terms of $Q(\nu)$ rather than $\cQ(\nu)$  will be clear only in section \ref{inhomo} where we extend the problem to an inhomogeneous one; at this level the two are completely equivalent.
}:
\be \label{tq}
\boxed{Q(\nu + 2i) + Q(\nu- 2i ) - 2Q(\nu) =   \frac{- 4\pi\lambda}{ \nu\coth(\frac \pi 2 \nu )  + \frac{2\alpha}{\pi} } \, Q(\nu) }\, 
\ee 

This illustrates the advantage of going into the $\nu$-space. Here the integral equation was recast into the finite-difference equation having the form of the so-called {\it TQ-Baxter} equation, which appears in the study of integrable systems. To each solution $\Psi(\nu)$ of \eqref{thooftnuspace}  decaying at least polynomially at $\abs{\Re\,\nu} \to \infty $, there is an associated solution $Q(\nu)$ of \eqref{tq} analytic in the strip $(-2i,2i)$, and 
growing slower than any exponential at $\abs{\Re\,\nu} \to \infty$. This asymptotic behavior, together with the quantization condition \eqref{eq:quanti}, determines the discrete spectrum of the 't Hooft equation. As we will discuss respectively in Section \ref{spectraldata} and \ref{analyticalwavefunctions}, this reformulation as the TQ-system makes possible to obtain asymptotic expansions for eigenvalues and as well as for the eigenfunctions. 

\paragraph{From TQ back to 't Hooft.}With the aforementioned asymptotic behavior, the converse is also true: 
any solution $Q(\nu)$ to the TQ-system subject to the asymptotic conditions  gives a solution to \eqref{thooftnuspace}. This can be proven\footnote{A similar argument is used in \cite{Tracy_1996} to prove Lemma 2 and 3. } by noting that if $Q(\nu)$ satisfies \eqref{tq}, the difference between the l.h.s.\ and r.h.s.\ of \eqref{inteqq},
\be h(\nu) = Q(\nu) - \lambda \sinh(\frac{\pi}{2}\nu) \infint \dd{\nu'} \frac{\pi(\nu-\nu')}{2\sinh(\frac{\pi}{2} (\nu - \nu'))}  \frac{Q(\nu')}{\sinh(\frac{\pi}{2} \nu') f(\nu')} \comma
\ee
 extends to a periodic function
\be
h(\nu+2i) + h(\nu-2i) = 0 \comma
\ee
 that is also entire and bounded in the strip as it does not have poles and decays at infinity; the only such function is a constant function and it has to be $h(\nu) = 0$ as from  $Q(0) = Q(\pm 2i)= 0$ we deduce that $h(\pm 2i) = h(0) = 0$.  This completes the proof of the equivalence between the Baxter-TQ equation \eqref{tq} and the integral equation determining the spectrum of mesons in the $\nu$-space \eqref{thooftnuspace}, and illustrates an interesting ``integrable'' structure underlining 't Hooft equation. 

\paragraph{Decay of the eigenfunctions.} As a simple application of the TQ-system presented above, let us  derive a stronger\footnote{Stronger then the polynomial decay.} condition on the decay of the eigenfunctions of the 't Hooft equation. For any $q(\nu)$ periodic with period $2i$ and growing less than exponentially, $Q(\nu) \asymp  e^{\pi k \nu} q(\nu)$ satisfies the $\nu\gg 1$ limit of \eqref{tq}:
\be\label{asymtq}
Q(\nu + 2i) +  Q(\nu - 2i) - 2Q(\nu) + \frac{4\pi \lambda}{\nu} Q(\nu) \asymp  -4 e^{\pi k \nu} q(\nu) \sin^2(k\pi) + \mathcal{O}(\nu^{-1})
\ee
as long as $k\in \mathbb{Z}$. Now,  given that $q(\nu)$ can be chosen arbitrarily, in order to guarantee that $Q(\nu)$ grows slower than an exponential, we must require $k \leq 0$. Henceforth, this already slightly more refined analysis demonstrates that requiring $Q(\nu)$ less than exponentially growing and $\Psi(\nu)$ decaying at infinity, automatically implies that the solutions $\Psi(\nu)$ decay exponentially at infinity. 


\paragraph{Potential relation to $W_{\infty}$-algebra.}
This connection to integrability opens up various promising directions for future exploration. Before proceeding further, let us remark that a hidden algebraic structure of the 't Hooft model was discussed long ago  based on the {\it collective field approach} \cite{Dhar_1994}. By reformulating the 't Hooft model in terms of bilocal fields, $M_{ij}(x,y) = \frac{1}{N} \tr \left[(\psi^a_i)^\dagger(x) \psi^a_j(y)\right]$, they found that mesons at large $N_c$ transform under representations of the algebra\footnote{A $W_\infty$ algebra in this context arises in the large $N$ limit as a limit of $W_N$ for any finite number of colors.} $W^L_\infty \otimes W^R_\infty \otimes U(N_f)$. This infinite-dimensional algebra acts as a spectrum-generating algebra of mesons since it does not commute with the Hamiltonian. In standard integrable systems, a similar role is played by Yangian or quantum group. It would be interesting to understand a connection between this $W_{\infty}$-algebra and the integrable structure that we observed here.

\subsection{Extending to an inhomogeneous problem}
\label{inhomo}
Solutions to the TQ-Baxter equation satisfying the quantization condition \eqref{eq:quanti} (and the aforementioned asymptotic behavior) exist only for the discrete values of $\lambda$ in the spectrum, and our task is solving simultaneously for $Q_n^{(\alpha)}(\nu)$ and $\lambda_n$. This, in general, is as difficult as the problem we started with in \eqref{thoofteq}. 

One way of making progress is to develop a systematic large $\lambda$ expansion, which refines and improves the WKB expansion at large quantum number. For this purpose, it is important to first extend $\lambda_n$ from discrete values to  arbitrary values of $\lambda \in \mathbb{R}$. The situation is analogous to the large spin expansion in conformal field theory, for which a systematic analysis was made possible by the Lorentzian inversion formula \cite{Caron-Huot:2017vep} that enables the analytic continuation of the conformal data away from integer spins.

To achieve this, we pursue the strategy presented in \cite{Fateev_2009}, and simply introduce an appropriate inhomogeneous term $F(\nu)$ (which we specify later) to the 't Hooft equation \eqref{thooftnuspace}. Then the solution $\phi(\nu|\lambda)$ (stressing now the dependence on the continuous parameter $\lambda$) satisfies the inhomogeneous Fredholm equation that extends \eqref{fred},
\be
\label{eq:extfred}
        \frac {F(\nu)}{\sqrt{f(\nu)}}=    \phi(\nu|\lambda) - \lambda \int_{-\infty}^{\infty}\dd{\nu'} \kappa(\nu,\nu') \phi(\nu'|\lambda),
\ee
Solutions to this equation exist for arbitrary $\lambda\in \mathbb{R}$, and they admit the Liouville-Neumann (L-N) series representation, which can be obtained by successive iterations of the inhomogeneous term:
\begin{subequations}
\label{eq:lineu}
\begin{align}
    \phi(\nu;\lambda) &= \sum_{n=0}^{\infty} \lambda^n \,\phi_n(\nu), \qquad \phi_n(\nu) = \int\dd{\mu} K_n(\nu,\mu) \, \frac {F(\mu)}{\sqrt{f(\mu)}}\; ,\\
    K_n(\nu,\mu) &= \int\dd{\rho_1} \cdots\dd{\rho_{n-1}} \kappa(\nu,\rho_1)\cdots \kappa(\rho_{n-1},\mu), \qquad K_0(\nu,\mu) := \delta(\nu-\mu)\, . 
\end{align}
\end{subequations}
To express the results more compactly, it is convenient to  introduce the {\it resolvent} $R(\mu,\nu|\lambda)$ (see also relevant discussions in \cite{Zamolodchikov_1994}), defined by $R:=\frac{\kappa}{1-\lambda \kappa}$. By definition, it satisfies the related integral equation (independent of the inhomogeneous term $F$):
\be
    R(\mu,\nu|\lambda) - \lambda \int_{-\infty}^{+\infty} \dd{\rho}  R(\mu,\rho|\lambda)\,\kappa(\rho,\nu) = \kappa(\mu,\nu) \, , \qquad  R(\mu,\nu|\lambda) = \sum_{n=0}^{\infty} \lambda^n K_n(\mu,\nu)\period
\ee
In terms of the resolvent, the solution $\phi(\nu|\lambda)$  is simply $\phi(\nu|\lambda) = \int_{-\infty}^{+\infty} \dd{\mu} R(\nu,\mu|\lambda)\,\frac {F(\mu)}{\sqrt{f(\mu)}}$.

Let us now specify the inhomogeneous term. As in \cite{Fateev_2009}, we choose the following basis of the inhomogeneous extensions, each of which has a definite parity under $\nu\to-\nu$:
\be
\label{inhomogeneousterm}
    F_+(\nu) = \frac{\nu}{\sinh(\frac \pi 2 \nu)}, \qquad F_-(\nu) = \frac{ 1}{\sinh(\frac \pi 2 \nu)}\period
\ee
It is interesting to note that they correspond to a quadratic (even) and a linear (odd) potential in the $[0,1]$ box: 
\be
        \mathcal{F}\left[F_+(\nu)\right] \propto x(1-2x), \qquad       \mathcal{F}\left[F_-(\nu)\right] \propto (-1 +2x)\; ,
\ee
As already said, each of the two, through the L-N series, drives a unique solution  of the extended eigenproblem $\Psi_\pm (\nu|\lambda) \left(= \pm \Psi_\pm (-\nu|\lambda) \right)$ (cf.~\eqref{eq:extfred}) : 
\be
\label{eq:exteig}
   F_\pm(\nu) = f(\nu)\,\Psi_\pm(\nu|\lambda)  - \lambda \fint_{-\infty}^{+\infty} \dd{\nu'}\frac{\pi (\nu - \nu') }{2 \sinh(\frac{\pi}{2}(\nu - \nu') )}\,\Psi_\pm(\nu'|\lambda) \period
\ee
The solutions to this inhomogeneous equation are (fast) decaying at $ \nu \to \pm \infty$ as well. A direct way to see this is by noting that each of the $\phi_n(\nu)$ decays as infinity given that both the $K_n$ and $F_\pm$ are exponentially suppressed.

At special values $\lambda=\lambda_n$, the inhomogeneous equation above has to reduce to the homogeneous one \eqref{thooftnuspace}. This implies that the inhomogeneous eigenfunctions $\Psi_{\pm}(\nu|\lambda)$ are singular at $\lambda=\lambda_n$ with $\Psi_n (\nu|\lambda_n)$ being the residue at $\lambda=\lambda_n$. More precisely, because of the parity, we have\footnote{As noted in \cite{Fateev_2009}, the coefficients $c_k$'s are proportional to the matrix elements of $\bar{\Psi}\gamma^{\mu}\Psi$ and $\bar{\Psi}\Psi$. We plan to go back to studying correlation functions of this model somewhere else.}
\beq
\Psi_{+}(\nu|\lambda)\sim \sum_{k=0}^{\infty}\frac{c_{2n}\Psi_{2n}(\nu)}{\lambda-\lambda_{2n}}\comma\qquad \Psi_{-}(\nu|\lambda)\sim \sum_{k=0}^{\infty}\frac{c_{2n+1}\Psi_{2n+1}(\nu)}{\lambda-\lambda_{2n+1}}\period
\eeq

\paragraph{TQ-system.}The inhomogeneous $Q_\pm(\nu|\lambda)$ defined as:
\be
Q_\pm(\nu|\lambda)= \sinh(\frac{\pi\nu}{2})\,f(\nu) \, \Psi_\pm(\nu|\lambda) = \mp Q_\pm(-\nu|\lambda)
\ee
satisfy the related inhomogeneous integral equation:
\be
\label{eq:inhointeqQ}
   \sinh(\frac{\pi}{2}\nu)  F_\pm(\nu) = Q_\pm(\nu)  -  \lambda \sinh(\frac{\pi}{2}\nu)  \int_{-\infty}^{+\infty} \dd{\nu'}\frac{\pi (\nu - \nu') }{2 \sinh(\frac{\pi}{2}(\nu - \nu') )}\,\frac{ Q_\pm(\nu')}{\sinh(\frac{\pi}{2}\nu') f(\nu')}
\ee
From \eqref{eq:inhointeqQ}, it immediately follows that
\be\label{eq:boundaryvalues}
Q_+(\pm 2i) = \pm 2i  \comma \qquad Q_{+}(0)=0\comma\qquad  Q_-(\pm 2i) =Q_{-}(0)=  1   \period
\ee
As is clear from these, the quantization conditions \eqref{eq:quanti} are not satisfied for the inhomogeneous extensions.
 
 An argument analogous to the one presented below \eqref{inteqq} implies that  $Q_\pm(\nu)$ grow less than an exponential in the entire strip $[-2i,2i]$. In addition, $Q_\pm(\nu)$ are analytic (without poles) in the strip. This is guaranteed by the $\sinh(\frac{\pi}{2}\nu)$ factor in the definition $Q(\nu) = \sinh(\frac{\pi}{2}\nu) \cQ(\nu)$, which cancels the poles in $F_{\pm}$, and is ultimately the reason behind using $Q(\nu)$ rather than $\cQ (\nu)$. 

Now, since $F_{\pm}$ satisfies \be \sinh(\frac{\pi(\nu +2i)}{2})F_\pm(\nu + 2i ) +\sinh(\frac{\pi(\nu -2i)}{2}) F_\pm(\nu - 2i ) - 2\sinh(\frac{\pi}{2}\nu)F_\pm (\nu) = 0\comma \ee one can follow the derivation of the previous subsection to show that $Q_\pm(\nu|\lambda)$  satisfies the same TQ-Baxter equation\footnote{This TQ-system admits exponentially growing solutions at $\nu\gg 1$ as well (see also the discussions around \eqref{asymtq}). However, they do not correspond to solutions of \eqref{eq:inhointeqQ}. The two problems are equivalent only under the asymptotic condition that $Q$ grows less than an exponential.} as in the homogeneous problem,
\be\label{tqinhom}
Q_\pm(\nu+2i|\lambda) + Q_\pm(\nu-2i|\lambda)- 2Q_\pm(\nu|\lambda) = \frac{-4\pi\lambda }{f(\nu) } Q_\pm(\nu|\lambda)
\ee
  but this time for any value of $\lambda\in \mathbb{R}$.

\paragraph{Wronskian.}
The Wronskian constructed out of the two independent solutions $Q_\pm(\nu|\lambda)$:
\be\label{eq:defwron}
W(\nu|\lambda) = Q_+(\nu + i|\lambda)Q_-(\nu - i|\lambda) - Q_+(\nu - i|\lambda)Q_-(\nu + i|\lambda),
\ee
results to be a  periodic function as following from the straightforward computation:
\be 
\frac{W(\nu+i |\lambda) }{W(\nu- i|\lambda )} = \frac{Q_+(\nu + 2i|\lambda) Q_-(\nu|\lambda) - Q_+(\nu|\lambda) Q_-(\nu+2i|\lambda)}{Q_+(\nu|\lambda) Q_-(\nu- 2i|\lambda) - Q_+(\nu - 2i|\lambda) Q_-(\nu|\lambda)} = 1\period
\ee
Furthermore the asymptotic growth condition at $|{\rm Re}\, \nu|\gg 1$ and the analyticity of $Q_{\pm}$ in the strip guarantee that $W(\nu|\lambda)$ is a constant, depending only on the normalization of  $Q_\pm(\nu|\lambda)$. 
With our normalization, fixed in \eqref{inhomogeneousterm} (see also \eqref{eq:boundaryvalues}), we have:
\be
\label{wrnorma}
W(\nu|\lambda)=W(i|\lambda) = Q_+(2i|\lambda)Q_-(0|\lambda) = 2i.
\ee

As mentioned before, at $\lambda=\lambda_n$, one of the solutions $Q_{\pm}$ develops a pole $\propto 1/(\lambda-\lambda_n)$. Owing to the identity $W(\nu|\lambda)=2i$, this in turn means that the residue of \eqref{eq:defwron} must be zero:
\beq
\begin{aligned}
&Q_{2n}(\nu+i)Q_{-}(\nu-i|\lambda_{2n})-Q_{2n}(\nu-i)Q_{-}(\nu+i|\lambda_{2n})=0\comma\\
&Q_+(\nu + i|\lambda_{2n+1})Q_{2n+1}(\nu - i) - Q_{+}(\nu - i|\lambda_{2n+1})Q_{2n+1}(\nu + i)=0\period
\end{aligned}
\eeq
Here $Q_{n}$ is the homogeneous $Q$-function for $\lambda=\lambda_n$. These equations imply that $Q_{2n}$ and $Q_{-}$ (and similarly $Q_{2n+1}$ and $Q_{+}$) coincide up to a multiplication of a periodic function when $\lambda$ is part of the discrete spectrum. Therefore, at these values of $\lambda$, we only have a single non-degenerate solution to the TQ-system satisfying the required asymptotic behavior.

\paragraph{Spectral determinants.}
We have extended the problem \eqref{thooftnuspace} to homogeneous one \eqref{eq:exteig} that can, in principle, be solved by means of the L-N series \eqref{eq:lineu} resulting in two independent solutions for each $\lambda\in\mathbb{R}$. Yet, the solutions obtained in that way are only indirectly related to the solutions of the original problem that has a discrete spectrum. Fortunately, there is a way to relate the two explicitly making use of the crucial property of the kernel $\kappa(\nu,\nu')$; it belongs to the class of \textit{``completely integrable kernels"} in the language of \cite{korepin, korepin_bogoliubov_izergin_1993}. For this class of kernels, it is known that the resolvent is given in a closed form  in terms of the two independent solutions. This is discussed in Appendix B of \cite{Fateev_2009}. In our case, this translates into the following expression:
\be
R(\mu,\nu|\lambda) = \frac{2 \sinh(\frac{\pi}{2}\mu)\sinh(\frac{\pi}{2}\nu)}{\pi \sinh(\frac{\pi}{2}(\mu-\nu))} \sqrt{f(\mu)\,f(\nu)} \left[\Psi_+(\nu|\lambda)\,\Psi_-(\mu|\lambda) -  \Psi_-(\nu|\lambda)\,\Psi_+(\mu|\lambda) \right].  
\ee
In turn, this implies that the two spectral determinants
\begin{equation}
\label{spectraldet}
D_+(\lambda) = \left( \frac{8\pi}{e} \right)^{\lambda} \prod_{n=0}^\infty \left( 1 - \frac{\lambda}{\lambda_{2n}} \right) e^{\frac{\lambda}{n+1}}, \quad
D_-(\lambda) = \left( \frac{8\pi}{e} \right)^{\lambda} \prod_{n=0}^\infty \left( 1 - \frac{\lambda}{\lambda_{2n+1}} \right) e^{\frac{\lambda}{n+1}}.
\end{equation}
satisfy the following non-trivial integral identities :
\small \be
\begin{split}
    \label{intformulas}
\partial_\lambda \log(D_+(\lambda)  D_-(\lambda)) = 
 2+ \int_{-\infty}^{+\infty}\dd{\nu} \frac{\pi}{2}\bigg[ &\frac{Q_+(\nu|\lambda)\,\partial_\nu Q_-(\nu|\lambda) - Q_-(\nu|\lambda)\,\partial_\nu Q_+(\nu|\lambda)}{f(\nu)} + \frac{\tanh(\frac \pi 2 \nu)}{\nu} \bigg] \, ,\\
\partial_\lambda \log(\frac{D_+(\lambda)}{D_-(\lambda)}) &= -\infint \dd{\nu} \frac{\pi^2}{2}\frac{Q_+(\nu|\lambda)\, Q_-(\nu|\lambda)}{\sinh(\pi\nu)f(\nu)},
    \end{split}
\ee
\normalsize
This follows from the facts that the logarithm of the spectral determinants are given by
\begin{align}\label{lambaexpspec}\log D_\pm(\lambda) &=  (\log(8\pi) -1)\lambda -  \sum_{s=1}^{+\infty} s^{-1}\, G^{(s)}_\pm\, \lambda^s \, \qquad G^{(s)}_+ = \sum_{n = 0}^{\infty} \frac{1}{\lambda^s_{2n}}, \qquad G^{(s)}_- = \sum_{n = 0}^{\infty} \frac{1}{\lambda^s_{2n+1}}\comma
\end{align}
and that the spectral sums $G_{\pm}^{(s)}$ are encoded in to the traces of the resolvent as (cf.\ eq (8.9) and (8.10) of \cite{Fateev_2009}, and Appendix \ref{localP1} for an analogous computation\footnote{The specific form of the $F_\pm(\nu)$ is paramount to  obtain these expressions (see discussion in Appendix B of \cite{Fateev_2009}, or the similar non-trivial trigonometric identities in \eqref{trigid} )}):
\be
\begin{split}
\sum_{s=1}^{\infty} \lambda^{s-1} \left( G_+^{(s)} + G_-^{(s)}\right) &= C + \infint \dd{\mu} \left[ R(\mu,-\mu|\lambda) +  R^{(0)}(\mu)\right] \comma\\
\sum_{s=1}^{\infty} \lambda^{s-1} \left( G_+^{(s)} - G_-^{(s)}\right) &=  \infint \dd{\mu} R(\mu,\mu|\lambda) \period
    \end{split}
\ee
With the choice of $R^{(0)}(\mu)$ in  \eqref{intformulas}, $\tanh (\pi\nu/2)/\nu$, one can check that the integrals are convergent using the asymptotic properties of the inhomogeneous $Q$-functions, and that the constant $C$ is $2$.

Using these expressions, we can extract the spectral data associated with the homogeneous problem from the inhomogeneous solutions $Q_\pm(\lambda)$. Specifically, we will illustrate in Section \ref{spectraldata} how to extract analytical expressions for the meson masses $\lambda_n$ as well as for spectral sums thereof \eqref{lambaexpspec}.  

\section{\boldmath Meson spectrum at complex quark mass and critical points }
\label{criticality}
In this section we discuss the properties of the eigenvalues $\lambda(\alpha)$ as the masses of the quarks are analytically continued to the complex plane $\alpha\in\mathbb{C}$\footnote{After the first version of this paper appeared in arXiv, we learned that partial results on numerical investigation of the meson spectrum at complex quark mass and criticality was presented by Zamolodchikov in a talk \cite{ZamoTalk}. See \href{https://indico.in2p3.fr/event/1886/sessions/3945/attachments/17798/21781/Zamolodchikov.pdf}{here}.}.
\subsection[Poles of $\Psi(\nu)$ and boundary conditions]{\boldmath Poles of $\Psi(\nu)$ and boundary conditions}
The key players in the following discussions are  zeros of $f(\nu)$ and associated poles of $\Psi(\nu)$. We thus first explain their basic properties.

The starting point is the defining functional equation \eqref{thooftnuspace} that we report here for convenience:
\begin{equation}
  f(\nu)\, \Psi(\nu) =\left(  \frac{2 \alpha}{\pi} + \nu \coth{\frac\pi 2 \nu} \right) \Psi(\nu) =  \lambda(\alpha) \int_{-\infty}^{+\infty}\dd{\nu'} \frac{\pi (\nu - \nu')}{ 2 \sinh{\frac\pi 2 (\nu-\nu')}} \Psi(\nu')
\end{equation}
For $\alpha>-1$, given that the kernel is continuous for any real value, the function $\Psi(\nu)$ has poles only in correspondence with the location of the zeros of $f(\nu)$:
\be \label{zeros}
f(\nu) = \frac{2\alpha}{\pi} + \nu\coth(\frac{\pi\nu}{2}) = 0\comma
\ee
and it is regular anywhere else in the strip $ \Im \, \nu \in [-2i,2i]$.
The zeros of $f(\nu)$ do not admit a closed form expression, but can be shown to lie for $\alpha >-1$ on the imaginary axis $\nu_k = \pm i  u_k$, with $2ki < \nu_k <2(k+1)i$, i.e.\ the the first zero lies in $\nu_0  \in (0,2i)$, the second $\nu_1 \in (2i,4i)$,  and so on. At each of these points $\pm \nu_k$, $ \Psi(\nu)$ displays simple poles. 

The closest poles of $\Psi(\nu)$ to the real axis regulate the exponential decay of the wavefunction in the Fourier dual space of $\nu$, that is the space of rapidities $\theta = \frac{1}{2}\log\frac{x}{1-x}$ as:
\be\label{polesfou}
\phi(\theta) \sim \cF\left[\frac{1}{\nu + i u_0} \pm \frac{1}{\nu - i u_0}\right] = -i \left(\Theta(\theta) \pm \Theta(-\theta)\right)\, e^{- u_0\,\abs{\theta}} \comma \qquad u_0 >0\comma
\ee
where $\Theta(x)$ is the Heaviside step function. The farther poles are subleading contributions. This is consistent with the boundary conditions in $x$-space as $\exp(-u_0\abs{ \theta})\sim x^{u_0/2}$ for $x\sim 0$ and $\exp(-u_0\abs{  \theta})\sim (1-x)^{u_0/2}$ for $x\sim 1$; then the comparison between \eqref{zeros} and \eqref{exponents} (determining the allowed boundary conditions $\beta_i$) demonstrates that  $\beta = \pm u_0/2$ (equal masses). Indeed we see that the $0 \leq\abs{\beta}\leq 1$ is consistent with $0 \leq\abs{u_0}\leq 2$. 
Thus, this argument provides an alternative derivation for the boundary conditions of the wavefunctions in position space, which we derived in \eqref{exponents}. From this perspective the condition $0\leq \beta \leq 1$ simply follows from the position of the closest (dominant) pole.

Instead, choosing non-regular boundary conditions for $\phi(x)$ corresponding to $\beta<0$, produce exponentially divergent wavefunctions $e^{u_0\,\abs{\theta}}$. These non-normalizable wavefunctions do not correspond to any solution of the spectral problem \eqref{thooftnuspace} (it is not possible to produce any such exponentially divergent solution by Fourier transform of any $\Psi(\nu)$). They correspond in $\nu$-space to solutions of the problem outside of $\mathbb{L}^2$.
\subsection{Criticality in the chiral limit}
In the limit where the chiral symmetry becomes exact $\alpha \to -1$, the poles of $\Psi(\nu)$ surface to the real axes and degenerate into a double zero,  $\nu_0(-1) = 0$:
\be
   \pm  \nu_0(\alpha \to -1) = \pm \frac{2i }{\pi}  \sqrt{\alpha}\, \sqrt{\alpha +1} \to 0 \period 
\ee
Hence, $\Psi(\nu)$, that was regular on the real axis for any $\alpha >-1$, now becomes singular at the origin in the chiral limit. Indeed, we already saw in section \ref{subsec:spectral} that the eigenfunction for the lowest eigenvalue is singular in the chiral limit being proportional to the $\delta(\nu)$ and is degenerate with $\delta'(\nu)$. We can also recover this results from the qualitative analysis of the previous subsection: the ground state $\Psi_0(\nu)$ must have no zeros on the real axis and parity even. So, taking the $+$ sign in \eqref{polesfou} and sending $u_0\to 0$, we indeed recover $\Psi_0(\nu) = \delta(\nu)$. To obtain the other solution $\delta^{\prime}(\nu)$, we use the fact that $f(\nu)$ now has a double zero at the origin and therefore $\Psi(\nu)$ can have a double pole. This leads to $\phi(\theta) \sim \cF\left[\frac{1}{(\nu - i \epsilon)^2} + \frac{1}{(-\nu + i \epsilon)^2}\right] =  \theta \, e^{-\epsilon\abs{ \theta}} $, corresponding to $\Psi(\nu) = \delta'(\nu)$, that is logarithmically divergent in $x$-space at the boundaries.
\begin{figure}[htbp]
  \centering
  \begin{subfigure}{0.3\linewidth}
    \includegraphics[width=\linewidth]{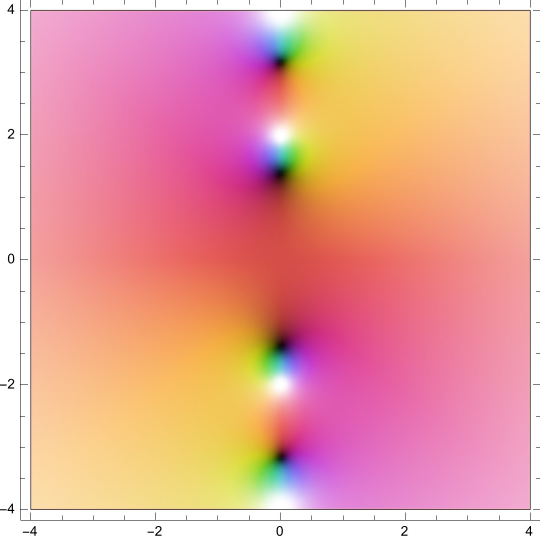}
    \caption{$\theta = 0$}
  \end{subfigure}
  \hfill
  \begin{subfigure}{0.3\linewidth}
    \includegraphics[width=\linewidth]{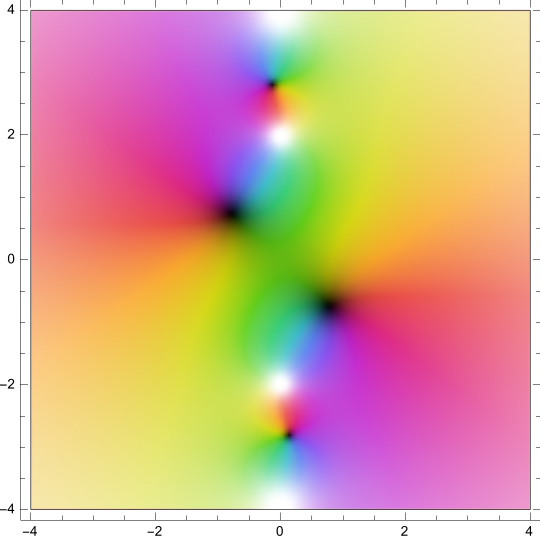}
    \caption{$\theta = 2/5$}
  \end{subfigure}
  \hfill
  \begin{subfigure}{0.3\linewidth}
    \includegraphics[width=\linewidth]{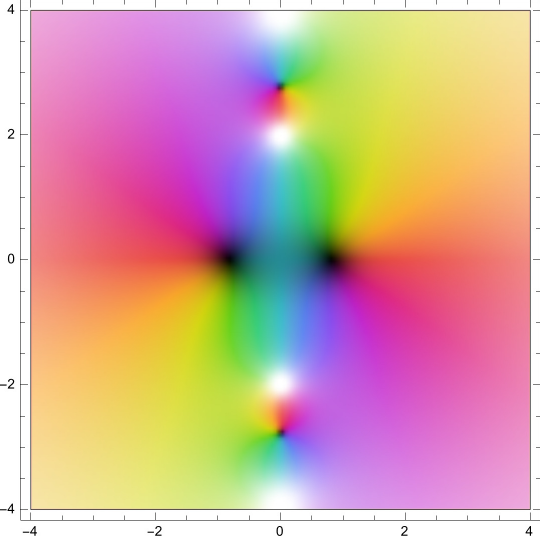}
    \caption{$\theta^* = 1/2$ 
    }
    \label{subfig:complex_plotsc}
  \end{subfigure}
\par\bigskip
  \begin{subfigure}{0.45\linewidth}
  \centering
    \includegraphics[width=0.67\linewidth]{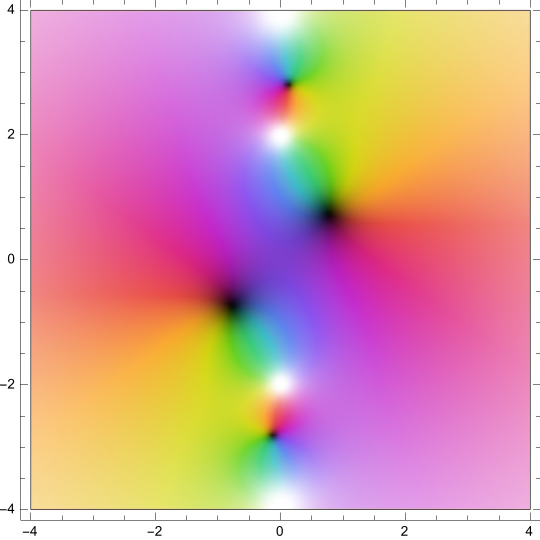}
    \caption{$\theta = 3/5$}
  \end{subfigure}
  \hfill
  \begin{subfigure}{0.45\linewidth}
  \centering
 \includegraphics[width=0.67\linewidth]{fig/complex_plot_alpha1.5_0}  
 \caption{$\theta = 1$} 
  \end{subfigure}
  \caption{Positions of the zeros (black) and poles (white) of $f(\nu) = 1.5 + \nu\coth(\frac{\pi\nu}{2})$ in the  $\nu$-complex plane along the curve $\gamma_{1.5}(\theta) = 1.5\, e^{2i \pi \theta}$. The contour encloses the point $\alpha = -1$ and indeed the two zeros are exchanged upon analytical continuation. In \figref{subfig:complex_plotsc} the zeros cross the real axis in $\nu^* = \pm 0.819864$\, .  } 
  \label{fig:complex_plots1.5}
\end{figure}

To further explore analytical properties of the eigenvalues close to this chiral point, it is useful to analytically continue the eigenproblem \eqref{thooftnuspace}, which was initially defined only for $\mathbb{R} \ni \alpha\geq -1$, to the complex plane of the masses $\alpha$. For any real $\alpha> -1$, we consider the analytic continuation performed along a closed contour in the complex plane:
\be
\gamma_\alpha(\theta): [0,2\pi]\to \mathbb{C}, \qquad  \gamma_{\alpha}(2\pi) =  \gamma_{\alpha}(0) = \alpha.
\ee
If $\gamma_\alpha(\theta)$ is such that it encloses the point $\alpha = -1$, the zeros of $f(\nu)$ in $\nu_0(\alpha) = \pm i u_0(\alpha)$ are exchanged along this contour 
\begin{equation}
   \widetilde{  u_0}(\alpha) =  u_0(\gamma_{\alpha}( 2\pi)) = -u_0(\gamma_{\alpha}(0)) =  -u_0(\gamma_\alpha(0)) =  -u_0(\alpha),\\
\end{equation}
where we have denoted by $\widetilde{g}$ the value of the function $g$ after the analytic continuation. In particular, this means that, at some time $\theta^*$,  the leading zeros of $f(\nu)$, $ \pm \nu_0= \pm iu_0(\alpha (\theta))$, cross the real axis. 
\begin{figure}[htbp]
  \centering
  \begin{subfigure}{0.3\linewidth}
    \includegraphics[width=\linewidth]{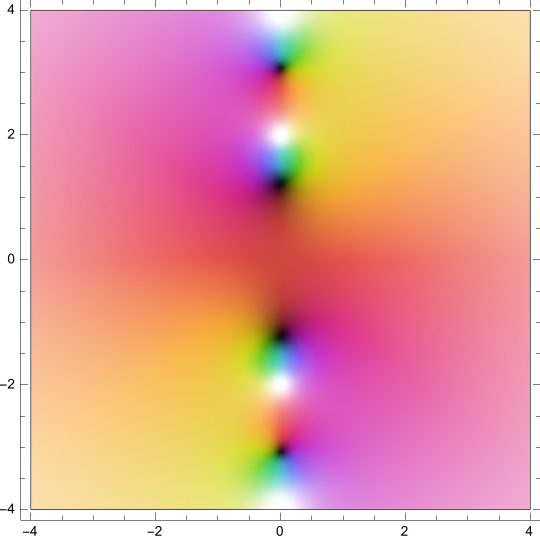}
    \caption{$\theta = 0$}
  \end{subfigure}
  \hfill
  \begin{subfigure}{0.3\linewidth}
    \includegraphics[width=\linewidth]{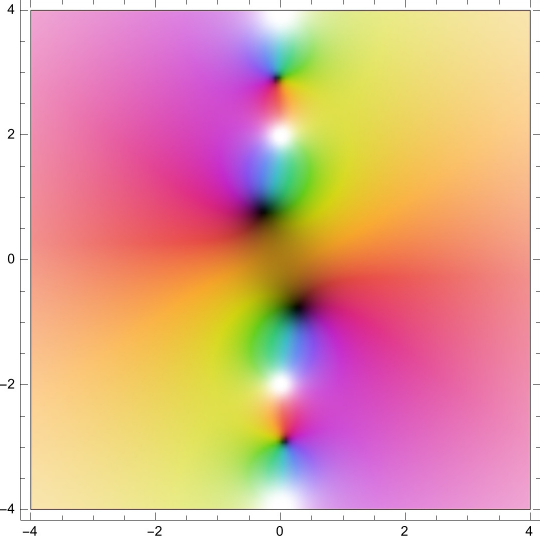}
    \caption{$\theta = 2/5$}
  \end{subfigure}
  \hfill
  \begin{subfigure}{0.3\linewidth}
    \includegraphics[width=\linewidth]{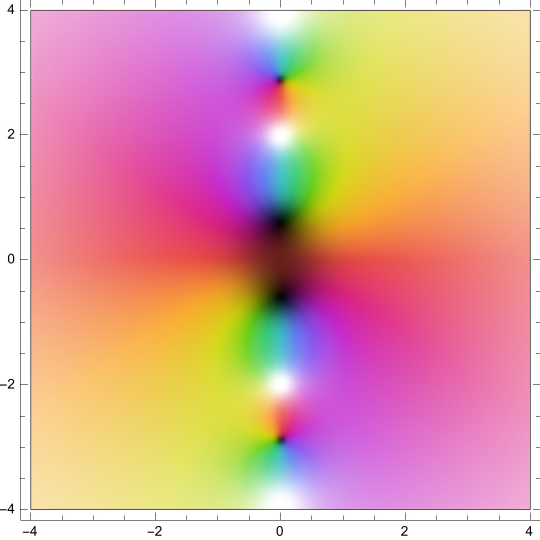}
    \caption{$\theta = 1/2$  }
  \end{subfigure}
\par\bigskip
  \begin{subfigure}{0.45\linewidth}
  \centering
    \includegraphics[width=0.67\linewidth]{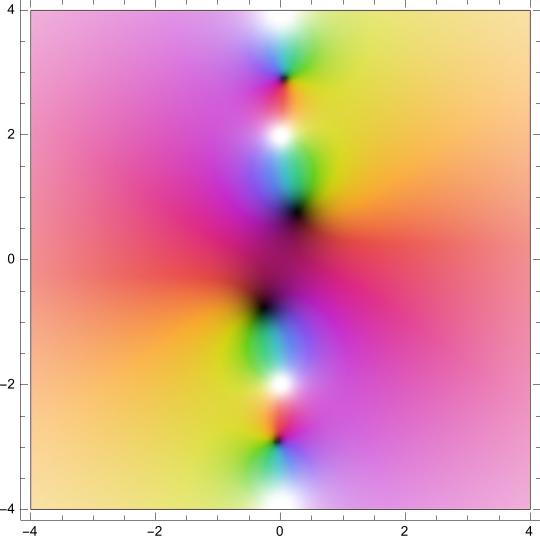}
    \caption{$\theta = 3/5$}
  \end{subfigure}
  \hfill
  \begin{subfigure}{0.45\linewidth}
  \centering
 \includegraphics[width=0.67\linewidth]{fig/complex_plot_alpha0.7_0}  
 \caption{$\theta = 1$} 
  \end{subfigure}
  \caption{Positions of the zeros (black) and poles (white) of $f(\nu) = 0.7 + \nu\coth(\frac{\pi\nu}{2})$ in the  $\nu$-complex plane along the curve $\gamma_{0.7}(\theta) = 0.7\, e^{2i \pi \theta}$. The contour does not enclose the point $\alpha = -1$ and the two zeros are not exchanged and do not cross the real axis.} 
  \label{fig:complex_plots0.7}
\end{figure}
 As a result, at $\theta = \theta^*$, the eigenfunction $\Psi(\nu; \alpha(\theta^*))$ will display a pair of simple poles\footnote{One can see this also from the fact  that zeros of $f(\nu)$ at generic complex values of $\alpha$ do not lie anymore on the imaginary axis, and for $\alpha < -1$, two of them are always reals, cf.\ \figref{subfig:complex_plotsc}.} on the real axis at $\nu = \pm \nu_0$.  
Instead, if the contour does not enclose $\alpha = -1$,  $ \widetilde{  u_0}(\alpha) = u_0(\alpha)$. The two situations are exemplified in \figref{fig:complex_plots1.5} and \figref{fig:complex_plots0.7}. 

\begin{figure}[htbp]
  \centering
  \begin{subfigure}{0.3\linewidth}
    \includegraphics[width=\linewidth]{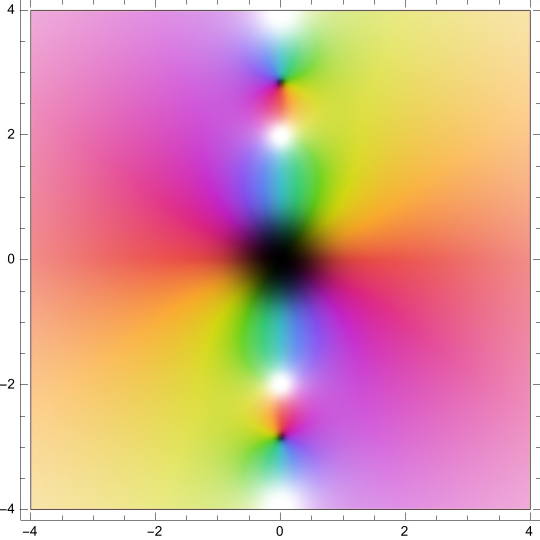}
    \caption{$\theta = 0$}
  \end{subfigure}
  \hfill
  \begin{subfigure}{0.3\linewidth}
    \includegraphics[width=\linewidth]{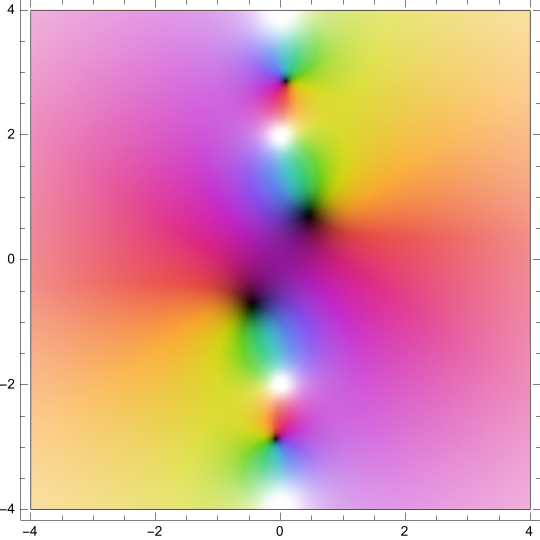}
    \caption{$\theta = 1/5$}
  \end{subfigure}
  \hfill
  \begin{subfigure}{0.3\linewidth}
    \includegraphics[width=\linewidth]{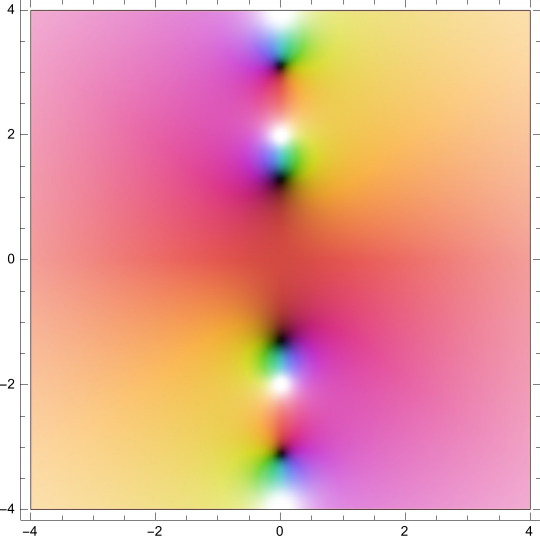}
    \caption{$\theta = 1/2$  }
  \end{subfigure}
\par\bigskip
  \begin{subfigure}{0.45\linewidth}
  \centering
    \includegraphics[width=0.67\linewidth]{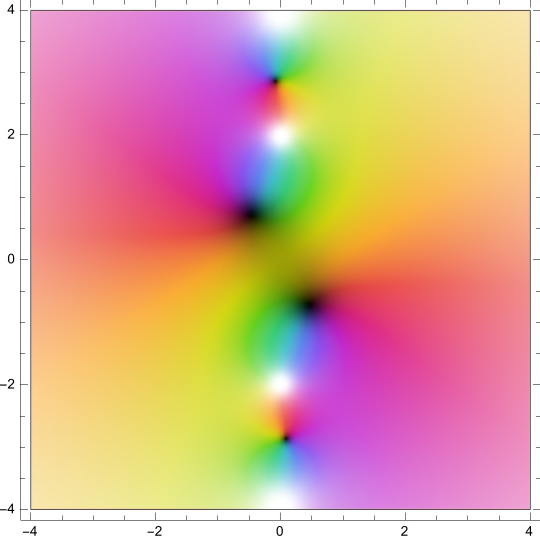}
    \caption{$\theta = 9/5$}
  \end{subfigure}
  \hfill
  \begin{subfigure}{0.45\linewidth}
  \centering
 \includegraphics[width=0.67\linewidth]{fig/complex_plot_alpha-1_0}  
 \caption{$\theta = 1$} 
  \end{subfigure}
  \caption{Positions of the zeros (black) and poles (white) of $f(\nu) = -1 + \nu\coth(\frac{\pi\nu}{2})$ in the  $\nu$-complex plane along the curve $\gamma_{-1}(\theta) = (-1) \, e^{2i \pi \theta}$. The zero in $\nu = 0$ is double. The analytical continuation along a closed contour starting and ending at $\alpha = -1$ do not exchange the position of the zeros. While for any $\alpha = -1 + \epsilon$, $\epsilon\in \mathbb{C}$, we have either the situation in \figref{fig:complex_plots1.5} or \figref{fig:complex_plots0.7}. } 
  \label{fig:complex_plots-1}
\end{figure}

Thus, as one analytically continue around the point $\alpha = -1$, the bound state equation is modified by picking up the pole contributions\footnote{This is quite analogous to excited state Thermodynamic Bethe ansatz equation discussed in \cite{Dorey:1996re}.} coming from $\nu = \pm \nu_0$: 
\begin{equation}
\begin{split}
\label{eq:falpha}
&\left(  \frac{2 \alpha}{\pi} + \nu \coth{\frac\pi 2 \nu} \right) \widetilde{\Psi}(\nu;\alpha ) = \widetilde{\lambda}(\alpha )\int_{-\infty}^{+\infty}\dd{\nu'} \frac{\pi (\nu - \nu')}{ 2 \sinh{\frac\pi 2 (\nu-\nu')}} \widetilde{\Psi}(\nu';\alpha ) +\\ &+  \widetilde{\lambda}(\alpha) \text{Res} \left( \frac{\pi (\nu - \nu')}{ 2 \sinh{\frac\pi 2 (\nu-\nu')}} \Psi(\nu';\alpha) , \nu' = -\nu_0\right) + \\ 
&-    \widetilde{\lambda}(\alpha) \text{Res} \left( \frac{\pi (\nu - \nu')}{ 2 \sinh{\frac\pi 2 (\nu-\nu')}} \Psi(\nu';\alpha) , \nu' = \nu_0\right)
\end{split}
\end{equation}
The different sign depends on the opposite orientation with which two poles cross the real axis. Furthermore, we can readily see, that performing yet another analytic continuation around $\alpha = -1$, the two poles cross the real axis swap again, this time with opposite phases. Henceforth the two-folded analytically continued bound state equation:
\be
    \left(  \frac{2 \alpha}{\pi} + \nu \coth{\frac\pi 2 \nu} \right)\widetilde{\widetilde{\Psi}}(\nu;\alpha) = \widetilde{\widetilde{\lambda}}(\alpha )\int_{-\infty}^{+\infty}\dd{\nu'} \frac{\pi (\nu - \nu')}{ 2 \sinh{\frac\pi 2 (\nu-\nu')}} \widetilde{\widetilde{\Psi}}(\nu'). 
\ee
has exactly the same form of the original one, and therefore upon analytic continuation twice around the point $\alpha = -1$, the eigenvalue $\lambda(\alpha) = \widetilde{\widetilde{\lambda}}(\alpha )$.  This means that as a function of complex $\alpha$, the energy of the bound state has a non-trivial two-sheeted analytic structure. 
In the limit $\alpha \to -1$, the two poles $\pm \nu_0$ collide on the real axis and the
two eigenvalues $\lambda (\alpha)$ and $\widetilde{\lambda}(\alpha)$ become degenerate (See \figref{fig:complex_plots-1}). This means that there is a square-root branch cut starting from $\alpha= -1$, and the lowest eigenvalue near $\alpha=-1$ behaves as 
\be \label{firsteig}\lambda_0(\alpha) \sim \sqrt{\alpha+1}\, .  \ee 
As already pointed out in \cite{Fateev_2009}, this behavior of the eigenvalue is reminiscent of critical points in phase transitions. In fact, as we see below, the infrared phase of the massless 't Hooft model ($\alpha=-1$) is described by conformal field theory, much like critical points in phase transitions. 

\subsection{Massless pion revisited}
\label{masselesspoint}
As we saw above, at $\alpha=-1$, the lowest meson has zero mass. In the literature, this is sometimes considered as a consequence of exact chiral symmetry and its spontaneous breaking. Strictly speaking, this is incorrect since the continuous global symmetry cannot be spontaneously broken in two dimensions \cite{Mermin:1966fe,Coleman:1973ci}. However, there are large $N$ examples \cite{Gross:1974jv} which exhibit the spontaneous continuous symmetry breaking in the strict $N\to \infty$ limit. In those examples, the ``Goldstone bosons" at large $N$ acquire mass at subleading orders in the $1/N$ expansion. Based on this fact, it was sometimes conjectured that the massless meson at $\alpha=-1$ will acquire mass because of the $1/N$ effects.


This conjecture is incorrect either. This follows from the recent analysis of the IR structure of 2d QCD  \cite{Delmastro:2021otj, AFFLECK1986448}. As shown there, 
 the IR phase of $SU(N)$ YM coupled to $N_f$ flavor of fundamental quarks can be described as a gauged WZW model where one gauges the $SU(N)_{N_f}$ symmetry acting on the $SO(N\, N_f)_1$ WZW model describing the free fermion Lagrangian:
\be
\cT_{IR} \equiv \frac{SO(N\, N_f)_1}{SU(N)_{N_f}} = U(N_f)_N\, .
\ee
In particular, setting $N_f=1$, their analysis predicts the persistence of a free massless boson realizing a $U(1)_N$ current algebra\footnote{ For $N_f>1$, a single $U(1)_N$ current  corresponding to a massless boson decouples from the the remaining interacting $SU(N_f)_N$  WZW model. This is reminiscent of the ``experimental" observations made in \cite{Krauth_1996} (also in \cite{BROWER1977131}) that 3-point (and higher) functions involving a massless meson and other massive mesons identically vanish. It would be interesting to provide a theoretical understanding for this observation.} at level $N$.

\subsection{Critical points on the second sheet}
\begin{figure}[htbp]
  \centering
  \begin{subfigure}{0.3\linewidth}
    \includegraphics[width=\linewidth]{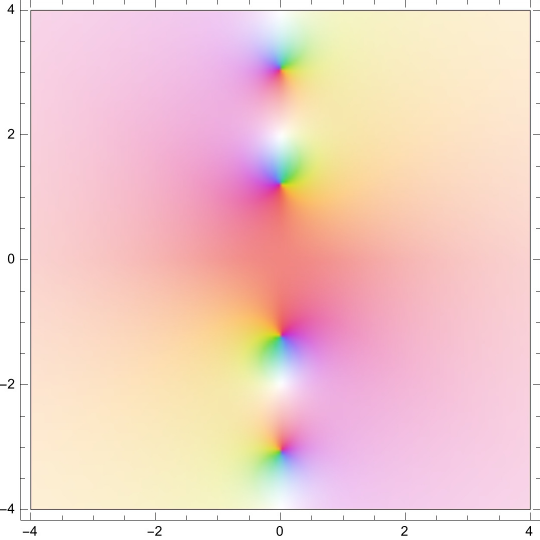}
    \caption{$\alpha = 0.7$}
  \end{subfigure}
  \hfill
  \begin{subfigure}{0.3\linewidth}
    \includegraphics[width=\linewidth]{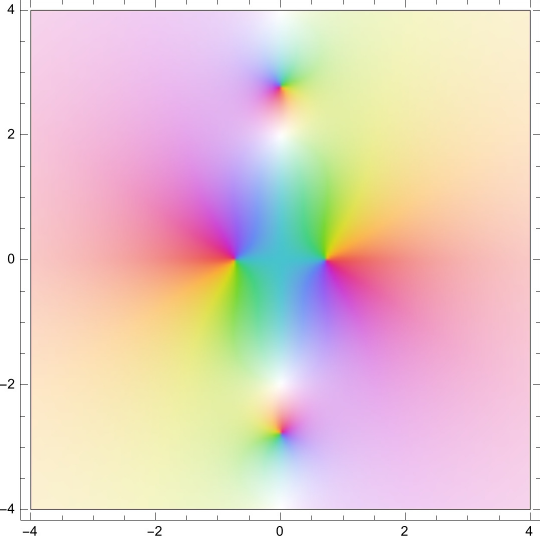}
    \caption{$\alpha = -1.4$}
  \end{subfigure}
  \hfill
  \begin{subfigure}{0.3\linewidth}
    \includegraphics[width=\linewidth]{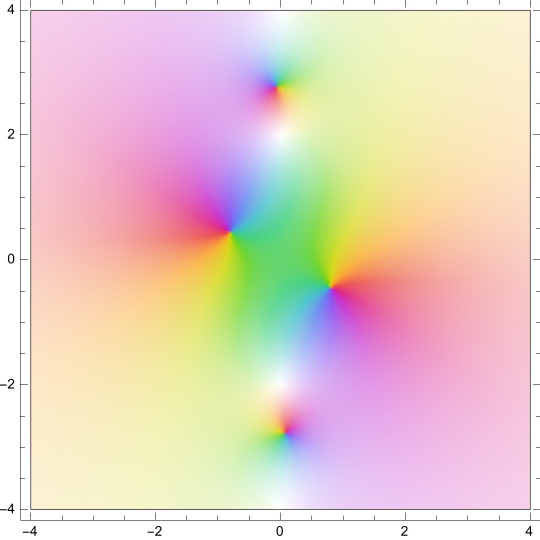}
    \caption{$\alpha = -1.4 + 0.5 i  $  }
  \end{subfigure}
\par\bigskip
  \centering
  \begin{subfigure}{0.3\linewidth}
    \includegraphics[width=\linewidth]{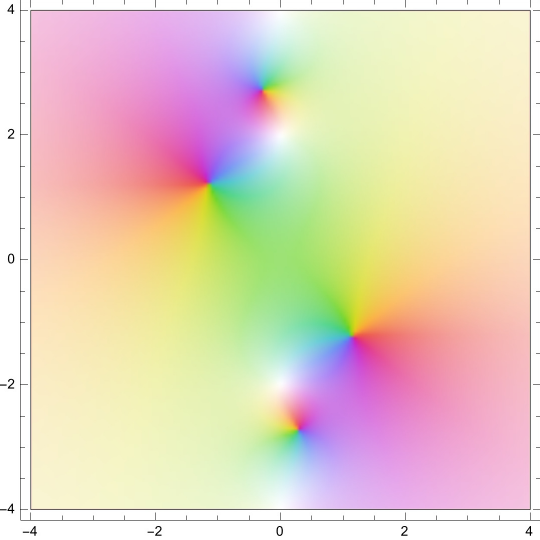}
    \caption{$\alpha = -1.67 + 1.9 i $}
  \end{subfigure}
  \hfill
  \begin{subfigure}{0.3\linewidth}
    \includegraphics[width=\linewidth]{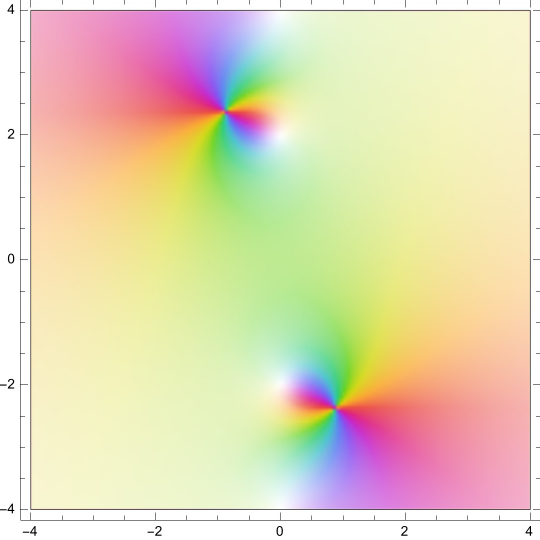}
    \caption{$\alpha = \alpha_1$}
    \label{subfig:alpha1}
  \end{subfigure}
  \hfill
  \begin{subfigure}{0.3\linewidth}
    \includegraphics[width=\linewidth]{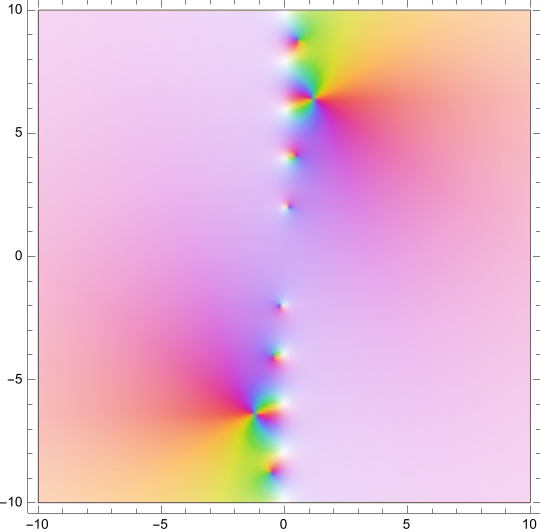}
    \caption{$\alpha= \alpha_3$  }
    \label{subfig:alpha3}
  \end{subfigure}
  \caption{Positions of the zeros (black) and poles (white) of $f(\nu) = \alpha + \nu\coth(\frac{\pi\nu}{2})$ in the  $\nu$-complex plane for different values of $\alpha$. As we vary $\alpha$, the first zeros in the upper and lower half plane wander in the complex plane and at the specific values $\alpha = \alpha_k$, they degenerate into double zeros. In \figref{subfig:alpha1} $\mp u_0$ degenerates with $\pm u_1$  at $\alpha = \alpha_1$ while in  \figref{subfig:alpha3},  at $\alpha = \alpha_3$,  $\mp u_0$ degenerates with $\pm u_3$  } 
  \label{fig:complex_plotshigher}
\end{figure}

\begin{figure}[htbp]
\centering
\begin{subfigure}[b]{0.23\linewidth}
\centering
    \includegraphics[width = \linewidth]{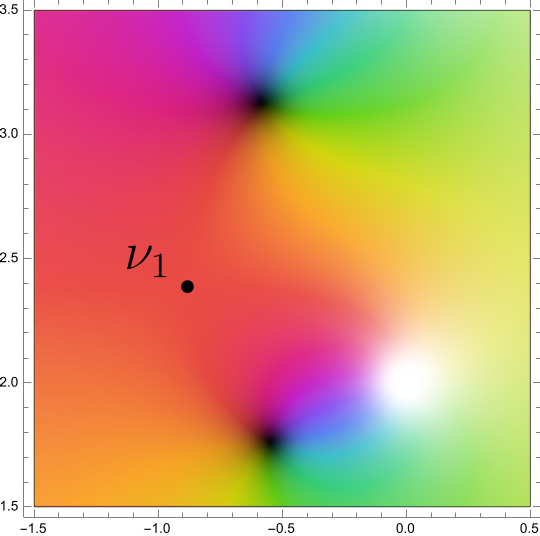}
    \caption{$\theta_1 = 0$}
\end{subfigure}\hfill
    \begin{subfigure}[b]{0.23\linewidth}
    \includegraphics[width = \linewidth]{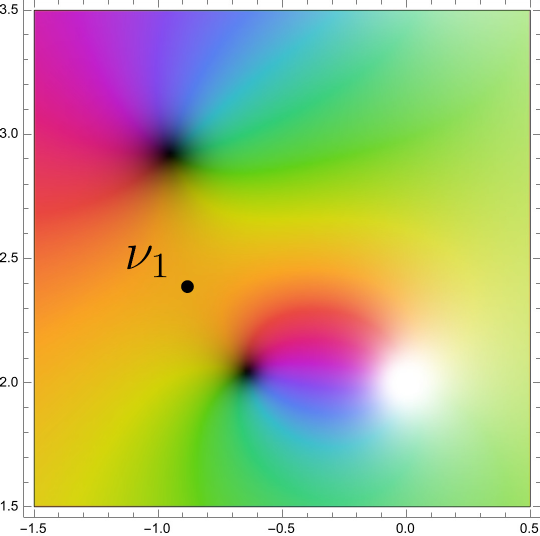}
    \caption{$\theta_1 = 2/5$}
\end{subfigure}\hfill
\begin{subfigure}[b]{0.23\linewidth}
    \includegraphics[width = \linewidth]{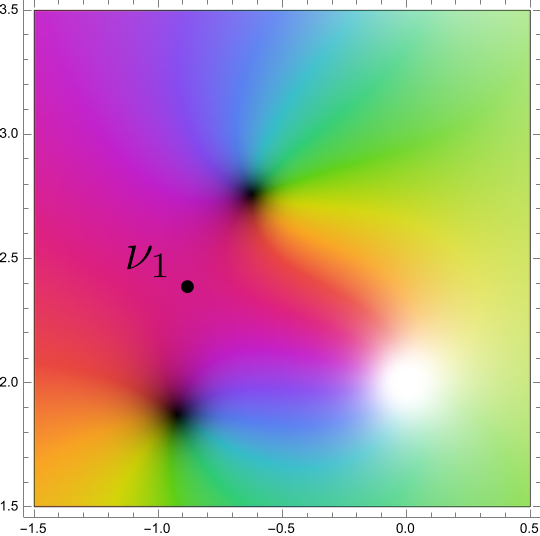}
    \caption{$\theta_1 = 3/5$}
\end{subfigure}\hfill
\begin{subfigure}[b]{0.23\linewidth}
    \includegraphics[width = \linewidth]{fig/complex_plot_alpha_second_pole_notex0}
    \caption{$\theta_1 = 1$}
\end{subfigure}\hfill
\par\bigskip
\centering
\begin{subfigure}{0.23\linewidth}
    \includegraphics[width = \linewidth]{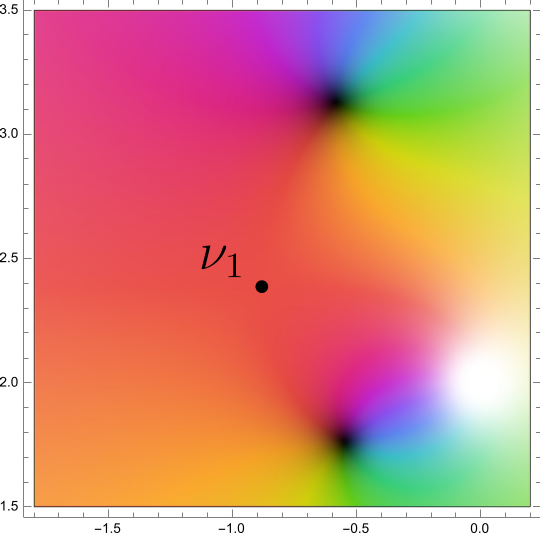}
    \caption{$\theta_2 = 0$}
\end{subfigure}\hfill
    \begin{subfigure}{0.23\linewidth}
    \includegraphics[width = \linewidth]{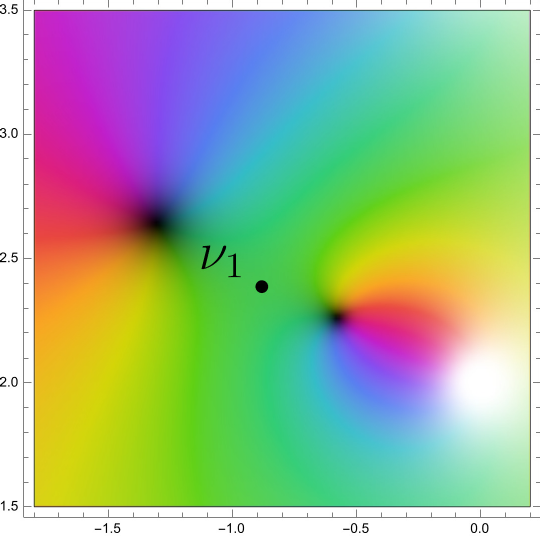}
    \caption{$\theta_2 = 9/20$}
\end{subfigure}\hfill
\begin{subfigure}{0.23\linewidth}
    \includegraphics[width = \linewidth]{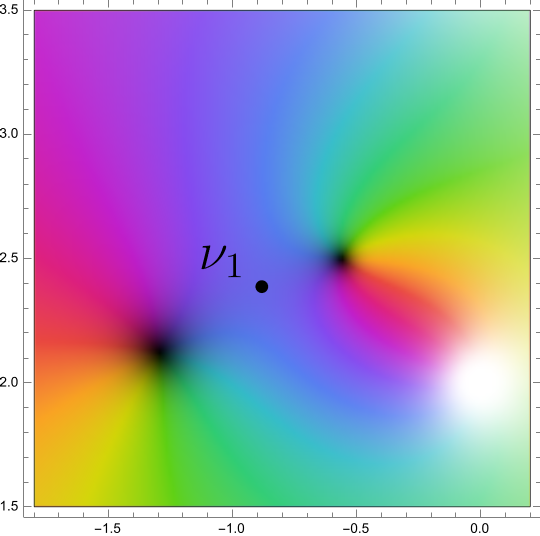}
    \caption{$\theta_2 = 11/20$}
\end{subfigure}\hfill
\begin{subfigure}{0.23\linewidth}
    \includegraphics[width = \linewidth]{fig/complex_plot_alpha_second_pole_ex0}
    \caption{$\theta_2 = 1$}
\end{subfigure}\hfill
\caption{Positions of the zeros (black) and poles (white) of $f(\nu) = a_i + \nu\coth(\frac{\pi\nu}{2})$ for $a_1 = \alpha_1 + 1$ and $a_2= \alpha_1 + 0.7$ in the  $\nu$-complex plane along the curve respectively  $\gamma_{a_1}(\theta_1) = a_1 +  \, 0.7\, e^{2i \pi \theta_1}$ (Figures 5a - 5d ) and  $\gamma_{a_2}(\theta_2) = a_2 +  \, \, e^{2i \pi \theta_2}$ (Figures 5e - 5h). The point marked with $\nu_1 = \frac{2}{\pi} \sqrt{\alpha_1}\, \sqrt{\alpha_1 +1}$ is the position of the double zero \eqref{nukdouble} corresponding with the critical value $\alpha_1 \simeq  -1.895 + 3.719\,i$.   The analytical continuation along a closed contour starting and ending at $a_i$ exchanges the position of the zeros if and only if the contour circulates around $\alpha_1$ as clear from the figure. }
\label{anacont}
\end{figure}

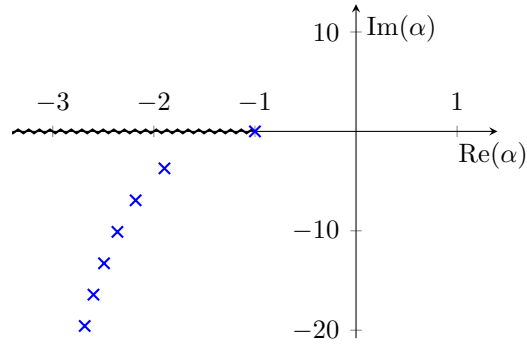
\begin{figure}[htbp]
  \centering
  \begin{tikzpicture}
    \begin{axis}[
      xlabel={$\Re (\alpha) $},
      ylabel={$\Im (\alpha)$},
      axis lines=middle,
      enlargelimits,
      width=8cm,
      height=6cm,
      font=\footnotesize,
      smooth,
      xmin=-3,
      xmax=1,
      ymin=-18,
      ymax=10,
      xticklabel style={
        yshift=4ex,
      },
      xlabel style={
        yshift=-3.5ex,
        xshift = 3ex,
      },
    ]
      \draw [decorate,thick, decoration={zigzag,segment length=4,amplitude=0.7,post=lineto,post length=0}] (-4,0) -- (-1,0);
      \pgfplotstableread[col sep=comma]{fig/points.csv}\datatable
            \addplot[blue, mark=x, only marks, mark size=3pt, mark options={thick}] table {\datatable};
    
    \end{axis}
  \end{tikzpicture}
  \caption{Square root branching points of $\lambda$ in the second sheet of the $\alpha$-plane in correspondence with the points $\alpha_k$ (in blue).}
   \label{fig:lambdacomplex}
\end{figure}

Interestingly, the point at $\alpha = -1$ is not the only singularly of $\lambda(\alpha\in\mathbb{C})$ in the complex plane of the masses. Indeed, $f(\nu)$ develops double  zeros whenever:
\begin{equation}\label{eq:defalknuk}
    \begin{cases}
        \alpha_k =  - \frac{\pi}{2} \nu_k \coth(\frac{\pi}{2}\nu_k)\\
        \pi \nu_k = \sinh(\pi \nu_k)
    \end{cases} \Longrightarrow \alpha_k = -\frac{1}{2} (1+ \cosh{\pi \nu_k})\, . 
\end{equation}
Those are infinite values of $\alpha_k$ all located on the second sheet of the $\alpha$-plane ($\arg{\alpha}> i \pi$ ), beyond the cut $(-\infty ,-1)$, see \figref{fig:lambdacomplex}. At each of these $\alpha_k$, there are two values\footnote{The equation \eqref{nukdouble} can be obtained by inverting the two equations on the left hand side of \eqref{eq:defalknuk}.} of $\nu_k= i u_k$
    \be
    \label{nukdouble}
     \frac{\pi}{2}\nu_k =  \frac{i\pi}{2} u_k= \pm \sqrt{\alpha_k}\,\sqrt{\alpha_k +1}\, ,
\ee
one in the upper half-plane and one in the lower half-plane, which degenerate to $\pm \nu_0$. See \figref{fig:complex_plotshigher}.
 An argument analogous to the one used for $\alpha_0 = -1$ implies that all those $\alpha_k$ are square-root branch cut singularities of the eigenvalues. Indeed, as before, one checks directly that upon analytically continuing around a closed curve, the zeros of the zeros $u_0$ and $u_{k\geq 1} $ are exchanged if and only if the curve circles around $\alpha_k$ an odd number of times. In \figref{anacont} we illustrate this for $\alpha = \alpha_1$.  

However, there is one point that needs to be taken into account: for odd $n$, $\Psi_n(\nu)$ is odd, and therefore its residues at $\nu\pm \nu_k$ have the same sign. Then, in this case, the residues produced by analytic continuation around $\alpha_k$ exactly cancel with each other. We thus conclude that as function of complex $\alpha$ they are single valued. By contrast, for even $n$,  the eigenvalues will have square-root type singularities:
\be
    \lambda_{2k}(\alpha) \sim \sqrt{\alpha - \alpha_k}
\ee
These singularities are all in the second sheet and accumulate towards $\infty$. So, as a function of complex $\alpha$, each of the even eigenvalues $\lambda_{2n}$ have a square root branch point in $\alpha_n$. For example in \figref{fig:lambdacomplex}, the complex plane of $\lambda$ is represented with a branch cut along $(-\infty,-1)$ in correspondence with the branch cut of $\lambda_0$. If we consider instead $\lambda_{2n}$, then there is  branch cut on the second sheet at $\alpha = \alpha_k$. 
We conclude that, in the complex plane of the masses we find infinitely many critical points where (one of the) even eigenvalues vanish as a square root (i.e.\ with the same ``critical exponent" of the point as $\alpha = -1$). The physical meaning of those critical point in the complex plane is yet to be fully understood, and later in this subsection, we try to gain some intuition by comparing with the case of Ising Field Theory.

\paragraph{A simple WKB analysis.}
The nature of those critical point in the second sheet is inherently connected with the mechanism of confinement in two dimensions. 
 Consider a simple toy model of a pair of quarks bounded by a linear potential in their center of mass frame. The Hamiltonian is (cf.\ \cite{fonseca2006ising} and references therein, or \cite{Lencs_s_2022} for a modern perspective) is
\be 
\mathcal{H} =\sqrt{p_1^2 + m^2} + \sqrt{p_2^2+ m^2} + \sigma \abs{x_1-x_2}\, ,  \qquad p_1 = -p_2 := p \period
\ee
The semiclassical analysis of this model follows from imposing the Bohr-Sommerfeld quantization on the  phase space $2 \oint p dq = 2\pi (2n  - 1/2)$ (fermions statistic) on the relative momentum $p$ parametrized as $\sqrt{p^2 + m^2} = 2m\cosh\theta$, one obtain the following WKB spectrum for the masses:
\be
M_n = m \cosh\theta_n , \, \qquad \sinh2\theta_n - 2\theta_n = \frac{\pi\sigma}{m^2} \left(n - \frac{1}{4} \right)\, .
\ee
As we see, there are infinitely many critical points in the complex $\frac{\sigma}{m^2}$ plane at which $\theta_n =\frac{i\pi}{2}$ and $M_n $ vanishes. The analysis of this paper goes beyond this simple WKB analysis, but confirm the qualitative features thereof. 

\paragraph{Comparison with Ising Field Theory.}
\label{IFTmesons}
 The small magnetic field approximation  of Ising Field Theory shares many qualitative features of the spectrum of mesons as the 't Hooft model \cite{fonseca2006ising}. Also in the case of IFT, we find that the mass of the lowest energy mesons has a square-root singularity of the form $M^2_1(\eta) \sim \sqrt{\eta- \eta^*}$ with $\eta = \frac{m}{\abs{h}^{8/15}}$ ($m$ is the mass of the quark in the IFT Lagrangian). If we go under the branch cut in the $\eta$-plane, there is a infinite tower of square root singularities at which the higher meson states become massless. 
 
To understand the physics at those critical points in either of the models, one should construct a systematic approximation of finite $N$ for QCD$_2$ and $h$ for IFT, and see what the fate of those critical points is. Such an analysis is yet to be performed for the 't Hooft model. Similarly, the corresponding analysis for IFT is also at an early stage. Still, it is well-known that the first of those singular points, the one at which $M_1$ is massless, corresponds in the full theory to the Yang-Lee edge singularity at $h = \pm i (0.1893\cdots) m^{15/8}$. In this case, the exponent in the full theory is different from the one in the small $h$-approximation: $M_1(\eta) \sim (\eta-\eta_{\rm YL})^{5/12}$. The CFT associated with the Yang-Lee critical point is the non-unitary minimal model $\cM_{2/5}$ (cf.\ \cite{Xu_2022} for a recent discussion). It will be important to identify CFTs corresponding to critical points on the second sheet for the 't Hooft model.
One possible scenario is that a refined analysis for finite number of colors will result in $M^2_{2k} \sim (\alpha - \alpha_k)^{1/2 + \mathcal{O}(1/N)}$, which would suggestt that the critical points are described by non-trivial interacting CFTs (this scenario was also contemplated at the end of \cite{Fateev_2009}). 
 
\section{Extracting spectral data}
\label{spectraldata}
In this section, we will discuss how to extract the spectrum of the eigenproblem \eqref{thoofteq}, i.e.\ the energy levels of the mesons,  from the reformulation of the latter in terms of the TQ-Baxter system \eqref{tqinhom}. In Section \ref{sec:spetrumana} will construct an asymptotic expansion for the eigenvalues $\lambda_n$, while in \ref{sec:spectralsums}, we will derive exact expressions for the spectral sums $G^{(s)}_\pm(\alpha)$ of \eqref{lambaexpspec}. 

\subsection{Solutions to TQ-system in terms of hypergeometric functions}
\label{nonanasol}
As the first step, we generalize the discussion in \cite{Fateev_2009} for $\alpha=0$ and demonstrate that one can construct analytic expression for two independent solutions to the TQ-system \eqref{tqinhom}
\be 
Q_\pm(\nu+2i|\lambda) + Q_\pm(\nu+2i|\lambda)- 2Q_\pm(\nu|\lambda) = \frac{-4\pi\lambda }{\frac{2\alpha}{\pi} + \nu\coth(\frac{\pi\nu}{2}) } Q_\pm(\nu|\lambda)
\ee
once we weaken the analyticity requirements.

The idea is to use the two standard solutions ${\rm U}(a,b,x)$, ${\rm M}(a,b,x)$ of the confluent hypergeometric differential equation (${\rm M}(a,b,x)$ is sometimes referred to as ${}_1 {\rm F}_1(a,b,x)$):
\begin{align}
 \label{eq:hym} \left(\nu + \frac{2\alpha z}{\pi}  \right) {\rm M}&\left(1+ \frac i 2 \left(\nu +  \frac{2\alpha z}{\pi} \right) , 2, - 2\pi i \lambda\, z \right)\; , \\   \label{eq:hyu} \Gamma\left(1+ \frac i 2 \left(\nu + \frac{2\alpha z}{\pi} \right)\right)\, {\rm U}&\left(1+ \frac i 2 \left(\nu + \frac{2\alpha z}{\pi}\right) , 2, - 2\pi i \lambda\,z \right)\; \period
\end{align}
Using the recurrence relations\footnote{See formulas (13.4.1) and (13.4.15) of \cite{abramowitz+stegun}, which we report here, specialized to our case for convenience:  
\begin{equation}
\begin{split}
    \frac{\Gamma(a)}{a-1} \Big[ {\rm{U}} (a-1,2,z)  + a(a-1) {\rm{U}} (a+1,2,z) -2 (a-1) {\rm U } (a,2,z) \Big] &= \frac{z}{a-1} \Gamma(a)\, {\rm U }(a, 2,z)\; ,\\
    (a-2) {\rm M} (a-1,2,z) + a {\rm M} (a+1,2,z) - 2(a-1) {\rm M}(a,2,z) &= \frac{z}{a-1} (a-1) {\rm M}(a,2,z)\; .
    \end{split}
\end{equation} } for these functions, one can show that they satisfy the functional relation \eqref{tqinhom} if we set $z$ to be
\be
z := z(\nu) = z(\nu  \pm 2i) = \tanh(\frac{\pi}{2}\nu)\; .
\ee
While the solution \eqref{eq:hym} is analytical everywhere on the real axis, \eqref{eq:hyu} is not, having a logarithmic branch-cut singularity at $z = 0$,  (formula (13.1.6) of \cite{abramowitz+stegun}): 
\be 
\label{branchhyu}
{\rm U } (a,2,z\sim 0) = \frac{(-1)^{2}}{2\Gamma(a-1)} {\rm M}(a,2,z) \log z  + (\text{reg.\ }) \, .
\ee
From the form of \eqref{branchhyu}, it is evident that an appropriate linear combination of the two functions, \eqref{eq:hym} and \eqref{eq:hyu}, will give a
 single-valued and analytic solution on the real axis. 
Imposing further that the solutions to be parity eigenstates, the parity-odd one is given by
\begin{equation}
\label{eq:mp}
\begin{split}
        {\rm M}_+(\nu|\lambda)&= \left(\nu + \frac{2\alpha z}{\pi}  \right)\,e^{i \pi \lambda z}\,{\rm M}\left(1+ \frac i 2 \left(\nu +  \frac{2\alpha z}{\pi}\right) , 2, - 2\pi i \lambda\, z \right) \\
     &= \left(\nu + \frac{2\alpha z}{\pi}  \right)\,e^{i \pi \lambda z}\, \sum_{n=0}^{\infty} \frac{\left(1+ \frac i 2 \left(\nu +  \frac{2\alpha z}{\pi}\right)\right)_n}{n! (n+1)!} (- 2\pi i \lambda\, z )^n
     \end{split}
\end{equation}
while the parity-even one is ($\psi(x)$ is the Polygamma function)\footnote{The $\log(16)$ is added to simplify the expressions below. It does not modify the analytical structure of the solution.}
\begin{equation}
\label{eq:mm}
\begin{split}
  \phantom{{\rm M}_-(\nu|\lambda)}
&\begin{aligned}
   \mathllap{   {\rm M}_-(\nu|\lambda)}&= \frac{1}{2}\left( e^{i \pi \lambda z}\,\Sigma(\nu|\lambda) +  e^{-i\pi\lambda z}\,\Sigma(-\nu|\lambda)\right)
    \end{aligned}\\
    &\begin{aligned}
   \mathllap{\Sigma(\nu|\lambda)}  = 1+ \sum_{n=1}^{\infty} \frac{\left(\frac i 2 \left(\nu +  \frac{2\alpha z}{\pi}\right)\right)_n}{n! (n-1)!} (- 2\pi i \lambda\, z )^n \bigg[ &\psi\left(\frac{1}{2}\right)  - \psi(n) - \psi(n+1)\\
    &+ \psi\left(n + \frac {i\nu}{2} +  \frac{i \alpha z}{\pi}\right) + \log(16) \bigg]
    \end{aligned}
    \end{split}
\end{equation}
The notation ${\rm M}_\pm$ for the respectively odd/even function may seem misleading but it is chosen so that the resulting $\Psi_\pm(\nu) = \sinh^{-1}\left(\frac{\pi\nu}{2}\right) f^{-1}(\nu) \, Q_\pm(\nu)$ would then be even/odd.

Although the solutions 
${\rm M}_\pm(\nu|\lambda)$ are regular on the real axis of $\nu$, they do not satisfy the required analyticity; they have singularities  in the strip $[-2i,2i]$ at $\nu = \pm i$ where $z(\nu)$ is singular. In particular, the last polygamma term in \eqref{eq:mm} introduces infinitely many singularities that accumulate at $z=\pm i$.
Hence, as they stand,  they do not provide a full solution of our TQ-system \eqref{tqinhom}. Nevertheless, they provide a starting point for our analysis. Below we explain how to systematically correct the solutions and compute spectral sums by expanding in powers of $\alpha$. 

\paragraph{Constructing the analytical solutions.}
Because of the accumulation of singularities introduced by the term $\psi\left(n + \frac {i\nu}{2} +  \frac{i \alpha z}{\pi}\right)$ in \eqref{eq:mm}, at any order in $\lambda$, we treat this polygamma as its further power expansion in the mass parameter:
 \be 
\psi\left(n + \frac {i\nu}{2} +  \frac{i \alpha z}{\pi}\right) = \sum_{n=0}^\infty \psi^{(k)}\(n+ \frac{i\nu}{2}\) \, \(\frac{iz \alpha}{\pi}\)^k \period
 \ee
Then, at any order in this expansion in $\alpha$ and $\lambda$, \eqref{seriesexp} the function:
\begin{equation}
\label{eq:mm2}
\begin{split}
  \phantom{{\rm M}^{(k)}_-(\nu|\lambda)}
&\begin{aligned}
   \mathllap{   {\rm M}^{(k)}_-(\nu|\lambda)}&= \frac{1}{2}\left( e^{i \pi \lambda z}\,\Sigma^{(k)}(\nu|\lambda) +  e^{-i\pi\lambda z}\,\Sigma^{(k)}(-\nu|\lambda)\right)
    \end{aligned}\\
    &\begin{aligned}
   \mathllap{\Sigma^{(k)}(\nu|\lambda)}  = 1+ \sum_{n=1}^{\infty} \frac{\left(\frac i 2 \left(\nu +  \frac{2\alpha z}{\pi}\right)\right)_n}{n! (n-1)!} (- 2\pi i \lambda\, z )^n \bigg[ &\psi\left(\frac{1}{2}\right)  - \psi(n) - \psi(n+1)\\
    &+ \log(16) + \sum_{k=0} \psi^{(k)}\(n+ \frac{i\nu}{2}\) \, \(\frac{z \alpha}{\pi}\)^k   \bigg]
    \end{aligned}
    \end{split}
\end{equation}
exhibits only poles of increasing order in $\nu = \pm i$. 

In order to construct solutions of the TQ-system with correct analyticity, we  consider the following linear combination:
\begin{equation}
\label{seriesexp}
    Q_\pm(\nu;\lambda) = A_\pm(\lambda,z)\,{\rm M}_\pm(\nu;\lambda) + 2 \pi i \lambda^{\frac{1\mp 1}{2}}\,z \, B_\pm(\lambda,z) \, {\rm M}_\mp(\nu;\lambda)
 \end{equation} 
 Although the coefficients depend on $\lambda$ and $z$, this is still a solution to  \eqref{tqinhom}. 

Then, we proceed by expanding both $\rm M_{\pm}$ and the coefficients in powers of $\lambda$ and $\alpha$. Since in the expansion in $\lambda$ at order $\lambda^n$, $\rm M_\pm$ contains already powers of $\alpha^{n+\frac{1\pm 1}{2}}$, we take the expansion in $\alpha$ to be at least of order  $n$ and $n+1$ respectively for $\rm M_+$ and $\rm M_-$.  At each other in this expansion, we fix the functions $A_\pm$ and $B_\pm$ such that the resulting expression is analytic in  $[-2i,2i]$ up to that order given order. This gives a well-defined problem at each order in $\lambda$ and $\alpha$ that can be uniquely solved up to a trivial scaling, fixed by requiring $A_\pm(\lambda,0) =1$ for any order in $\alpha$.  Let us denote:
\be 
A_\pm = \sum_{l=0}^\infty A_\pm^{(l)} \lambda^{l} \comma \qquad B_\pm = \sum_{l=0}^\infty B_\pm^{(l)} \lambda^{l}\comma
\ee
then, each coefficient $A_\pm(l)$, $B_\pm(l)$ is itself given by an infinite series in $\alpha$:
\be 
A_\pm^{(l)} = \sum_{k=0} A^{(l,k)} \alpha^k\comma \quad B_\pm^{(l)} = \sum_{k=0} B^{(l,k)} \alpha^k\period
\ee
In the following we report the first few coefficients in the $\lambda$ expansion up to order $\alpha^4$. Up to $\lambda^2$ and $\alpha^5$ we have, and defining $\tau = \frac{\pi^2}{4}\tanh{\frac{\pi\nu}{2}}$ :
\be 
\begin{split}
A_\pm^{(0)} &= 1 \comma \qquad A^{(1)}_- = \frac{8\alpha \tau}{\pi^2}\comma \\
A_+^{(1)} &= -\frac{8 \alpha \tau}{\pi^2} 
+ \frac{112 \alpha^2 \tau \zeta(3)}{\pi^4} 
+ \alpha^3 \left( \frac{224 \tau^2 \zeta(3)}{\pi^6} 
- \frac{1488 \tau \zeta(5)}{\pi^6} \right) \\
&\qquad \qquad \qquad + \alpha^4 \left( 
\left( -\frac{992 \tau}{3 \pi^6} - \frac{3968 \tau^2}{\pi^8} \right) \zeta(5) 
+ \frac{20320 \tau \zeta(7)}{\pi^8} \right)\\
A_-^{(2)} &= -\frac{8 \alpha  \tau }{\pi ^2}-\tau + \alpha ^2 \(\frac{16 \tau ^2}{\pi ^4}+\frac{56 \tau  \zeta (3)}{\pi ^4} \) +\alpha ^3 \left(\frac{112 \tau  \zeta (3)}{3 \pi ^4}-\frac{992 \tau  \zeta (5)}{\pi ^6}\right) + \\
&\qquad \qquad\qquad \qquad +\alpha ^4 \left(\frac{15240 \tau  \zeta (7)}{\pi ^8}-\frac{992 \tau  \zeta (5)}{\pi ^6}\right) \\
A^{(2)}_+ &= \tau 
+ \alpha \left( 
-\frac{24 \tau}{\pi^2} - \frac{28 \tau \zeta(3)}{\pi^2} 
\right) 
+ \alpha^2 \left( 
-\frac{80 \tau^2}{\pi^4} 
+ \left( \frac{392 \tau}{\pi^4} - \frac{56 \tau^2}{\pi^4} \right) \zeta(3) 
+ \frac{372 \tau \zeta(5)}{\pi^4} 
\right) +\\
&+ \alpha^3 \left[
\left( \frac{112 \tau}{\pi^4} + \frac{896 \tau^2}{\pi^6} \right) \zeta(3) 
- \frac{1568 \tau \zeta(3)^2}{\pi^6} 
+ \left( -\frac{5952 \tau}{\pi^6} + \frac{248 \tau}{3 \pi^4} + \frac{992 \tau^2}{\pi^6} \right) \zeta(5) 
- \frac{5080 \tau \zeta(7)}{\pi^6} 
\right] \\
&+ \alpha^4 \Bigg[ 
\left( -\frac{3136 \tau}{3 \pi^6} - \frac{3136 \tau^2}{\pi^8} \right) \zeta(3)^2 
+ \left( -\frac{10912 \tau}{3 \pi^6} - \frac{13888 \tau^2}{\pi^8} + \frac{992 \tau^3}{\pi^8} \right) \zeta(5) +\\
&+ \zeta(3) \left( \frac{896 \tau^3}{\pi^8} + \frac{48608 \tau \zeta(5)}{\pi^8} \right) 
+ \left( \frac{86360 \tau}{\pi^8} - \frac{2540 \tau}{\pi^6} - \frac{15240 \tau^2}{\pi^8} \right) \zeta(7) 
+ \frac{71540 \tau \zeta(9)}{\pi^8} 
\Bigg]
\end{split}
\ee
\be
\begin{split}
    B_+^{(0)} &= -\frac{\alpha }{\pi ^2} 
\\B_-^{(0)} &=-\frac{7 \alpha \zeta(3)}{\pi^2} 
+ \alpha^2 \left(-\frac{14 \tau \zeta(3)}{\pi^4} + \frac{93 \zeta(5)}{\pi^4}\right) 
+ \alpha^3 \left(\left(\frac{62}{3 \pi^4} + \frac{248 \tau}{\pi^6}\right) \zeta(5) - \frac{1270 \zeta(7)}{\pi^6}\right) +\\
& \qquad\qquad  + \alpha^4 \left(\frac{248 \tau^2 \zeta(5)}{\pi^8} + \left(-\frac{635}{\pi^6} - \frac{3810 \tau}{\pi^8}\right) \zeta(7) + \frac{17885 \zeta(9)}{\pi^8}\right)
\\
B_+^{(1)} &= \frac{1}{4} - \frac{2 \alpha}{\pi^2} + \alpha^2 \left(-\frac{4 \tau}{\pi^4} + \frac{14 \zeta(3)}{\pi^4}\right) + \alpha^3 \left(\frac{28 \zeta(3)}{3 \pi^4} - \frac{248 \zeta(5)}{\pi^6}\right) + \\ &\qquad\qquad \qquad+ \alpha^4 \left(-\frac{248 \zeta(5)}{\pi^6} + \frac{3810 \zeta(7)}{\pi^8}\right)\\
B_-^{(1)} &= 2 
+ \alpha \left(\frac{8 \tau}{\pi^2} - \frac{14 \zeta(3)}{\pi^2}\right) 
+ \alpha^2 \left(\left(-\frac{28}{3 \pi^2} - \frac{56 \tau}{\pi^4}\right) \zeta(3) + \frac{248 \zeta(5)}{\pi^4}\right) +\\
&+ \alpha^3 \left(-\frac{112 \tau^2 \zeta(3)}{\pi^6} + \left(\frac{248}{\pi^4} + \frac{744 \tau}{\pi^6}\right) \zeta(5) - \frac{3810 \zeta(7)}{\pi^6}\right) +\\
& + \alpha^4 \left(\left(\frac{248}{5 \pi^4} + \frac{496 \tau}{3 \pi^6} + \frac{1984 \tau^2}{\pi^8}\right) \zeta(5) + \left(-\frac{5080}{\pi^6} - \frac{10160 \tau}{\pi^8}\right) \zeta(7) + \frac{57232 \zeta(9)}{\pi^8}\right)\\
B_+^{(2)} &= \frac{2}{3} 
+ \alpha \left(-\frac{2}{\pi^2} + \frac{5 \tau}{3 \pi^2} - \frac{7 \zeta(3)}{\pi^2}\right) 
+ \alpha^2 \left(-\frac{4}{3 \pi^2} - \frac{8 \tau}{\pi^4} + \left(\frac{56}{\pi^4} - \frac{14}{3 \pi^2}\right) \zeta(3) + \frac{124 \zeta(5)}{\pi^4}\right) +\\
& + \alpha^3 \left(-\frac{16 \tau^2}{3 \pi^6} + \left(\frac{56}{\pi^4} + \frac{56 \tau}{\pi^6}\right) \zeta(3) - \frac{196 \zeta(3)^2}{\pi^6} + \left(-\frac{1116}{\pi^6} + \frac{124}{\pi^4}\right) \zeta(5) - \frac{1905 \zeta(7)}{\pi^6}\right) +\\
& + \alpha^4 \Bigg[-\frac{784 \zeta(3)^2}{3 \pi^6} + \left(-\frac{1488}{\pi^6} + \frac{124}{5 \pi^4} - \frac{992 \tau}{\pi^8}\right) \zeta(5) + \zeta(3) \left(\frac{56}{5 \pi^4} + \frac{112 \tau}{3 \pi^6} + \frac{6944 \zeta(5)}{\pi^8}\right) + \\
&\qquad\qquad+\left(\frac{18288}{\pi^8} - \frac{2540}{\pi^6}\right) \zeta(7) + \frac{28616 \zeta(9)}{\pi^8}\Bigg]
\end{split}
\ee

\subsection{Spectral sums}
\label{sec:spectralsums}
Once we have an order-by-order expansion of \eqref{seriesexp}, it is straightforward to apply at each order in $\lambda$ the integral formulas \eqref{intformulas} to compute  the spectral sums $G^{(s)}_\pm(\lambda)$, $s\geq 2$ through eq.\ \eqref{spectraldet}:
\be\label{spectralsums}
\partial_\lambda \(\log(D_+(\lambda)) \pm \log(D_+)(\lambda)\) = (1\pm 1) (\log(8\pi)-1) - \sum_{s=1}^{\infty}\(G^{(s)}_-(\alpha) \pm G^{(s)}_+(\alpha) \) \lambda^s
\ee
that is, for $s>1$:
\small \be
\begin{split}
    \label{intformulassums}
G^{(s)}_\pm(\alpha) + G^{(s)}_\pm(\alpha) &= \frac{1}{s-1}\Bigg[  \pdv[s-1]{\lambda}
\int_{-\infty}^{+\infty}\dd{\nu} \frac{\pi}{2} \frac{Q_+(\nu|\lambda)\,\partial_\nu Q_-(\nu|\lambda) - Q_-(\nu|\lambda)\,\partial_\nu Q_+(\nu|\lambda)}{f(\nu)} \Bigg]_{\lambda = 0} \\
G^{(s)}_\pm(\alpha) - G^{(s)}_\pm(\alpha) &= \frac{-1}{s-1} \Bigg[  \pdv[s-1]{\lambda}\infint \dd{\nu} \frac{\pi^2}{2}\frac{Q_+(\nu|\lambda)\, Q_-(\nu|\lambda)}{\sinh(\pi\nu)f(\nu)} \Bigg]_{\lambda= 0}
    \end{split}
\ee
\normalsize
For $s=1$, using that:
\be 
Q_+(\nu) = 1+ \mathcal{O}(\lambda)\comma  \qquad Q_-(\nu) = \nu + \mathcal{O}(\lambda)
\ee
one gets directly the formulas:
\be \begin{split}
\left.\partial_\lambda\log(D_+(\lambda) D_-(\lambda))\right|_{\lambda=0}&=2-\infint\left[\frac{\tanh \left(\frac{\pi  \nu }{2}\right)}{\nu }-\frac{1}{\frac{2 \alpha }{\pi }+\nu  \coth \left(\frac{\pi  \nu }{2}\right)}\right]\dd{\nu} \\&= 2-\infint\frac{ \alpha  \tanh \left(\frac{\pi  \nu }{2}\right)}{\frac{\pi  \nu}{2} \left(\frac{2 \alpha }{\pi }+\nu  \coth \left(\frac{\pi  \nu }{2}\right)\right)}\dd{\nu}\\
&=2-\frac{\alpha}{2}\infint\dd{\nu}\frac{\pi}{2} \frac{\pi  \alpha  \sinh (\pi  \nu ) \sinh \left(\frac{\pi  \nu }{2}\right)}{\frac{\pi  \nu}{2}  \cosh ^2\left(\frac{\pi  \nu }{2}\right) \left(\alpha  \sinh \left(\frac{\pi  \nu }{2}\right)+\frac{1}{2} \pi  \nu  \cosh \left(\frac{\pi  \nu }{2}\right)\right)}\period
\end{split}
\ee
\be\begin{split}
    \left.\partial_\lambda\log(\frac{D_+(\lambda)}{D_-(\lambda)})\right|_{\lambda=0}&= -\infint\dd{\nu}\frac{\pi  \nu\,  \text{csch}(\pi  \nu )}{\frac{2 \alpha }{\pi }+\nu  \coth \left(\frac{\pi  \nu }{2}\right)}\\
    &=\alpha \infint\dd{\nu}\frac{\pi}{2}\frac{\text{sech}^2\left(\frac{\pi  \nu }{2}\right)}{\left(\alpha+\frac{\pi\nu}{2}  \coth \left(\frac{\pi  \nu }{2}\right)\right)}-\infint\dd{\nu}\frac{\pi }{\cosh (\pi  \nu )+1}\\
    &=\alpha \infint\dd{\nu}\frac{\pi}{2}\frac{\text{sech}^2\left(\frac{\pi  \nu }{2}\right)}{\left(\alpha+\frac{\pi\nu}{2}  \coth \left(\frac{\pi  \nu }{2}\right)\right)}-2
\end{split}
\ee
Then, using the result above, and applying the relations \eqref{spectralsums} at lowest order in $\lambda$, one obtains directly:
\be 
G_{\pm}^{(1)}(\alpha) = \log(8\pi) - 2 \pm 1 - \frac{\alpha}{4} \int_{-\infty}^\infty 
\frac{\sinh(t) \big(\sinh(2t) \pm 2t\big)}{t \cosh^2(t) \big(\alpha \sinh(t) + t \cosh(t)\big)} \dd{t}
\ee
already reported in \cite{Fateev_2009}.

To obtain the spectral sums $G_{\pm}^{(2)}$ , we need to consider the terms of $Q_{\pm} (\nu)$ up to the first order in $\lambda$. For this we only need $A_{\pm}$ up to order $\lambda$,  $B_+$ at order  $\lambda$, and $B_{-}$ at order $\lambda^{0}$. These polynomials  as well as ${\rm M}_\pm$ are only determined as power series in $\alpha$, for any desired order thereof. Hence, in order to extract the spectral sums from the integral expressions \eqref{intformulassums}, we expand the integrand in power of $\alpha$, up to the desired order. We can then integrate them term by term in the $\alpha$ expansion.

Integrating numerically term-by-term we obtain:
\be 
\begin{split}
G^{(2)}_+(\alpha) &= 8.41439832 -11.5719695\alpha + 13.7055746\alpha^2 -15.2501917\alpha^3 + 16.4273895\alpha^4 + \mathcal{O}(\alpha^5)\\
G^{(2)}_-(\alpha) &= 2.0000000  -1.3333333\alpha+ 0.95349463\alpha^2 -0.71898694\alpha^3 +0.56448932\alpha^4 + \mathcal{O}(\alpha^5)\\
\end{split}
\ee
that can be recognized\footnote{With the help of numerical methods e.g. PSLQ algorithm implemented in Mathematica: see \href{https://library.wolfram.com/infocenter/MathSource/4263/}{this URL}. } to coincide numerically with:
\small
\be 
\begin{split}
G^{(2)}_+(\alpha) &= 7 \zeta_3+ \alpha\left[ 
\frac{8}{3} + \frac{14 \zeta_3}{3} - \frac{56 \zeta_3}{\pi^2} - \frac{124 \zeta_5}{\pi^2}\right] +\alpha^2 \left[-\frac{224 \zeta_3}{3 \pi^2} + \frac{196 \zeta_3^2}{\pi^4} + \frac{1240 \zeta_5}{\pi^4} - \frac{124 \zeta_5}{\pi^2} + \frac{1905 \zeta_7}{\pi^4}\right]+\\
&+\alpha^3\left[-\frac{112 \zeta_3}{5 \pi^2} + \frac{784 \zeta_3^2}{3 \pi^4} + \frac{1984 \zeta_5}{\pi^4} - \frac{124 \zeta_5}{5 \pi^2} - \frac{6944 \zeta_3 \zeta_5}{\pi^6} - \frac{21336 \zeta_7}{\pi^6} + \frac{2540 \zeta_7}{\pi^4} - \frac{28616 \zeta_9}{\pi^6}\right] \\
&+\alpha^3\Bigg[\frac{784 \zeta_3^2}{9 \pi^4} + \frac{14384 \zeta_5}{15 \pi^4} - \frac{34720 \zeta_3 \zeta_5}{3 \pi^6} + \frac{61504 \zeta_5^2}{\pi^8} - \frac{40640 \zeta_7}{\pi^6} + \frac{2921 \zeta_7}{3 \pi^4} + \frac{106680 \zeta_3 \zeta_7}{\pi^8} + \\ &+ \frac{343392 \zeta_9}{\pi^8} - \frac{143080 \zeta_9}{3 \pi^6} + \frac{429870 \zeta_{11}}{\pi^8}\Bigg] + \mathcal{O}(\alpha^5)\\
G^{(2)}_-(\alpha) &= 2 -\frac{4\alpha}{3}+ \alpha^2 \left[ \frac{56 \zeta_3}{3 \pi^2} - \frac{124 \zeta_5}{\pi^4}\right] + \alpha^3\left[ \frac{56 \zeta_3}{5 \pi^2} - \frac{496 \zeta_5}{\pi^4} + \frac{3048 \zeta_7}{\pi^6}\right]       \\
&+\alpha^4\left[-\frac{5704 \zeta_5}{15 \pi^4} + \frac{10160 \zeta_7}{\pi^6} - \frac{57232 \zeta_9}{\pi^8}\right]+ \mathcal{O}(\alpha^5)
\end{split}
\ee
\normalsize
that are the expansions of the formulas derived in \cite{Litvinov:2024riz} up to the same order.

For $s=3$ we find:
\be 
\begin{split}
G^{(3)}_+(\alpha) &= 20.4981208 -45.6613977\alpha + 72.1676608\alpha^2-98.7405496\alpha^3 + 124.869920\alpha^4 + \mathcal{O}(\alpha^5)\\
G^{(3)}_-(\alpha) &= 1.71982418 -2.03864075\alpha + 1.03864075\alpha^2 -1.75397929\alpha^3 + 1.55184859\alpha^4 + \mathcal{O}(\alpha^5)
\end{split}
\ee
These match perfectly with the respective expressions from \cite{Litvinov:2024riz} expanded up to $\alpha^4$:
\small
\be \begin{split}
G^{(3)}_+(\alpha) &= -\frac{4 \pi^2}{3} + 28 \zeta_3 + \alpha \left[
4 + \left(42 - \frac{84}{\pi^2}\right) \zeta_3 - \frac{147 \zeta_3^2}{\pi^2} - \frac{651 \zeta_5}{\pi^2}\right] +\\
&+\alpha^2\left[\frac{12}{5} - \frac{196 (-6 + \pi^2) \zeta_3^2}{\pi^4} - \frac{1116 (-2 + \pi^2) \zeta_5}{\pi^4} + \zeta_3 \left(14 - \frac{224}{\pi^2} + \frac{5208 \zeta_5}{\pi^4}\right) + \frac{11430 \zeta_7}{\pi^4}\right]+\\
&+\alpha^3\Bigg[-\frac{196 (-36 + \pi^2) \zeta_3^2}{3 \pi^4} - \frac{2744 \zeta_3^3}{\pi^6} + \left(\frac{5952}{\pi^4} - \frac{8618}{15 \pi^2}\right) \zeta_5 - \frac{46128 \zeta_5^2}{\pi^6} + \frac{1524 (-28 + 15 \pi^2) \zeta_7}{\pi^6} +\\
&+\zeta_3 \left(-\frac{2212}{15 \pi^2} + \frac{1736 (-27 + 5 \pi^2) \zeta_5}{\pi^6} - \frac{80010 \zeta_7}{\pi^6}\right) - \frac{186004 \zeta_9}{\pi^6}  \Bigg] +\\
&+\alpha^4\Bigg[-\frac{5488 \zeta_3^3}{\pi^6} - \frac{92256 (-5 + \pi^2) \zeta_5^2}{\pi^8} + \zeta_3^2 \left(\frac{22736}{15 \pi^4} + \frac{145824 \zeta_5}{\pi^8}\right) + \frac{508 (-240 + 31 \pi^2) \zeta_7}{\pi^6} +\\
&+\zeta_5 \left(-\frac{620 (-56 + \pi^2)}{7 \pi^4} + \frac{1417320 \zeta_7}{\pi^8}\right) - \frac{61320 (-12 + 7 \pi^2) \zeta_9}{\pi^8} +\\
&+\zeta_3 \left(-\frac{24}{\pi^2} + \frac{1736 (-310 + 13 \pi^2) \zeta_5}{5 \pi^6} - \frac{32004 (-24 + 5 \pi^2) \zeta_7}{\pi^8} + \frac{1201872 \zeta_9}{\pi^8}\right) + \frac{2947680 \zeta_{11}}{\pi^8}\Bigg]\\
G^{(3)}_-(\alpha) &= \frac{4}{9}(-6 + \pi^2)+ \alpha \left[\frac{8}{3} - \frac{28 \zeta_3}{3} + \frac{62 \zeta_5}{\pi^2}\right] + \alpha^2\left[-\frac{8}{5} - \frac{28 \zeta_3}{5} + \frac{248 \zeta_5}{\pi^2} - \frac{1524 \zeta_7}{\pi^4}\right]+ \\
&+\alpha^3\left[ \frac{1288 \zeta_3}{45 \pi^2} + \frac{124 (-40 + 23 \pi^2) \zeta_5}{15 \pi^4} - \frac{1016 (-1 + 5 \pi^2) \zeta_7}{\pi^6} + \frac{28616 \zeta_9}{\pi^6}\right]+ \\
&+\alpha^3\left[\frac{16 \zeta_3}{\pi^2} + \frac{248 (-392 + 15 \pi^2) \zeta_5}{105 \pi^4} - \frac{2032 (-15 + 7 \pi^2) \zeta_7}{3 \pi^6} + \frac{8176 (-12 + 35 \pi^2) \zeta_9}{3 \pi^8} - \frac{491280 \zeta_{11}}{\pi^8}\right]
\end{split}
\ee
\normalsize
\subsection{Asymptotic expansion}
\label{sec:spetrumana}
In this subsection we construct an asymptotic expansion in $\abs{\lambda} \to \infty$ directly (rather than using the solutions \eqref{eq:mp} and \eqref{eq:mm}  constructed above). To do that, we first observe that the expansion in $\abs{\lambda}\to \infty$ of the two solutions, \eqref{eq:mp} and \eqref{eq:mm}, contains a prefactor $(-\lambda)^{-\frac{i\nu}{2}-\frac{i\alpha z}{\pi}}$ times a further asymptotic series in powers of $\lambda^{-1}$\cite{abramowitz+stegun}. Hence, it is natural to look for asymptotic solutions of \eqref{tq} with the following structure around $\lambda\to-\infty$,
\begin{equation}
\label{eq:asymans}
    S(\nu|\lambda) = (-\lambda)^{-\frac{i\nu}{2}-\frac{i\alpha z}{\pi}}\,\sum_{n=0}^{\infty} S_n(\nu)\,\lambda^{-n}\, ,
\end{equation}
 Note that $\lambda\to-\infty$ is not the physical region of the eigenvalues of 't Hooft equation and we will eventually analytically continue to the positive $\lambda$.

\label{asympstep1}
\paragraph{General strategy.}
By plugging the ansatz \eqref{eq:asymans} into \eqref{tq} we obtain:
\begin{equation}
\label{eq:series}
  \sum_{k=0}^{\infty} S_k(\nu + 2i)\, \lambda^{-k+1} + \sum_{k=0}^{\infty} S_k(\nu - 2i)\, \lambda^{-k-1} + 2 \sum_{k=0}^{\infty} S_k(\nu)\, \lambda^{-k} = \frac{4\pi\lambda}{\frac\nu z + \frac{2\alpha}{\pi}}\sum_{k=0}^{\infty} S_k(\nu )\, \lambda^{-k}
\end{equation}
that has to be satisfied at each order in $\lambda$. 
Remarkably, the order-$n$ term is recursively fixed by the previous ones. So we just need to solve for the single function $S_0(\nu)$ to reconstruct the full series. Let us illustrates this in the simple case of $S_1(\nu)$ since the generalization is trivial. Assuming that the solution $S_0(\nu)$ at order $\mathcal O(\lambda^{1})$  ($k = 0$) is known, we can plug the ansatz $S_1(\nu) = S_0(\nu)\, s_1(\nu)$ into the order $\mathcal{O}(\lambda^0)$ ($k=1$) of \eqref{eq:series}:
\begin{equation}
    S_1(\nu + 2i) + 2S_0(\nu) = \frac{4\pi}{\frac{\nu}{z} + \frac{2\alpha}{\pi}} S_0(\nu)\quad \Rightarrow\quad     s_1(\nu + 2i) - s_1(\nu) = -\frac{1}{2\pi z} \left(\nu + \frac{2\alpha z}{\pi}\right),
\end{equation}
that is solved uniquely by the function:
\begin{equation}
    f_1(\nu) = \frac{1}{2\pi i  z}\left(1 + \frac{i\nu}{2}+\frac{i\alpha z}{\pi} \right)\left( \frac{i\nu}{2} + \frac{i\alpha z}{\pi}\right) . 
\end{equation}
up to an arbitrary constant that we can always set to zero. We can then generalise to any order $k$ by using the ansatz  $S_k(\nu) = s_k(\nu) S_0(\nu)$ leading to:
\begin{equation}
\label{eq:fk}
        s_k(\nu) = \frac{1}{k! \, (2\pi i  z)^k}\left(1 + \frac{i\nu}{2}+\frac{i\alpha z}{\pi} \right)_k\left( \frac{i\nu}{2} + \frac{i\alpha z}{\pi} \right)_k, 
\end{equation}
where we used the Pochhammer Symbol: $(a)_k = \Gamma(a+k)/\Gamma(a)$. Thus, as anticipated, the whole series \eqref{eq:s0} is determined just by the single function $S_0(\nu)$:
\begin{equation}
\label{eq:asymseries0}
 S(\nu|\lambda) \asymp (-\lambda)^{-\frac{i\nu}{2}-\frac{i\alpha z}{\pi}}\,S_0(\nu)\,\sum_{n=0}^{\infty} s_n(\nu)\,\lambda^{-n}.
\end{equation}
As a preliminary remark, note that the series in \eqref{eq:asymseries} is divergent and it has to be understood only as an asymptotic expansion for the solutions of \eqref{tq} (and therefore the symbol $\asymp$ has been used).

Solutions constructed in this way, in particular the coefficients $s_k(\nu)$, do not have the required analyticity in the strip. To remedy this, we build two independent solutions by taking appropriate linear combinations of $S(\pm\nu|\lambda)$ using the $\nu$-dependent coefficients $P_\pm(z,\lambda)$:
\begin{equation}
\label{eq:qsolf}
\begin{split}
Q_\pm(\nu|\lambda) &= P_\pm(\nu,\lambda)\, S(\nu|\lambda) \mp  P_\pm(-\nu,\lambda)\, S(-\nu|\lambda); \\
P_\pm(\nu,\lambda) &= C_\pm(\lambda) \sum_k P_\pm^{(k)}\Big(z,z^{-1},\log(-\lambda)\Big) \, \lambda^{-k}
\end{split}
\end{equation}
Here $P_\pm(z,\lambda)$ are functions of $\nu$ and $\lambda$, invariant under $\nu \to \nu \pm 2i$, which can be taken to have polynomial coefficients of the invariant variables  $z$ and $\frac{1}{z}$. We anticipate that this coefficients will contain a $\log(-\lambda)$ dependence. For convenience we factor out an overall normalization $C(\lambda)$, which is independent of $\nu$. 

The singularities at $\nu =i$ generated by the prefactor $(-\lambda)^{-\frac{i\alpha z}{2}}$ can be removed by a redefinition of the polynomial $P_\pm(z,\lambda)\to (-\lambda)^{\frac{i\alpha z}{2} }P_\pm(z,\lambda)$ given that $(-\lambda)^{\frac{i\alpha z}{2} }$ is invariant under $\nu \to \nu \pm 2i$. In fact, this could have been done already at the beginning by redefining the ansatz \eqref{eq:asymseries0} to:
\begin{equation}
\label{eq:asymseries}
    S(\nu|\lambda) = (-\lambda)^{-\frac{i\nu}{2}}\,S_0(\nu)\,\sum_{n=0}^{\infty} s_n(\nu)\,\lambda^{-n}\, .  
\end{equation}
\paragraph{\boldmath Constructing $S_0(\nu)$.}
We now construct $S_0(\nu)$. This amounts to determining the solutions at order $\mathcal{O}(\lambda^1)$ of \eqref{eq:series}:
\begin{equation}
\label{eq:s0}
    S_0(\nu + 2i)  = \frac{4 \pi^2  z}{\pi\nu + 2\alpha z} S_0(\nu).
\end{equation}
A solution of \eqref{eq:s0} with $\alpha = 0$ was already discussed in \cite{Fateev_2009} and it is expressed in terms of Barnes G-functions:
\begin{equation}
    S_0(\nu)\eval_{\alpha = 0} = (2\pi)^{-\frac{1}{2} - \frac{i\nu}{2}}\,\frac{G\left(2 + \frac{i \nu}{2}\right)\, G\left(\frac{1}{2} - \frac{i \nu}{2}\right)}{G\left(1 - \frac{i \nu}{2}\right)\, G\left(\frac{3}{2} + \frac{i \nu}{2}\right)}\comma 
\end{equation}
which is analytical everywhere inside the strip $(-2i,2i)$. 
Having a solution of \eqref{eq:s0} with $\alpha = 0$, we can build a solution for $\alpha \neq 0$ by using again an ansatz of the form $S_0(\nu) = \left(S_0(\nu)\eval_{\alpha = 0}\right)\,\sigma(\nu)$ with $\sigma(\nu)$ satisfying
\begin{equation}
\label{sigmarecurrence}
    \sigma(\nu+2i) = \left(1 + \frac {2\alpha}{\pi} \frac{z}{\nu}\right)^{-1}\,\sigma(\nu). 
\end{equation}
The recurrence equation \eqref{sigmarecurrence} is solved exactly by:
 \be
\sigma(\nu) =  \frac{\sqrt{\pi}}{\Gamma\left(\frac{1}{2} +  \frac {i \alpha z}{\pi} \right)}\,\frac{\Gamma\left(1 + \frac{i\nu}{2} + \frac {i \alpha z}{\pi}\right)}{\Gamma\left(1 + \frac{i\nu}{2}\right)}, \qquad \sigma(i) = 1 \ee
Yet, the resulting $S_0(\nu) = S_0(\nu;0)\, \sigma(\nu)$ is not a solution of the TQ-System because of the poles at the zeros of $\left(\frac{1}{2} +  \frac {i \alpha z}{\pi} \right)$ accumulating at $\nu=i$.
In analogy to what done in the previous section, we proceed to expand $\sigma(\nu)$ in a formal power of $\alpha$:
\be
\label{expansionsigmatilde}
\sigma(\nu) = \sum_{k=0}^\infty \sigma_k(\nu,z)\, \alpha^k \period
\ee
This function display unwanted poles of increasing order at $\nu = \pm i$, where $z$ is singular. At any order in $\alpha$, to remove then we can just multiply by polynomial in power of $\alpha$ fixed term by term by cancellation of poles:
\be 
S_0^{(k)}(\nu) = A^{(k)}(z)\, S_0(\nu)\comma \qquad A^{(k)}(\alpha,z)= \sum_{l=0}^k a^{(l)} \, \alpha^l
\ee
Let us introduce for simplicity: $\chi = i \pi \tanh{\frac{\pi\nu}{2}}$.  $A^{(k)}(z)$ up to  order $\alpha^4$ is:
\small
\be\begin{split}
A^{(4)}(z)&= 1 + \alpha\chi\frac{  (\gamma + \log 4)}{\pi^2} 
+ \alpha^2 \Bigg(\chi^2
    -\frac{ (\pi^2 - 2 (\gamma + \log 4)^2)}{4 \pi^4} 
    - \chi \frac{7 \zeta_3}{\pi^4}
\Bigg) \\
&+ \alpha^3 \Bigg[
    \chi^3\frac{ \big((\gamma + \log 4) (-3 \pi^2 + 8 \log^2 2 + \gamma (2 \gamma + \log 256)) + 28 \zeta_3\big)}{12 \pi^6} \\
    &\qquad\quad+ \chi^2 \frac{(\pi^4 - 42 (\gamma + \log 4) \zeta_3)}{6 \pi^6} 
    + \chi\frac{62  \zeta_5}{\pi^6}
\Bigg] \\
&+ \alpha^4 \Bigg[
  \chi^4\Big(\frac{
        4 \gamma^4 - \pi^4 + 32 \gamma^3 \log 2 - 48 \pi^2 \log^2 2 + 128 \gamma \log^3 2 + 64 \log^4 2 
        - 12 \gamma^2 (\pi^2 - 8 \log^2 2)}{{96 \pi^8} } \\
        &\qquad \qquad\quad+\frac{- 16 \gamma \pi^2 \log 8 + 224 (\gamma + \log 4) \zeta_3}{{96 \pi^8} }
    \Big) \\
    &\qquad \quad-\chi^3 \frac{ \big(-2 \gamma \pi^4 - 4 \pi^4 \log 2 + 21 (-\pi^2 + 2 (\gamma + \log 4)^2) \zeta_3 + 372 \zeta_5\big)}{12 \pi^8} \\
    &\qquad \quad - \chi^2\frac{ (\pi^6 - 147 \zeta_3^2 - 372 (\gamma + \log 4) \zeta_5)}{6 \pi^8} + \chi \Bigg(
        \frac{31 \zeta_5}{3 \pi^6} - \frac{635 \zeta_7}{\pi^8}
    \Bigg)
\Bigg]
\end{split}
\ee
\normalsize

Then, at any order $k$ in $\alpha$, the candidate for the two independent solutions of the the TQ-system in the asymptotic regime $\lambda\to -\infty$ are:
\begin{equation}
\begin{split}
\label{eq:sol}
    S^{(k)}(\nu|\lambda) &= (-\lambda)^{-\frac{i\nu}{2}}S_0^{(k)}(\nu)\sum_{n=0}^{\infty} s_n(\nu)\,\lambda^{-n}\, ,   \\
     Q^{(k)}_\pm(\nu|\lambda) &= P^{(k)}_\pm(\nu,\lambda)\, S^{(k)}(\nu|\lambda) \mp  P^{(k)}_\pm(-\nu,\lambda)\, S^{(k)}(-\nu|\lambda)\, , \\
P^{(k)}_\pm(\nu,\lambda) &= C^{(k)}_\pm(\lambda) \sum_{n=0}^\infty P_\pm^{(k,n)}\Big(z,z^{-1},\log(-\lambda)\Big) \, \lambda^{-n}
\end{split}
\end{equation}
\paragraph{Removing the singularities.}
As mentioned before, the coefficient functions $s_n(\nu)$'s do not have a required analyticity in the strip $(-2i,2i)$: they have poles of order $(n-1)$ at the point $\nu = 0$ and of order $n$ at $\nu = \pm i$. 
To remove these singularities, we choose the polynomials $P^{(k)}_\pm\Big(z,z^{-1},\log(-\lambda)\Big)$ so that they cancel all those singularities and any given order in $\lambda$ and $\alpha$. As we will see, this can be done uniquely. 
The contributions can be split as follows :
\begin{equation}
\begin{split}
\label{eq:qsol}
 P^{(k)}_\pm(z,\lambda) = &C^{(k)}_\pm(\lambda) \bigg[\sum_{n=0} R_\pm^{(n,k)}(z^{-1};\alpha,\log(-\lambda))\,\lambda^{-n} \bigg]\,  \bigg[\sum_{n=0} T^{(n)}(z;\alpha)\,\lambda^{-n} \bigg]\\
 := &C_\pm(\lambda) \, R^{(k)}_\pm(z^{-1};\alpha,\lambda) \,  T(z;\alpha,\lambda)\, . 
\end{split}
\end{equation}
The functions $R^{(k)}_\pm$ and $T$ are polynomials in their first variable with $\alpha$ and $\lambda$ appearing as parameters (also polynomially) in the coefficients thereof and they can be determined independently from the other. In the function $T$   we dropped the subscript ${(k)}$ turns out to be exact in $\alpha$ (and indeed it matches perfectly with the one presented in \cite{Litvinov:2024riz}). 

We proceed as follows:
\begin{itemize}
\item 
For any fixed $k$ being the order of expansion in $\alpha$, at every order in $\lambda$, we expand $Q_\pm(\nu|\lambda)$ in \eqref{eq:qsolf} around $\nu = 0$ and fix the coefficients \\$r^{(l)}_\pm(\alpha, \log(-\lambda))$ in the polynomials: 
\begin{equation}
R^{(n,k)}_\pm\left(z^{-1};\alpha,\log(-\lambda) \right) = \sum_{l=1}^{n-1} r^{(l,n,k)}_\pm(\alpha, \log(-\lambda)) \, z^{-l}    
\end{equation}
so that they cancel all the poles of the form $1/\nu^k$ in $Q^{(k)}_\pm(\nu|\lambda)$. This fixes uniquely the function $R_\pm^{(n,k)}(z,\lambda)$ up to an overall $\lambda$-depending constant $C^{(k)}_-(\lambda)$ that we fix upon requiring the normalisation $Q_-(0;\lambda) = 1$ at all orders in $\lambda$. This automatically fixes also $C^{(k)}_+(\lambda)$ up to an overall scaling  thanks to the relation presented in (4.12) of \cite{Fateev_2009}.

The resulting expression can be conveniently expressed through the combinations:
\begin{equation}
        L = \log(-2\pi\lambda) + \gamma_E , \qquad c = i\pi z^{-1} , \qquad \beta = \alpha +1 \, ,
\end{equation}
and just to give a flavor of the resulting expressions, we report the result up to $\mathcal{O}\left(\frac{\log^2(\lambda)}{\lambda^4}\right)$ and $\alpha^3$: 
\begin{equation}
\label{eq:rpm}
\begin{split}
  &R^{(3)}_\pm(c,L) -1  =  \pm \frac{\beta c}{\lambda^2} \bigg[ \frac{1}{4 \pi^4} + \frac{1}{24 \pi ^6 \lambda} \Big( 6c+\beta (6\mp1)-12L + \\
  &\qquad \qquad + 12\alpha \left(1+\log(4)\right) -\frac{84}{\pi^2}  \alpha^2  \zeta_3\Big) +31 \alpha^3 \zeta_5/\pi^{10}  \bigg] + \mathcal{O}\left(\frac{\log^2(\lambda)\alpha^4}{\lambda^4}\right)
\end{split}
\end{equation}
that matches perfectly with what reported in \cite{Litvinov:2024riz} up to the order considered here.
\item Analogously, as we expand the functions $Q_\pm(\nu ;\lambda)$ near $\nu = i$ (only $\nu = i$ is enough given the parity of $Q_\pm(\nu|\lambda)$), we exhibit poles coming from $s_k(\nu)$ which cancellation determines uniquely the numbers
 $c_{m,l}$  that specify $T^{(k)}(z;\alpha)$:
\begin{equation}
T^{(k)}(z;\alpha) = \sum_{l= 1}^k \sum_{m = k+1 }^{2k} c_{m,l}\, \alpha^m \, z^l   
\end{equation}
 In principle, one could have a different functions for $Q_{\pm}$, but it turns out that the same function works for both: $T^{(k)}_+(z;\alpha) = T^{(k)}_-(z;\alpha)$. We therefore  suppressed the index. The determination of $T^{(k)}(z;\alpha)$ is completely independent from the one of $R_\pm(c,L)$. The overall scaling $\kappa$ that is left unfixed is determined by imposing the Wronskian relation $Q_+(i;\lambda)\, Q_-(i;\lambda) = i$.  In \eqref{eq:rpm}, we have already presented the $C_\pm(\lambda)$ with this normalization imposed. 
 
Using the same variable as above $\chi = i \pi z$ for convenience, the first orders are:
  \begin{equation}
  \begin{split}
  \label{eq:T}
   T(\chi;\alpha,\lambda) =  1 - \frac{\alpha^2 \chi}{2 \pi ^4\lambda }   + \frac{\alpha^3 \chi}{8 \pi ^8\lambda^2}\left(\alpha\chi -  2 \pi^2\right) - \frac{\alpha^4 \chi}{48 \pi^{12} \lambda^3} \left(  \alpha^2\chi^2 - 6\pi^2 \alpha\chi + 10\pi^4\right) + \\
   + \frac{\alpha^5 \chi}{384 \pi ^{16} \lambda^4 } \left(\alpha^3\chi^3-12 \pi ^2 \alpha^2\chi^2+52 \pi^4 \alpha\chi-84 \pi^6\right) + \mathcal{O}\left(\lambda^{-5}\right)
  \end{split}
  \end{equation}  
\end{itemize}

\section{'t Hooft model in the complex-mass plane: further results}\label{sec:further}
In this section, we further explore the properties of the bound state equation in the complex plane of the masses both numerically and analytically.  A thorough numerical study of 't Hooft equation in the complex plane goes beyond the scope of this work. Instead in this section, we make use of numerical methods as a mean to test the analytical results we have obtained.  
In the past, many efforts have been devoted to study 't Hooft equation and its solutions numerically, employing the spectral form 
\eqref{thooftnuspace}. While the original 't Hooft equation \eqref{thoofteq} is a highly singular integral equation, its spectral form \eqref{thooftnuspace}  presents a perfectly regular kernel, and we can solve it numerically via discretization and direct diagonalization. This presents various advantages as we analytically continue in the first sheet of the $\alpha$-plane \footnote{Another advantage is that we do not have to worry about the different boundary conditions in position space when $\alpha = -1$, cfr.\ e.g.\ section 4.2 of \cite{Kaushal:2023ezo} or the discussion in \cite{Anand:2021qnd}.}. This is because, by solving the eigenproblem via the direct diagonalization, we do not need to find boundary conditions of the eigenfunctions for unphysical values of the masses; instead we can just find the eigenvalues as determined by \eqref{thooftnuspace} for $\alpha\in\mathbb C$.
\begin{figure}[htbp]
\centering
\begin{subfigure}{0.5\linewidth}
        \centering
    \includegraphics[width=\linewidth]{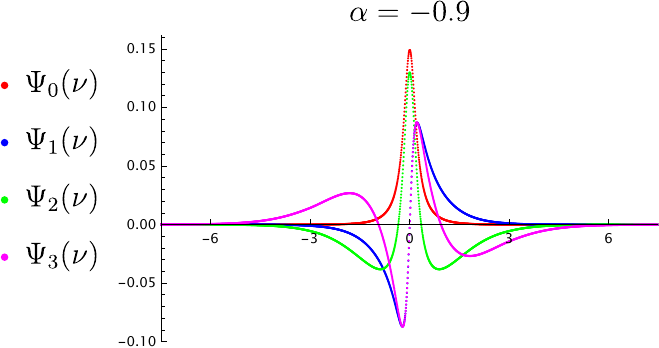}
\end{subfigure}\hfill
\begin{subfigure}{0.5\linewidth}
        \centering
    \includegraphics[width=\linewidth]{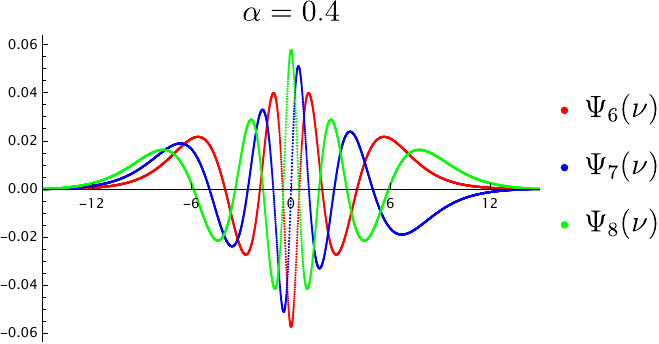}
\end{subfigure}
    \caption{Numerical eigenfunctions for different values of $n$ and $\alpha$. Here $N=4000, L = 15$. }
    \label{fig:numeigen}
\end{figure}

In the following the numerical computation of the eigenproblem \eqref{thooftnuspace} is performed by cutting off the integral to the finite-length interval\footnote{This approximation rapidly converges as we increase $L$ since we have already that $\frac{\pi \nu}{ 2 \sinh(\frac{\pi \nu}{2})}\eval_{  \nu = 10} = 4.73443 \times 10^{-6}$.} $[-L,L]$  and approximating it with the Gaussian quadrature method and discretizing on a $N$ point lattice. Then, eigenvalues and eigenfunctions are readily computed by diagonalizing the resulting $N\times N$ matrices\footnote{Let us remark that the choice of $L$ has to be commensurate with the energy level we are interested in computing. As we increase the energy level (and/or $\alpha$), the support of the eigenfunction, that are picked around the origin,  grows more and more. For instance, for $\alpha = 1$, while for the first $4$ or $5$ eigenfunctions a small value of $L\sim 5$ is sufficient, already for the $6$th energy level, a bigger interval $L\sim 10$  is needed to  cover the support of the eigenfunctions. On the other hand, as we increase the value of $L$, we need to increase $N$ accordingly to guarantee a sufficient sample density in the region where the eigenfunctions has support. Here, we avoid discussing thoroughly these kind of issues, and we just perform diagonalization in the most direct way, remarking once again that there is large room for improvement of these numerical methods.}.

Before proceeding, let us check that this very rapid but crude method to numerically solve the eigenproblem gives accurate solutions. To do that, we set $\alpha = 0$ so that we can compare our numerical solutions with the results obtained by diagonalizing the position space problem using a Chebyschev polynomials basis. For $N= 2000,  L = 10$, it takes $\sim 15$ seconds to generate and diagonalise the resulting $2000 \times 2000$ matrices for the first 40 eigenvalues and eigenfunctions on a 4 core machine using a Mathematica script. In Table \ref{tab:numvsmum}, we compare the values of the first few eigenvalues with the two numerical methods. In \figref{fig:numeigen} some examples of numerical eigenfunctions for two different values of $\alpha$. Note that, as in a quantum mechanical problem in a  potential well, the $n$-th eigenfunction has exactly $n$ nodes and that, as we know already, the even (odd) eigenfuntions are even (odd).
\begin{table}[htbp]
    \centering
    \begin{tabular}{c|c||c}
       $n$&  $2\lambda^{(\rm num)}_n$ &  $2\lambda^{(\rm num)}_n$ from \cite{Fateev_2009} \\
       \hline
         0 & $\underline{0.738}363$ & $\underline{0.737}06174629269 $\\
        1 & $\underline{1.76}381$ & $\underline{1.75}37313369175$\\
        2 & $\underline{2.75}8$ & $\underline{2.748}1609123706$\\
        3 & $\underline{3.78}37$ & $\underline{3.75}10575817054$    \end{tabular}
    \caption{Comparison of the first 4 eigenvalues obtained by direct discretization with $N=4000$, with the numerical results of \cite{Fateev_2009}. The values of $L$ depends on the level $n$ as the support of the eigenfunction grows as we increase the energy level. For all but the ground state we set $L=10$ while for $n = 0$ we set $L=5$.  }
    \label{tab:numvsmum}
\end{table}
\subsection{Asymptotic approximation of the mesons wavefunctions}
\label{analyticalwavefunctions}
\begin{figure}[htbp]
  \centering
  \begin{subfigure}{0.3\linewidth}
    \includegraphics[width=\linewidth]{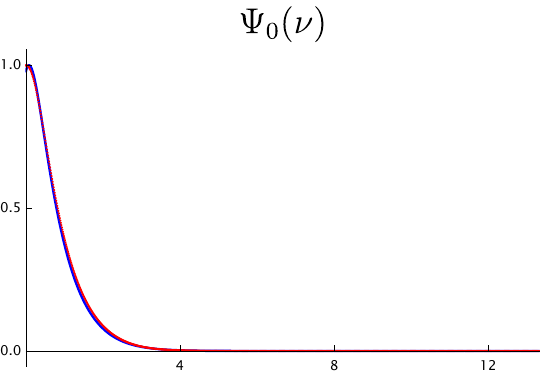}
  \end{subfigure}
  \hfill
  \begin{subfigure}{0.3\linewidth}
    \includegraphics[width=\linewidth]{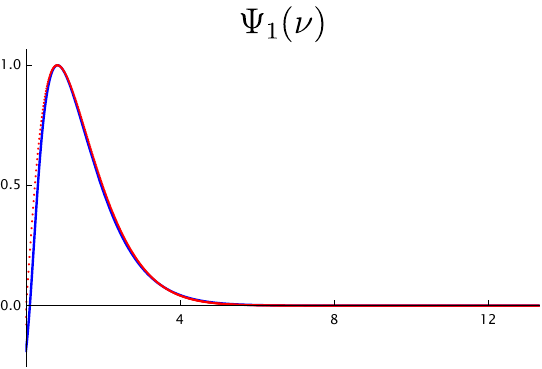}
  \end{subfigure}
  \hfill
  \begin{subfigure}{0.3\linewidth}
    \includegraphics[width=\linewidth]{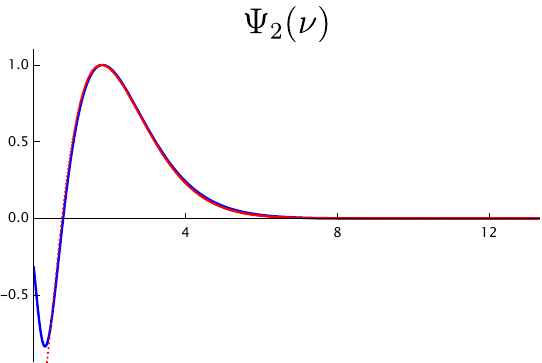}
  \end{subfigure}
\par\bigskip
  \centering
  \begin{subfigure}{0.3\linewidth}
    \includegraphics[width=\linewidth]{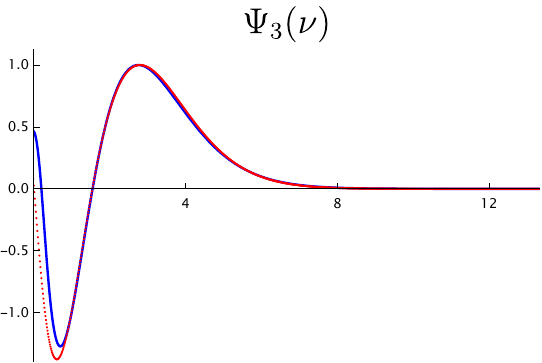}
  \end{subfigure}
  \hfill
  \begin{subfigure}{0.3\linewidth}
    \includegraphics[width=\linewidth]{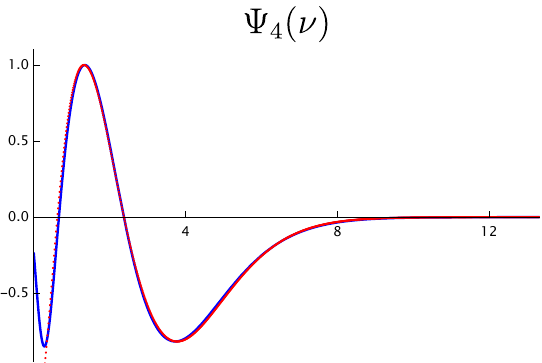}
  \end{subfigure}
  \hfill
  \begin{subfigure}{0.3\linewidth}
    \includegraphics[width=\linewidth]{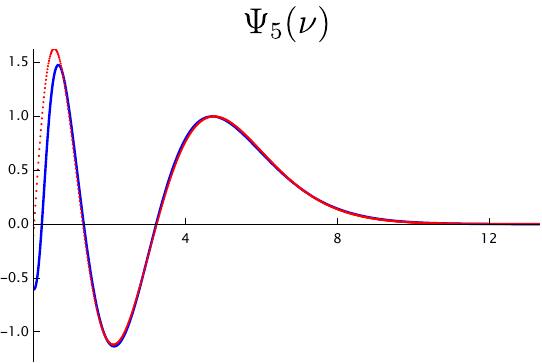}
  \end{subfigure}
  \caption{Comparison between numerical solutions (red dots) and analytical asymptotic predictions (blue line) for $\alpha =1$.} 
  \label{fig:numvsana1}
\end{figure}

\begin{figure}[htbp]
  \centering
  \begin{subfigure}{0.3\linewidth}
    \includegraphics[width=\linewidth]{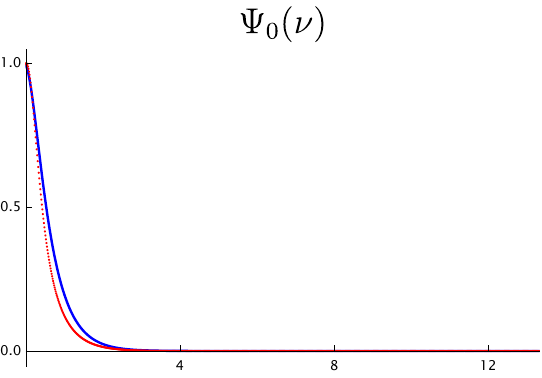}
  \end{subfigure}
  \hfill
  \begin{subfigure}{0.3\linewidth}
    \includegraphics[width=\linewidth]{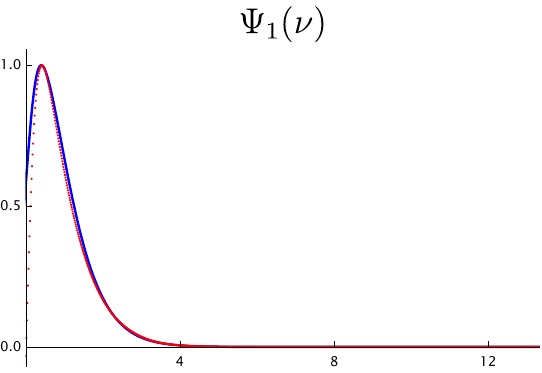}
  \end{subfigure}
  \hfill
  \begin{subfigure}{0.3\linewidth}
    \includegraphics[width=\linewidth]{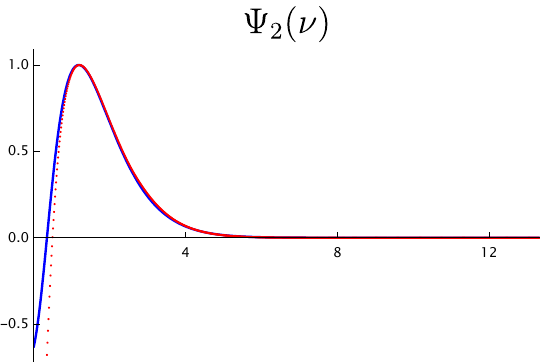}
  \end{subfigure}
\par\bigskip
  \centering
  \begin{subfigure}{0.3\linewidth}
    \includegraphics[width=\linewidth]{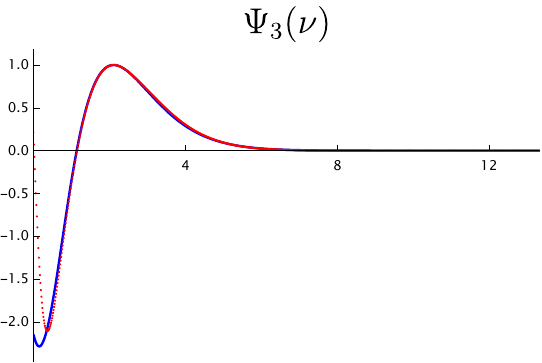}
  \end{subfigure}
  \hfill
  \begin{subfigure}{0.3\linewidth}
    \includegraphics[width=\linewidth]{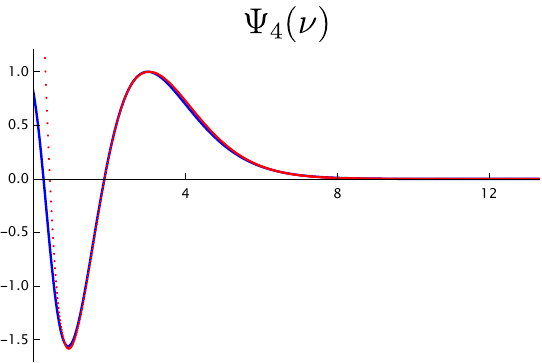}
  \end{subfigure}
  \hfill
  \begin{subfigure}{0.3\linewidth}
    \includegraphics[width=\linewidth]{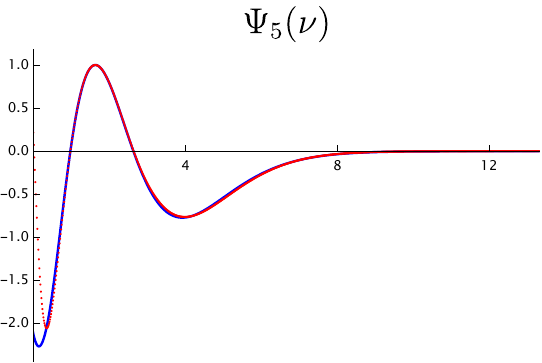}
  \end{subfigure}
  \caption{Comparison between numerical solutions (red dots) and analytical asymptotic predictions (blue line) for $\alpha =-0.6$.} 
  \label{fig:numvsana06}
\end{figure}

In section \ref{spectraldata} we described a procedure to obtain the asymptotic expansion of $Q_\pm(\nu|\lambda)$ in various regimes and to extract the spectrum from them. It is also possible to construct an analytic expansion of the solutions $Q(\nu)$ of the homogeneous problem as well. 
To do that,  we start with the regular solution  ${\rm M}_{\pm}$ \eqref{eq:mp} of the inhomogeneous TQ-system; this function does not have the required analyticity as it stands. However in the limit $\nu\gg 1$ ($z\to 1$), the resulting function is real and analytic in a neighborhood of the real line:
\be 
\label{curlyq}
Q_\pm(\nu|\lambda) \asymp \cQ(\nu|\lambda) = \left(\nu + \frac{2\alpha}{\pi} \tanh(\frac{\pi\nu}{2}) \right)\,  e^{i \pi \lambda} \,   {\rm M}\left(1 + \frac{i\nu}{2} + \frac{i \alpha}{\pi} \tanh(\frac{\pi\nu}{2}), 2 , -2 i\pi \lambda\right)\; . 
 \ee
Then the eigenfunctions $\Psi_n(\nu\gg 1)$,  can retrieved in this approximation through \eqref{qdefinition} just by evaluating $\lambda$ at the spectral values $\lambda_n$:
 \be 
 \begin{split}
 \label{asymeigenfunc}
\Psi_n(\nu) &\asymp \frac{\cQ(\nu|\lambda_n)}{\alpha\sinh(\frac{\pi\nu}{2})+ \frac{\pi\nu}{2}\cosh(\frac{\pi\nu}{2})}\; ,  \qquad \nu \gg 1 \; ,
\end{split}
 \ee
for negative values of $\nu$, we can just impose the parity conditions $\Psi_n(-|\nu| ) \asymp (-1)^n \Psi_n(\nu)$.
It turns out that \eqref{asymeigenfunc} furnishes an accurate approximation of the eigenfunction even at small value of $\nu$, as one can verify by comparing it with the result of the numerical diagonalization\footnote{Alternatively, we could use directly the series expansion of $Q_\pm(\nu)$ we constructed in \ref{sec:spectralsums} including arbitrary higher orders in $\lambda$. This would probably lead to a better approximation of the solutions for $\nu \sim 0$, but the agreement with the numeric results already for this very crude approximation, are satisfactory as a check of the validity of our analytical results. }. 
To test \eqref{asymeigenfunc} against the numerical solutions, we set $N=4000$ and $L= 40$ (with these settings it takes $\sim 30$ seconds on a 4 core machine to run the entire diagonalisation routine), and we evaluate \eqref{asymeigenfunc} at the corresponding values of $\lambda_n$ and we compare the analytical predictions with the numerical eigenfunctions. Note that the numerical and analytical eigenfunctions have different normalisations, so to compare them, we normalise the odd (even) ones such that the value at the first (second) local maximum is 1 for both of them.  
\begin{figure}[htbp]
  \centering
  \begin{subfigure}{0.3\linewidth}
    \includegraphics[width=\linewidth]{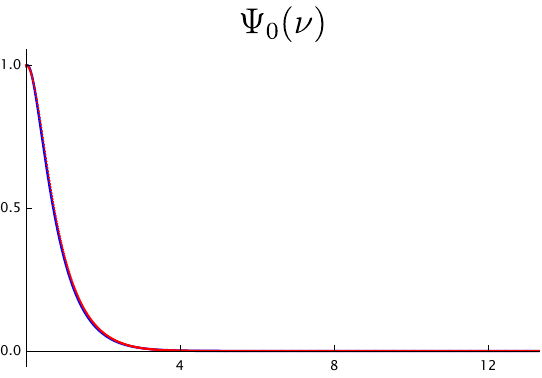}
  \end{subfigure}
  \hfill
  \begin{subfigure}{0.3\linewidth}
    \includegraphics[width=\linewidth]{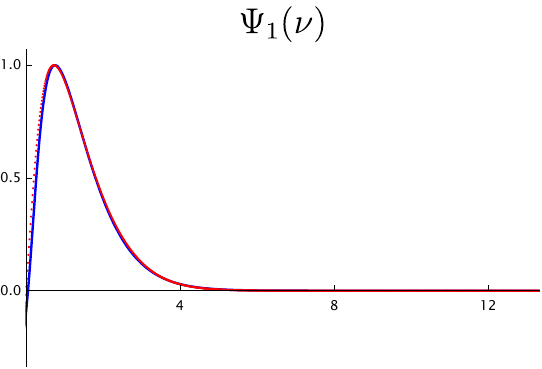}
  \end{subfigure}
  \hfill
  \begin{subfigure}{0.3\linewidth}
    \includegraphics[width=\linewidth]{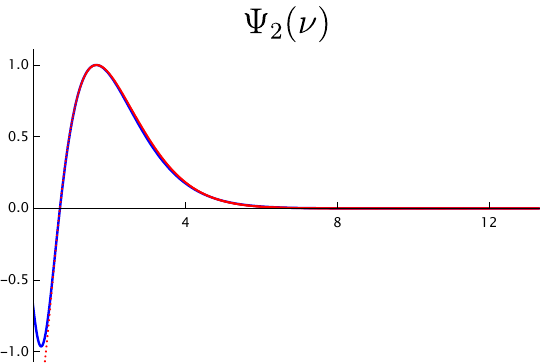}
  \end{subfigure}
\par\bigskip
  \centering
  \begin{subfigure}{0.3\linewidth}
    \includegraphics[width=\linewidth]{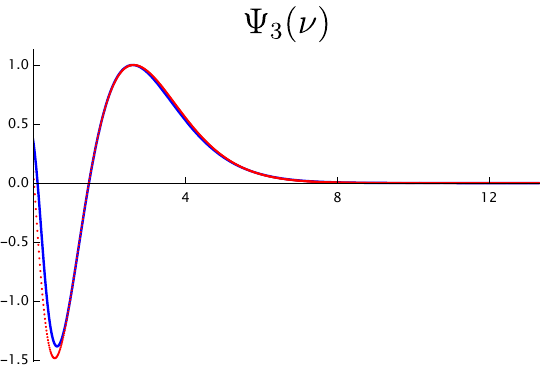}
  \end{subfigure}
  \hfill
  \begin{subfigure}{0.3\linewidth}
    \includegraphics[width=\linewidth]{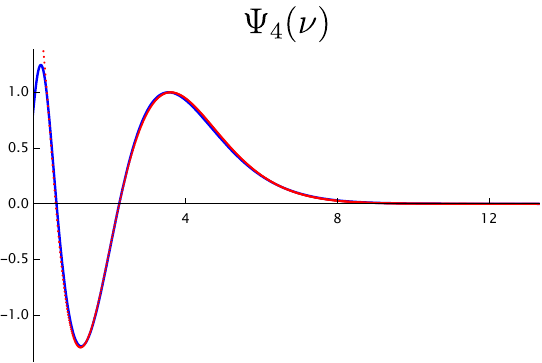}
  \end{subfigure}
  \hfill
  \begin{subfigure}{0.3\linewidth}
    \includegraphics[width=\linewidth]{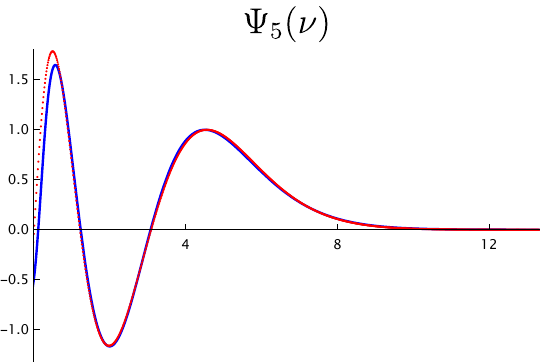}
  \end{subfigure}
  \caption{Comparison between numerical solution (red bullets) and analytical asymptotic predictions (blue line) for $\alpha =\frac1 2$.} 
  \label{fig:numvsana05}
\end{figure}
In \figref{fig:numvsana05}, \ref{fig:numvsana1} and \ref{fig:numvsana06} we plot both the numerical solutions and the analytical predictions for $\nu >0$ (the negative part is just determined by parity) for different values (positive and negative) of $\alpha$. As anticipated the agreement between the analytical predictions and the numerical solution is remarkable, especially considering that \eqref{asymeigenfunc} was supposed to approximate the solution for $\nu \gg 1$., Of course, if $\nu$ gets closer to $0$, it eventually approximates less and less the solution. 

As $\alpha$ approaches the chiral limit, $\alpha\to -1$,  the first eigenfunction has support only in the region $\nu  < 1$ (recall that in the limit $\alpha = -1$ $\Psi_0\to \delta(\nu)$ ), far away from the asymptotic regime $\nu \gg 1$. Therefore, for such values of $\alpha$, one should expect that the approximation becomes worse; indeed, this is evident, for instance,  in \figref{fig:numvsana06}. 


Nonetheless,it is worth emphasizing that, even for such values of $\alpha$, the analytical predictions stay rather close to the numerical solution. This highlights a practical merit of our analytic asymptotic expression. In the future, we plan to use these asymptotic solutions in order to evaluate scattering amplitudes of mesons.

\subsection[Spectrum for complex masses]{ \boldmath Spectrum for complex masses}
\begin{figure}[htbp]\begin{minipage}{.57\textwidth}
     \begin{tabular}{c|c|c}
 $n$ & $\alpha = 1.4   \pm  i $ & $\alpha = -2 \pm 0.7 i$ \\
\hline
 $0$ & $1.50212 \pm 0.500954\,i$ & $-0.249776 \pm 0.64655 \, i$ \\
 $1$ & $2.76496 \pm 0.612707 \, i$ &$0.570509 \pm 1.08639 \,i $ \\
 $2$ & $3.89594 \pm 0.667206 \, i$ & $1.45422 \pm 1.24904 \, i$\\
 $3$ & $5.01248 \pm 0.710913 \, i$ & $2.35834 \pm 1.35136 \, i $ \\
 $4$ & $6.09972 \pm 0.741821 \, i$ & $3.2913 \pm 1.40555 \, i$ \\
    \end{tabular}
    \captionof{table}{First 5 energy levels for $4$ distinct values of the masses, $L=10$, $N=1000$.  }
    \label{tab:complexmasses}
\end{minipage}%
\hfill
\begin{minipage}{.4\textwidth}
        \centering
        \includegraphics[width = 0.6 \textwidth]{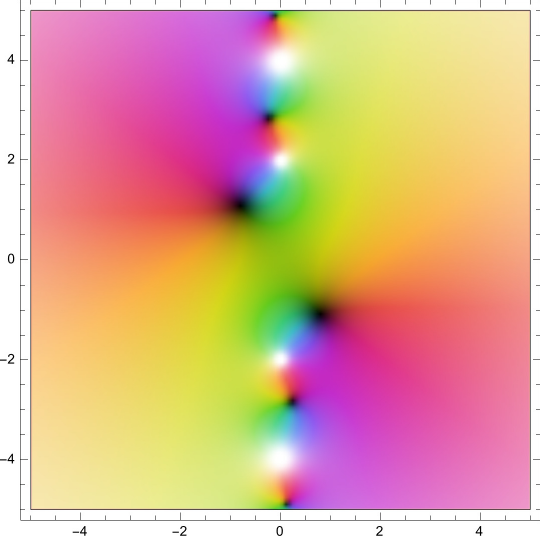}
        \captionof{figure}{The zeros of $f(\nu) = \frac{2\alpha}{\pi}+\nu\coth(\frac{\pi\nu}{2})$ do not lie on the real axis for $\alpha \not\in \mathbb{R}$. Here $\alpha = -1 + \frac{3}{2} i $}
        \label{complexzerosalpha}
\end{minipage}
\end{figure}

\begin{figure}[htbp]
 \centering
  \begin{subfigure}{0.45\linewidth}
    \includegraphics[width=\linewidth]{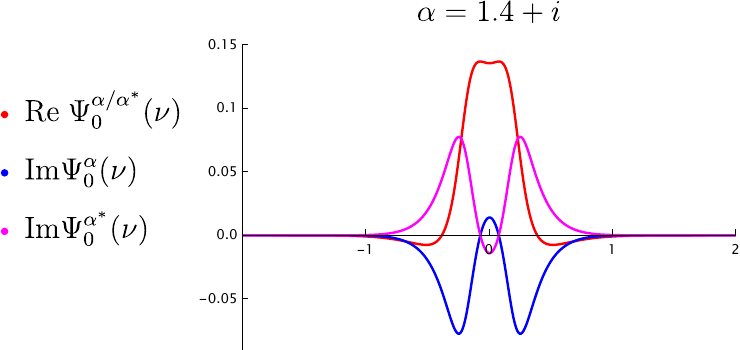}
  \end{subfigure}
  \hfill
  \begin{subfigure}{0.45\linewidth}
    \includegraphics[width=\linewidth]{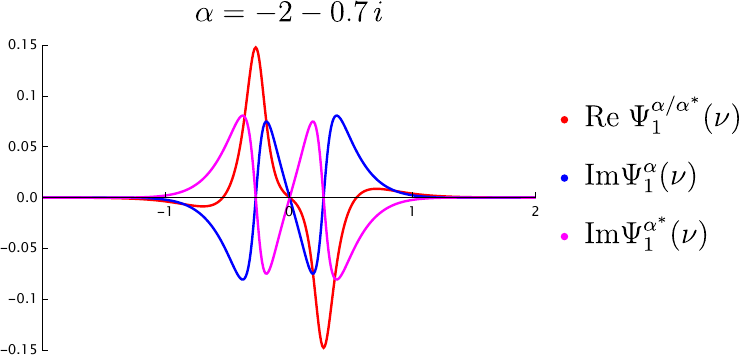}
  \end{subfigure}
  \hfill
    \caption{Eigenfunctions for some pairs of complex conjugate masses.}
    \label{fig:complexmasses}
\end{figure}

As we remarked at the beginning of this section, one of the advantages of studying the equation using the spectral representation is that it makes straightforward to study the problem in the plane of complex masses. 
Consider 't Hooft equation with complex masses $\alpha\in\mathbb C$: as discussed in Section \ref{t Hooft model} the Hamiltonian is no longer Hermitian for  complex masses and the eigenvalues will be complex. If $\Psi^\alpha_m(\nu)$ is an eigenstate with energy $\lambda_n$, then the $\cC$-conjugated solution $\left(\Psi^{\alpha^*}_m\right)^* (\nu)$ will no longer be an eigenstate of the same Hamiltonian, rather it will be an eigenfunction with energy $\lambda^*$ of the 't Hooft equation with complex mass $\alpha^*$. In  Table \ref{tab:complexmasses} we illustrate this for 2 different pairs of complex conjugated masses. In this case, the eigenfunctions do not have poles as there are no real zeros of $\alpha +\frac{\pi\nu}{2}\coth(\frac{\pi\nu}{2})$ for complex $\alpha$ (cfr.\ \figref{complexzerosalpha}), and one can check that indeed the expectation $\Psi^{\alpha}_n = \left(\Psi^{\alpha^*}_n\right)^{*}$ is verified by the numerical solutions (e.g.\ \figref{fig:complexmasses}). 

\subsubsection[Imaginary masses and spontaneous symmetry breaking of $\cP\cT$]{\boldmath Imaginary masses and spontaneous symmetry breaking of $\cP\cT$}
\label{immass}
\begin{figure}\begin{minipage}{.57\textwidth}
\scalebox{0.85}{
    \begin{tabular}{c|c||c|c}
    $n$ & $\lambda$ & $n$ & $\lambda$ \\ 
    \hline
         $0/\overline{0}$&    $-0.0735993 \pm 0.10608\, i$ &    $3/\overline{3}$&     $2.35968 \pm 0.327709\, i $ \\
         $0_+$ & $0.337355$&   $3_+$ &   $2.15487 $\\
        $0_-$&   $-0.997462$ &  $3_-$ &   $-2.79341 $ \\
          $1/\overline{1}$ &   $0.624888 \pm 0.249851\, i$ &  $4/\overline{4}$ &   $3.23552 \pm 0.24321 \, i$\\
           $1_+$&   $0.515636$  &$4_+$ &  $3.25385 $\\
             $1_-$&   $-1.57807$  &  $4_-$& $-3.42428$\\
              $2/\overline{2}$&   $1.40231 \pm 0.424763\, i $  &$5/\overline{5}$ &    $4.24429 \pm 0.214925 \,i$ \\ 
               $2_+$&   $1.98711$  & $5_+$ &$3.2985$\\
                $2_-$&   $-2.18286$ & $5_-$ &$-4.06832$ \\
    \end{tabular}}
     \captionof{table}
{First $24$ numerical eigenvalues for $\alpha = -1.2$ $L=10$, $N= 1000$. }
    \label{tab:num-1.2}
\end{minipage}%
\hfill
\begin{minipage}{.4\textwidth}
        \centering
        \includegraphics[width = 0.7 \textwidth]{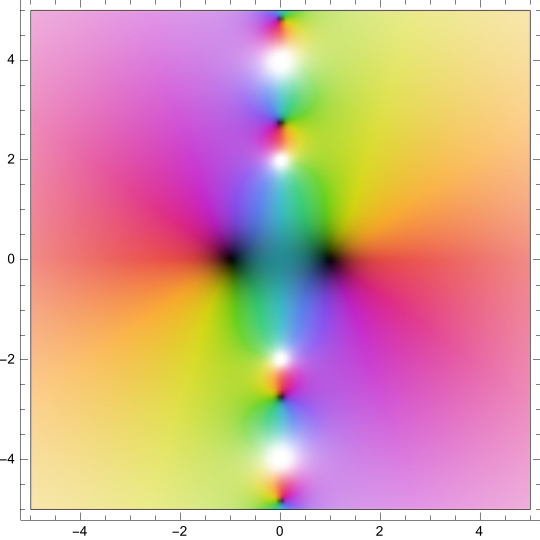}
        \captionof{figure}{The zeros of $f(\nu) = \frac{2\alpha}{\pi}+\nu\coth(\frac{\pi\nu}{2})$ lie on the real axis for $\alpha <-1$. Here $\alpha = -1.7$}
        \label{realzerosalpha}
\end{minipage}
\end{figure}

\begin{figure}
  \centering
  \begin{subfigure}{0.3\linewidth}
    \includegraphics[width=\linewidth]{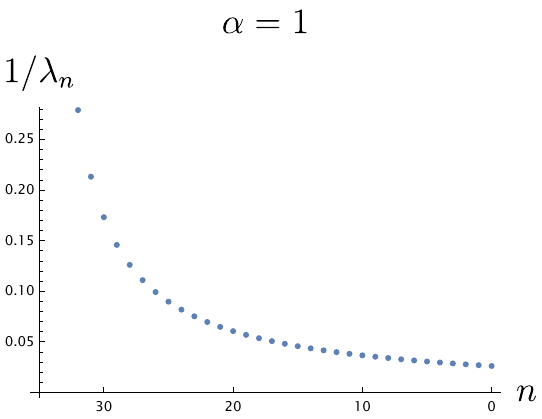}
  \end{subfigure}
  \hfill
  \begin{subfigure}{0.3\linewidth}
  \includegraphics[width=\linewidth]{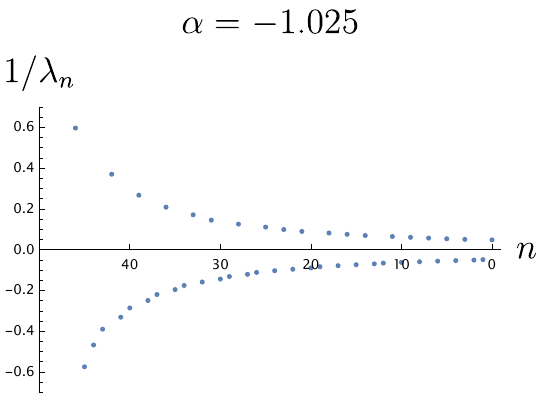}
  \end{subfigure}
  \hfill
    \begin{subfigure}{0.3\linewidth}
  \includegraphics[width=\linewidth]{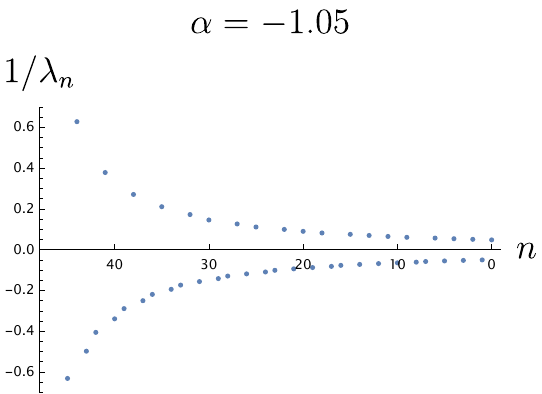}
  \end{subfigure}
  \hfill
      \caption{Inverse of the real eigenvalues of one positive and two negatives values of $\alpha$.    }
    \label{fig:realeigen}
\end{figure}

\paragraph{Boundary conditions.}
In the special case of tachyonic masses $\mathbb R \ni \alpha < -1$, the boundary conditions in the $[0,1]$ box $x^{\beta}, (1-x)^\beta$  are complex, given that the solutions of \eqref{exponents} are purely imaginary. Furthermore, for $\alpha < -1$  the eigenfunctions are expected to display simple poles at the positions corresponding to the two real  (see \figref{realzerosalpha}) simple zeros of  $f(\nu) = \alpha + \frac{\pi\nu}{2}\coth(\frac{\pi\nu}{2})$ (cf.\ with discussion at the beginning of Section \ref{criticality}), that are exactly $\pm 2 i \beta$. Henceforth, the basis of function solving the spectral problem \eqref{thooftnuspace} for imaginary masses is no longer in the class of $\mathbb {L}^2$ normalizable functions. For $\alpha>-1$, the normalizability of the wavefunctions restricted the value of $\beta$ to $\Re\, \beta>0$. By contrast, for $\alpha<-1$, there is no physical principle which selects one of the boundary conditions and  both $\pm\beta$ (that are purely imaginary) are equally valid boundary conditions. This leads to a two-fold spectrum for $\alpha < -1$. Continuing them back to $\alpha>-1$ through the complex-mass plane, one of them becomes a regular normalizable solution while the other is divergent near the boundary of $x\in [0,1]$, cf.\ \figref{fig:realeigen}. This is clear in particular for $\alpha=-1$, where  $\beta=0$ is doubly degenerated, and the ground state with $\lambda=0$ has one regular solution $\phi(x)=1$, and one irregular solution $\phi(x)=\log\frac{x}{1-x}$.

In what follows, we study this extended eigenproblem with the two-fold spectrum.

\begin{figure}[htbp]
  \centering
  \begin{subfigure}{0.45\linewidth}
    \includegraphics[width=\linewidth]{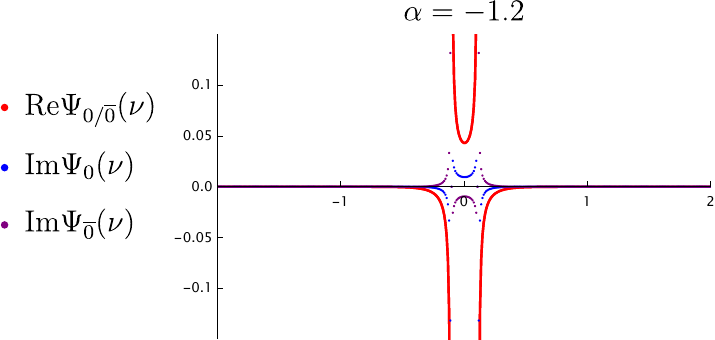}
  \end{subfigure}
  \hfill
  \begin{subfigure}{0.45\linewidth}
    \includegraphics[width=\linewidth]{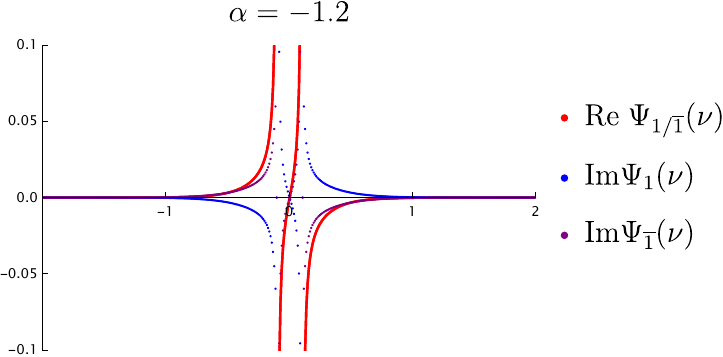}
  \end{subfigure}
  \hfill
\par\bigskip
  \centering
  \begin{subfigure}{0.45\linewidth}
    \includegraphics[width=\linewidth]{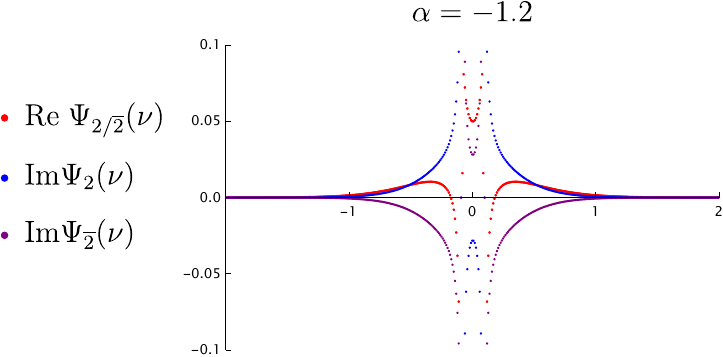}
  \end{subfigure}
  \hfill
  \begin{subfigure}{0.45\linewidth}
    \includegraphics[width=\linewidth]{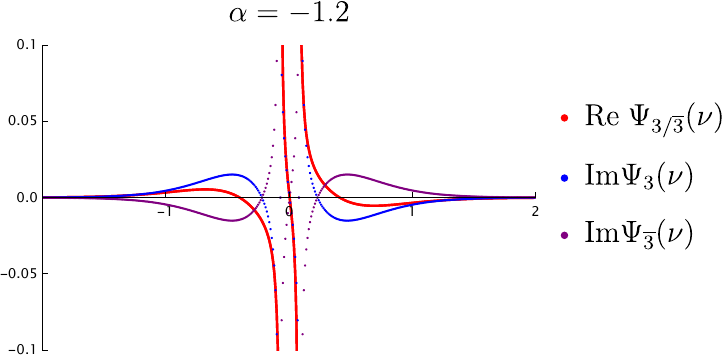}
  \end{subfigure}\hfill
  \par\bigskip
  \centering
  \begin{subfigure}{0.45\linewidth}
    \includegraphics[width=\linewidth]{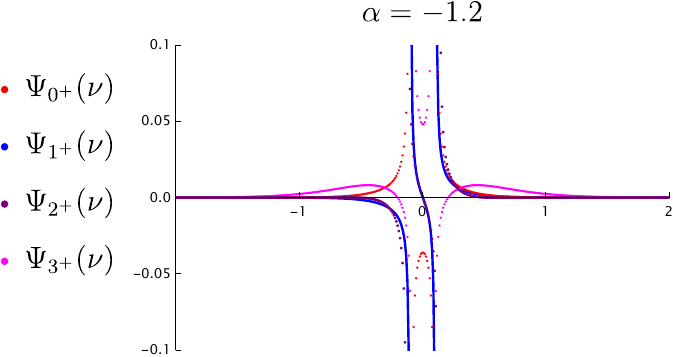}
  \end{subfigure}
  \hfill
  \begin{subfigure}{0.45\linewidth}
    \includegraphics[width=\linewidth]{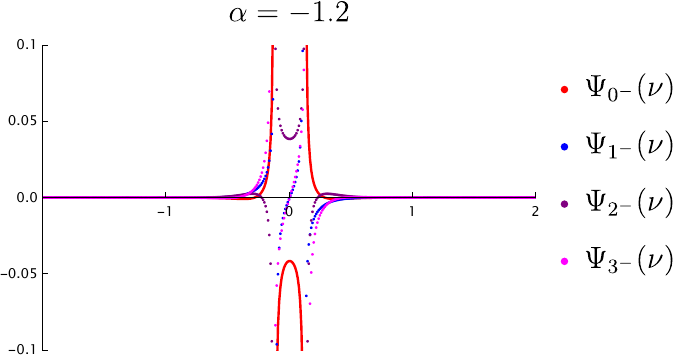}
  \end{subfigure}
  \caption{First $16$ numerical eigenfunctions for $\alpha = -1.2$. $L=10, N=1000$. Solid lines are for asymptots visualisation.} 
  \label{fig:a18}
\end{figure}

\begin{figure}
    \centering
    \includegraphics{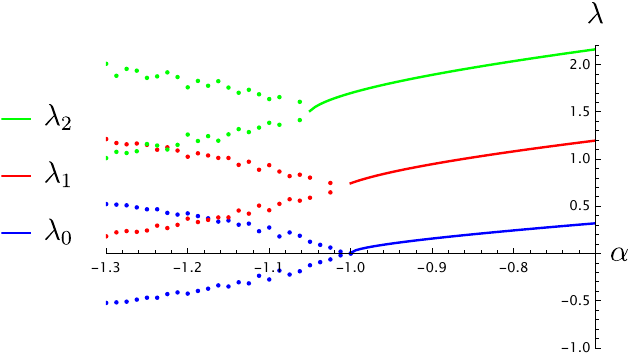}
    \caption{First $3$ energy levels as function of $\alpha$. As soon as the eigenvalues become complex, we only plot the imaginary parts thereof.}
    \label{fig:evolution}
\end{figure}

\paragraph{Numerical solutions.}
In  Table \ref{tab:num-1.2}, we report the first $20$ eigenvalues for $\alpha = -1.2$. As we can see in \figref{fig:a18}, the spectrum of the extended eigenproblem is naturally organized into 1.~complex conjugate pairs $m$ and $\bar{m}$ and 2.~real pairs $m_{+}$ and $m_{-}$.
  Wavefunctions corresponding to $m_{\pm}$ are real while those of $m$ and $\bar{m}$ are complex. One can also check that the position of the poles of the wavefunctions in \figref{fig:a18} correspond to real zeros of $f(\nu)$ as expected. Furthermore the wavefunctions are even (odd) for even (odd) level $m$, and the number of nodes is equal to the level $m$ for both real and imaginary parts. 
Interestingly, in the specific case of $\alpha = -1.2$, we see that from the $6/\overline{6}$-th level onward, the eigenvalues are always real. This is a general feature of the model: the number of non-reals eigenvalues  increases as $\alpha (<-1)$ decreases: at $\alpha = -1$, we have exactly two real degenerate (one physical, one unphysical) solutions $\delta(\nu), \delta'(\nu)$, and they become complex conjugate pairs as we further decrease $\alpha$. As we go towards more and more negative values of  $\alpha$ this happens, one by one, also to all the higher energy levels,  as can see from \ref{fig:evolution}.
In \figref{fig:realeigen}, we plot the inverse of the real eigenvalues for one positive and two negatives values of $\alpha$ illustrating the ``doubling" of the spectrum already observed in the previous  paragraph.

\paragraph{\boldmath $\cP\cT$-symmetry.} The existence of a finite number of complex eigenvalues can be understood as a consequence of the spontaneous $\cP\cT$-symmetry breaking. For $\alpha<-1$, the Hamiltonian is no longer Hermitian, but it is still invariant under the $\cP\cT$-symmetry \cite{Bender_1998,Bender_2005,Bender:2004sa}. Whenever $\cP\cT$-symmetry is exact (and not spontaneously broken), the eigenfunctions and eigenvalues are real even if the Hamiltonian is non-Hermitian. The fact that also non-hermitian Hamiltonians can have real eigenvalues is an interesting subject on its own and it has been studied in depth in the literature (cf.\ \cite{Bender_1998,Bender_2005,Bender:2004sa} and references therein). However, depending on the parameters of the Hamiltonian, this symmetry can be spontaneously broken, resulting in a finite number of complex eigenvalues, as we saw above for the 't Hooft model with $\alpha<-1$.

\paragraph{\boldmath $\cP\cT$-breaking in IFT.}
Let us also comment on yet another similarity with Ising Field Theory. In the Bethe-Salpeter approximation, the square-root branch points that correspond to $\alpha = -1$ are pushed to infinity $f_0/m^2 = \eta \to \infty$ and are virtually invisible. However, in the full IFT, they are located at finite point and exhibit the same qualitative features as the 't Hooft model. As shown recently in \cite{Lencses:2022ira}, the \textit{Truncated
free fermionic space approach} (see references therein) applied to the IFT in an imaginary magnetic field in finite volume revealed that the model possesses two different phases distinguished by the spontaneous symmetry breaking of $\cP\cT$ symmetry. For $\frac{\abs{h}}{\abs{m}^{15/8}}=\abs{\eta} < \abs{\eta_{\rm YL}}$, (i.e.\ for ``\textit{small}"  magnetic field) the energy levels of the Hamiltonian are all real (even though the IFT Hamiltonian is not Hermitian) and $\cP\cT$-Symmetry is preserved, for $\abs{\eta} =\abs{\eta_{\rm YL}}$ the first excited state degenerates with the ground state and the theory is critical (Yang-Lee CFT). More importantly for us, as one increases $\abs{\eta} > \abs{\eta_{\rm YL}}$, i.e.\ beyond the Bethe-Salpeter approximation where $\eta_{\rm YL}\to \infty$, the model enters a spontaneously broken phase of $\cP\cT$-symmetry with exactly the same qualitative behavior as we observed in the 't Hooft model: the two degenerate levels at $\eta_{\rm YL}$ become complex conjugate at $\eta> \eta_{\rm YL}$, and one by one the higher states will eventually do so as well. This provides evidence that the features that we have seen in the 't Hooft model might not be a large $N$ artifact and might persist beyond the large $N$ limit.

\section{Application to non-perturbative topological string}
\label{topstring}
In recent years, an interesting relation has been established  between partition functions associated to topological strings on a toric Calabi-Yau threefold $X$ and the spectral problem following from the quantization of the curve encoding its local mirror Calabi-Yau $\widehat{X}$. This is sometimes referred to as the topological string/spectral theory (TS/ST) correspondence and was used to define a non-perturbative completion of the topological string partition function, which, in its original form, is defined only through perturbation theory. Reviewing this correspondence in full details goes beyond the scope of this paper, and we refer to  \cite{grassi2015topological, marino2018wavefunctions, kallen2014instanton, kallen2014spectral} and references therein for a comprehensive discussion. Instead, here we comment on connections and applications of the techniques employed in this paper to that context.
\subsection{Elements of TS/ST correspondence}
\label{eltsst}
The TQ-systems similar to the one considered in this paper \eqref{tq} arise from the quantization of a curve associated to a mirror dual of a toric Calabi-Yau threefold:
$W_X\left(e^x,e^p\right) = 0$, 
that is a Riemann surface embedded in $\mathbb {C} ^* \times \mathbb {C} ^*$. Upon quantizing the curve $W_X\left(e^x,e^p\right)$ by promoting $x,p$ to canonically conjugated variables $[\widehat {x} , \widehat {p}\,] = i\hbar$, it is promoted to  the spectral problem\cite{grassi2015topological}:
\be
W_X\left(e^{\widehat{x}},e^{\widehat{p}}\right) \ket{\phi} = \cO(\widehat{x}, \widehat {p}) \ket{\phi} - \lambda \ket{\phi} = 0
\ee
for the self-adjoint quantum operator $\cO(\widehat{x}, \widehat {p})$. When the mirror curve is elliptic (genus $1$), 
the operator $\cO(\widehat{x}, \widehat {p})$ has a discrete and positive spectrum. Thus, the spectral problem above gives the well-defined eigenvalue problem :
\be
\cO(\widehat{x}, \widehat {p}) \ket{\psi_n} - e^{E_n} \ket{\psi_n} = 0 \period
\ee
Because of the positivity of the spectrum, the operator $\cO(\widehat{x}, \widehat {p})$ can be thought of as an exponential of an Hamiltonian $\cH(\widehat{x}, \widehat{p})$. In fact, it is more convenient to consider the spectral problem associated to its density matrix,
\be
\rho(\widehat{x}, \widehat{p}) = \cO^{-1}(\widehat{x}, \widehat {p}) = e^{- \cH(\widehat{x}, \widehat {p}) }
\ee
since it is of the so called \textit{trace-class}, i.e.\ such that all the spectral sums
\be
Z_s = \sum_{n=0}^{\infty} e^{-s E_n} = \sum_{n=0}^{\infty} \frac{1}{\lambda^s_n}, \qquad s =1,2, \cdots
\ee
are convergent. Furthermore all the spectral data for the problem can be collected into the following spectral determinant (\textit{Fredholm determinant}):
\be\begin{split}
\Xi(\lambda, \hbar) &= \det(1+ \lambda \rho) = \prod_{n=0}^{\infty} \left( 1 + \lambda e^{-E_n}\right) = \prod_{n=0}^{\infty} \left( 1 + \frac{ \lambda} {\lambda_n} \right)  \comma
\\
\log\Xi(\lambda, \hbar) &= - \sum_{s=1}^\infty Z_s \frac{(-\lambda)^s}{s}\comma
\end{split}
\ee
defined for any $\lambda\in \mathbb{R}$ and having zeros at the location of the eigenvalues of $\rho(\widehat{x},\widehat{p})$. It is clear that the quantities $Z_s$ and $\Xi(\lambda, \hbar)$ are closely related (cfr.\ eq.\ \eqref{spectraldet}, \eqref{lambaexpspec}) to the spectral sums $G^{(s)}_\pm$ and the spectral determinants $D_\pm(\lambda)$ defined in this paper. The crucial content of the TS/ST correspondence is that $\log\Xi(\lambda, \hbar) $ is conjectured to be computed exactly by the non-perturbative topological string free energy that can in turn be computed using, among others, the tools from the enumerative geometry\cite{grassi2015topological}. This allows, for instance, to determine exact non-perturbative quantization conditions determining the eigenvalues of $\rho(\widehat{x}, \widehat {p})$ analogous to the ones presented in this paper and in \cite{Fateev_2009,Litvinov:2024riz}.
\subsection[Local $\mathbb{P}^1\times\mathbb{P}^1$: spectral determinants from inhomogeneous extension]{\boldmath Local $\mathbb{P}^1\times\mathbb{P}^1$: spectral determinants from inhomogeneous extension}
\label{localP1}
Here we discuss the possibility of applying the methods developed in this paper to this context, and present some preliminary results. Concretely, let us consider the case of the  local $\mathbb{P}^1\times\mathbb{P}^1$ considered in \cite{kallen2014instanton}. In this case, the spectral problem associated with the density matrix is
\be \begin{split}
\phi(x) &= \lambda \infint\dd{y} \rho(x,y) \phi(y) , \qquad \lambda =\frac{e^E}{4\pi k}\comma \\
\rho(x,y) &= \frac{1}{\sqrt{2 \cosh(\frac x 2 )}\sqrt{2 \cosh(\frac y 2 )}} \frac{1}{\cosh(\frac{x-y}{2k})}
\end{split}
\ee
where $2\pi k =\hbar$. As expected, $\rho$ gives an Hermitian Hilbert-Schmidt operator with discrete positive spectrum. It is convenient for the following to reformulate the integral problem as: 
\be \begin{split}
f(x) \Psi(x) &= \lambda \infint \dd{y} S(x-y)\,\Psi(y) \comma \\
S(x) &= \frac{1}{2\cosh(\frac{x}{2k})}, \qquad f(x) = 2\cosh(\frac{x}{2}), \qquad \phi(x) = \Psi(x) \sqrt{f(x)}  \period  
\end{split}\ee
Now, following the analysis for the 't Hooft equation in $\nu$-space \eqref{thooftnuspace} in Section \ref{TQsec}, we define
\be
Q(x) = \sinh \left(\frac{x}{k}\right)f(x) \Psi(x)\comma
\ee
which satisfies the related integral equation
\beq
Q(x) = \lambda \sinh\left(\frac{x}{k}\right)\infint \dd{y} S(x-y)\,\frac{Q(y)}{\sinh \left(\frac{y}{k}\right)f(y)}\period
\eeq
Then the integral spectral problem becomes equivalent to the following TQ-system:
\be 
\label{spectralcurve}
Q(x+ i\pi k) + Q(x-i\pi k) = -2\pi k \lambda \frac{Q(x)}{f(x)}
\ee
Upon imposing that the $\cQ(x):=Q(x)/\sinh(x/k)$ decays in the strip $[-i\pi,+i\pi]$, one finds that the solution to this equation exists only for discrete values of $\lambda$. Thus, the equation \eqref{spectralcurve}, together with the asymptotic decaying condition on $\cQ(x)$, gives  the quantization of the curve relative to the local $\mathbb{P}^1 \times \mathbb{P}^1 $, as already discussed in \cite{kallen2014instanton}. It also satisfies a ``quantization" condition similar to \eqref{eq:quanti}:
\beq
Q(0)=Q(\pm i\pi k)=0\period
\eeq

Now, to study the equation \eqref{spectralcurve} for any value of $\lambda\in\mathbb{C}$, we follow the analysis in the main text, and consider the following {\it inhomogeneous Fredholm equation}
\be f(x)\, \Psi(x|\lambda) - \lambda\infint\dd{y} S(x-y)\, \Psi(y|\lambda) = F(x|\lambda)\period
\ee
 The inhomogeneous term $F(x|\lambda)$ spans a two-dimensional space, and in accordance with the symmetry $x\to -x$ of the integral equation, we take the following function $F_\pm$ as a basis for this space:
\be
F(x|\lambda) = q_+(\lambda) F_+(x) + q_-(\lambda) F_-(x), \qquad F_+(x) = \frac{1}{\cosh(\frac{x}{2k})}, \quad  F_-(x) = \frac{1}{\sinh(\frac{x}{2k})}
\ee
Then, the solutions $\Psi_\pm$ satisfy the following integral eigenproblems:
\begin{align}
     \, f(x)\, \Psi_+(x|\lambda) - \lambda\infint\dd{y} S(x-y)\, \Psi_+(y|\lambda) &= F_+(x|\lambda)\\
     \, f(x)\, \Psi_-(x|\lambda) - \lambda\fint_{-\infty}^{+\infty}\dd{y} S(x-y)\, \Psi_-(y|\lambda) &= F_-(x|\lambda)
\end{align}
Where the principal part is needed due to the inhomogeneous term in $x=0$. Then the $Q(x)$ satisfy the  integral equation
\be
\label{eq:inhointeqQ2}
    \,\sinh\left(\frac{x}{k}\right) F_\pm(x) = Q_\pm(x)  -  \lambda  \sinh\left(\frac{x}{k}\right) \fint_{-\infty}^{+\infty} \dd{y}\frac{1}{\cosh( \frac{x - y}{2k} )}\,\frac{ Q_\pm(y)}{ \sinh\left(\frac{y}{k}\right)f(y)}\comma
\ee
for any $\lambda \in \mathbb{C}$. Thanks to the identity, $\sinh(\frac{x + i k \pi}{k})F_\pm(\nu +i k \pi) + \sinh(\frac{x - i k \pi}{k})F_\pm(\nu - i k \pi) = 0$, $Q_{\pm}$ satisfy the same TQ-system:
\be\begin{split}
&Q_\pm(x+i \pi k) + Q_\pm(x - i \pi k) = -2\pi k \lambda \frac{Q_\pm(x)}{f(x)}\period
\end{split}
\ee
As in the case of 't Hooft model, they do not satisfy the ``quantization" condition:
\beq
Q_{+}(0)=0\comma\quad Q_{+}(i \pi k)=-Q_{+}(-i\pi k)=2i\comma\qquad Q_{-}(0)=2\comma \quad Q_{-}(\pm i\pi k)=0\period
\eeq

Now, consider the Liouville-Neumann series giving the exact solutions $\Psi_\pm(x|\lambda)$:
\begin{equation}
\begin{split}
        f(t) \Psi_+(t|\lambda) &= \sum_{n=0}^{\infty} \infint F_+(x|\lambda) \prod_{j = 1}^n \frac{\dd{t_j}}{f(t_j)} S(t_j - t_{j-1}),\\
  f(t) \Psi_-(t|\lambda) &= \sum_{n=0}^{\infty} \fint_{-\infty}^{+\infty} F_-(x|\lambda) \prod_{j = 1}^n \frac{\dd{t_j}}{f(t_j)} S(t_j - t_{j-1})\comma
\end{split}
\end{equation}
where $t_0 = t$. Then, upon considering the following combination, we obtain
\be
\begin{split}
&f(t)f(t') \Big[\Psi_+(t'|\lambda)\Psi_-(t|\lambda) - \Psi_-(t'|\lambda)\Psi_+(t|\lambda)\Big] =\\
&\sum_{l,m = 0}^{\infty} \lambda^{l+m} \fint_{-\infty}^{+\infty} \cosh(\frac{t'_l - t_m}{2k}) \left(\frac{1}{\cosh(\frac{t'_l}{2k})\sinh(\frac{t_m}{2k})}- \frac{1}{\sinh(\frac{t'_l}{2k})\cosh(\frac{t_m}{2k})} \right)\\
&\times S(t'_l - t_m)\, \prod_{j = 1}^l \frac{\dd{t'_j}}{f(t'_j)} S(t'_j - t'_{j-1})\, \prod_{i = 1}^m \frac{\dd{t_i}}{f(t_i)} S(t_i - t_{i
-1})\comma
\end{split}
\ee 
with $t_0 = t, t'_0 = t'$. Now, uniforming the notation for integration variables to:
\be (\tau_1,\cdots,\tau_m, \tau_{m+1},\cdots, \tau_{m+k}) = (t_1,\cdots,t_m, t'_{k},\cdots, t'_{1})
\ee
we get
\be
\begin{split}
&f(t)f(t') \Big[\Psi_+(t'|\lambda)\Psi_-(t|\lambda) - \Psi_-(t'|\lambda)\Psi_+(t|\lambda)\Big] =\\
&\sum_{l= 1}^{\infty} \lambda^{l} \fint_{-\infty}^{+\infty} \Bigg[\sum_{m= 0}^l \cosh(\frac{\tau_{m+1} - \tau_m}{2k}) \left(\frac{1}{\cosh(\frac{\tau_{m+1}}{2k})\sinh(\frac{\tau_{m}}{2k})}- \frac{1}{\sinh(\frac{\tau_{m+1}}{2k})\cosh(\frac{\tau_{m}}{2k})} \right)  \Bigg] \\
& \hspace{5mm}\times \prod_{j = 1}^l \frac{\dd{\tau_j}}{f(\tau_j)}  \prod_{j = 1}^{l+1} S(\tau_j - \tau_{j-1})
\end{split}
\ee 
The inhomogeneous function $F_\pm$ had been chosen such that sum over $m$ could be performed analytically by means of the following identity:
\be 
\begin{split}\label{trigid}
&\sum _{m=0}^l \left(\frac{\cosh (t_{m+1}-t_m)}{\sinh (t_{m+1}) \cosh (t_m)}-\frac{\cosh (t_{m+1}-t_m)}{\sinh (t_m) \cosh (t_{m+1})}\right) =\\
& = \left(\frac{\cosh (t_{l+1}-t_0)}{\sinh (t_{l+1}) \cosh (t_0)}-\frac{\cosh (t_{l+1}-t_0)}{\sinh (t_0) \cosh (t_{l+1})}\right)\comma
\end{split}
\ee
which we can use, together with $\tau_{l+1} = t'_0 = t'$ and $\tau_{0} = t_0 = t$, to rewrite our identity as:
\small
\be
    \Xi(t,t')f(t)f(t')\Big[\Psi_+(t'|\lambda)\Psi_-(t|\lambda) - \Psi_-(t'|\lambda)\Psi_+(t|\lambda)\Big] =\sum_{l= 1}^{\infty} \lambda^{l}\infint \prod_{j = 1}^l \frac{\dd{\tau_j}}{f(\tau_j)}\prod_{j = 1}^{l+1} S(\tau_j - \tau_{j-1}) \comma
    \ee\normalsize\be
    \frac{1}{\Xi(2k t, 2k t')} = \frac{\cosh (t'-t)}{\sinh (t') \cosh (t)}-\frac{\cosh (t'-t)}{\sinh (t) \cosh (t')}\period
\ee
\normalsize
Remarkably, the r.h.s.\ is proportional to the Liouville-Neumann series for the resolvent associated with our integral equation that, by definition, satisfies the following problem:
\be 
R(t,t'|\lambda) - \lambda \infint \dd{\tau}    \frac{S(t-\tau)}{\sqrt{f(t)f(\tau)}}\, R(\tau,t'|\lambda) =  \frac{S(t-t')}{\sqrt{f(t)f(t')}}
\ee
We thus arrive at the main formula:
\be
R(t,t'| \lambda ) = \Xi(t,t') \sqrt{f(t)f(t')}\,\Big[\Psi_+(t'|\lambda)\Psi_-(t|\lambda) - \Psi_-(t'|\lambda)\Psi_+(t|\lambda)\Big]
\ee
This shows that the kernel $S$ for this problem is of the \textit{completely integrable} type as well. 

Using this identity, we can express the spectral determinants $D_\pm(\lambda)$ defined as in \eqref{spectraldet}
in terms of $Q$-functions: 
\be
\begin{split}
\label{integrap1}
\partial_\lambda\log \frac{D_+(\lambda)}{D_-(\lambda)} &= \infint \dd{t} R(t,-t|\lambda) = \infint \dd{t} \frac{1}{2 f(t) \sinh \left(\frac{2 t}{k}\right)} Q_+(t) Q_-(t) \comma\\
\partial_\lambda\log D_+(\lambda)D_-(\lambda) &= \infint \dd{t} \left[ R(t,t|\lambda) - R^{(0)}(t)\right]\\ &=\infint \dd{t} \left[\frac{2k \left(Q_+(t) Q_-'(t)-Q_-(t) Q_+'(t)\right)}{2 \pi  f(t)} - R^{(0)}(t)\right]\comma
\end{split}
\ee
where $R^{(0)}(t)$ has to be chosen to guarantee convergence of this integral depending on the value of $k$. Those expressions are exactly the analogous to  \eqref{intformulas} and are the key consequence of the complete integrability of the kernel.  Starting from these expressions, one can in principle proceed as in Section \ref{spectraldata} to extract spectral sums and eigenvalues associated to this spectral problem. We hope to get back to study this in the future, as this could provide useful information on the convergence properties of the expansions compared with the exact one of \cite{kallen2014instanton}.

\section{Conclusion}
\label{conclusions}
\paragraph{Summary.}In this work, we examined analytical properties of mesons in large $N_c$ QCD$_2$. We showed that  the  eigenproblem (' t Hooft equation) that determines the discrete mass spectrum of the mesons, is equivalent to finding solutions to a TQ-Baxter system. Rather than solving the latter directly, we discussed how the specifics of this problem makes possible to determine the spectral data associated to the TQ-system through the analysis of its inhomogeneous generalization. Unlike the original homogeneous problem, which has solutions only in correspondence with the discrete spectrum of the 't Hooft model, the resulting inhomogeneous TQ-system is solvable analytically by the (asymptotic) series expansion in the continuous spectral parameter and the mass parameter $\alpha$. From those, we extract the spectral sums and the meson wavefunctions, which were shown to approximate very well the numericcal solutions.  

Another result of this paper is illustrating the analytical structure of the spectrum the mesons as we analytically continue the complex plane of the masses. The model exhibit a very rich and interesting structure in the complex plane: there is a square-root branch point in the chiral limit where there is the emergence of a massless meson associated to the unbroken chiral symmetry; more interestingly we provide evidence for the presence of infinitely many more square root branching points in the second sheet of the complex plane corresponding to critical points where one of the even eigenvalues turns to zero. Understanding the fate of this fixed point as we go beyond the large $N_c$ approximation, would be of paramount importance. We hope to go back to this point in the future. 

There are also several points of contact between this model and Ising Field Theory \cite{fonseca2006ising}, another prominent example of a confining theory in 2 dimensions. As we discussed in the text, this model also admits a Bethe-Salpeter approximation where the spectrum of the theory is approximated by two-quark mesons. The analytical properties of the spectrum in the two models are strikingly similar and suggest that the peculiar  analytical structure we found could be an universal feature of confining theories in two dimensions, as also suggested already by the WKB analysis of linear potentials we commented in Section \ref{criticality}. 
\bigskip

As already announced, this work is the first part of a bigger project in which we analyse mesons in a larger class of two dimensional QCD\textit{-like} theories. In the second part of this project \cite{Ambrosino:2024prz}, we will address the same questions for the large $N_c$ generalized Yang Mills theories:
\be 
\mathcal{L} = \frac{N}{8\pi} \Tr B \wedge F -  \frac{N}{4\pi}V(B) + \overline{\psi}_a \left(i \gamma^\mu D_\mu -m_a\right)\psi_a \; .
\ee 
Also for this very large class of  models, we show that the associated Bethe-Salpeter equation can be reformulated into a TQ-system, and many general features regarding the analyticity structure persist.

\paragraph{Future directions.}
In addition to the study of generalized Yang-Mills that we have already announced, there are many interesting future directions worth pursuing in the future:
\begin{itemize}
    \item A natural continuation of this work is a thorough study of the consequence of this emergent integrability structure at the level of form factors and scattering amplitudes of mesons, and correlation functions of conserved currents. The asymptotic expansions of wavefunctions developed in this paper are expected to be useful for the evaluation of these quantities.
    We will address this problem elsewhere. 
    \item Another direction we hope to pursue in the future is extending the study of this paper to QCD$_2$ theories coupled to matter in other representations of the gauge group and/or with different reality conditions (Majorana). This is a subject that has attracted a lot of attention in recent years, prominently for what regards the case of QCD coupled to adjoint quarks \cite{Dalley:1992yy,Kutasov:1993gq,Boorstein:1993nd,Bhanot:1993xp,Demeterfi:1993rs,Smilga:1994hc,Lenz:1994du,Katz:2013qua,Katz:2014uoa,Cherman:2019hbq,Komargodski:2020mxz,Dempsey:2021xpf,Popov:2022vud,Dempsey_2023,Dubovsky:2018dlk,Donahue:2019adv,Donahue:2019fgn,Donahue:2022jxu,Asrat:2022aov}. It would be interesting to study whether, at least in some particular limits, there is a similar structure as the one of the 't Hooft model. Of particular interest would be the limits in which quark loops are suppressed (such as the large quark mass limit) since the relevant diagrams in the limits are of the ``fish-net" type, to which the integrability techniques are known to be applicable \cite{Zamolodchikov:1980mb,Gurdogan:2015csr,Caetano:2016ydc}.
    \item One of the key points of our analysis was the use of the inhomogeneous Fredholm equation that allowed one to analytically continue meson masses from discretized values to continuous ones. This analytic continuation is reminiscent of the analytic continuation of spin of the $TQ$-Baxter equation for the $SL(2)$ spin chain (see e.g.\cite{Janik:2013nqa}). There are similarities and differences from our approaches and it would be nice to have a clear understanding of the relation between the two\footnote{We thank Pedro Vieira for asking this question.}. 
    \item It would be very interesting to systematically include finite $N_c$ corrections to 't Hooft equation to check, up to what extent the integrable structure that we describe here, is an emergent feature of the large $N_c$ expansion. Moreover, as already discussed in the main text, adding systematic correction in $1/N_c$ would be crucial for understanding the nature of the critical points in the second sheet of the complex-mass plane, and in particular the fate of \textit{critical exponents}, that could help to identify the associated CFT in the second sheet in the finite $N_c$ QCD$_2$.   
    \item With regards to generalization to finite $N_c$ QCD$_2$, some promising results are provided by the recent discussion of $SU(2)$ QCD coupled to fermions in large representation $J$ of the gauge group \cite{Kaushal:2023ezo}. The techniques developed in this paper apply trivially also to that case, given that the integral equation determining the spectrum of the mesons in the double scaling limit where we send $J \to \infty$ while taking $g^2J$ fixed, is formally identical to the 't Hooft equation studied here. It would be very interesting to extend this large spin analysis to the higher rank case as this could be a very promising way to analyze the role of finite $N_c$ corrections.
    \item As already discussed in the main text, another crucial aspect to explore is uncovering the fundamental origin of the integrable structures we have uncovered in this paper. In particular, it would be important to establish a connection with the $W_{\infty}$-algebra, discussed in the existing literature \cite{Dhar_1994}.
    \item It would be worth studying whether similar techniques apply also to other two dimensional models.  One could expect that multicritical versions of the Ising field theory might admit a systematic approximation analogous to the Bethe-Salpeter equation of the Ising field theory (e.g.\ a semiclassical approximation of the perturbed Tricritical Ising model is discussed in \cite{Lencs_s_2022}). Writing down such equations and analyzing them using the techniques discussed here would be an interesting future direction. 
    \item Regarding the connection to topological string discussed in Section \ref{topstring}, it is important to clarify the relation between the techniques employed in this paper and the ones in the topological string literature. For this purpose, it would be useful to complete the analysis, which we initiated in section \ref{topstring}, and compute the spectral sums and the asymptotic expansion of the spectrum and eigenfunctions. Another important direction to purse is to apply our techniques to other toric CY backgrounds including the ones for which the mirror curve is of higher genus.
    \item It has been pointed out in the literature (see e.g.~\cite{Marino:2016rsq,marino2018wavefunctions}) that open topological string wavefunctions satisfy the $TQ$-Baxter equation associated with the mirror curve. It is therefore important to  understand their precise relation to our $Q_{\pm}(x)$ (which satisfy the $TQ$-system for the inhomogeneous problem).
    \item A key question is whether one can find a CY geometry producing our TQ-system for the 't Hooft model \eqref{tq} as the quantization of their mirror curve. Such a connection, if it exists, would allow one to compute eigenfunctions and the spectrum of the 't Hooft model using topological string.
\end{itemize}
We leave all these investigations to future work.

\subsection*{Acknowledgement} 
We thank Ofer Aharony, Giulio Bonelli, Andrea Cavagli\`a, Diego Delmastro, John Donahue, Jaume Gomis, Nikolay Gromov, Davide Gaiotto, Tal Sheaffer, Pedro Vieira, Spenta Wadia, and Hao-Lan Xu for useful discussions and comments.
	\appendix
\section{Alternative derivation of the TQ relation}\label{appderivation}
In this appendix we present an alternative derivation of the equivalence between the homogeneous 't Hooft equation in $\nu$-space \eqref{thooftnuspace} and the TQ-equation \eqref{tq}-
Compared to the one presented in the main text in Section \ref{TQsec} (cfr.\  \eqref{defcq} - \eqref{tq}), the proof provided here has the advantage to be more direct of  highlighting the role played by the analytical structure of the integral equation's kernel\footnote{Cfr.\ also with the 
discussion in \cite{Vegh:2023snc}}. 

The starting point is the integral equation \eqref{inteqq} satisfied by $Q(\nu) = \sinh(\frac{\pi\nu}{2}) f(\nu) \Psi(\nu)$ that we report here for convenience:
\be
       Q(\nu) = \lambda\sinh(\frac\pi 2 \nu) \int_{-\infty}^{+\infty} \dd{\nu'} \frac{\pi (\nu - \nu') }{2 \sinh(\frac{\pi}{2}(\nu - \nu') )} \frac{Q(\nu')}{\sinh(\frac \pi 2 \nu')f(\nu')} \period
\ee
From this, we can readily compute:
\begin{align}
\nonumber
        Q(\nu \pm 2 i) = &-\lambda \sinh(\frac\pi 2 \nu) \int_{-\infty}^{+\infty} \dd{\nu'} \frac{\pi (\nu - \nu') }{2 \sinh(\frac{\pi}{2}(\nu - \nu'\pm 2 i) )} \frac{Q(\nu')}{\sinh(\frac \pi 2 \nu')f(\nu')}\,+ \\ \label{eq:der1} &-\lambda \sinh(\frac\pi 2 \nu) \int_{-\infty}^{+\infty}\dd{\nu'} \frac{\pm 2\pi i }{ 2\sinh(\frac\pi 2(\nu-\nu' \pm 2i))} \frac{Q(\nu')}{\sinh(\frac \pi 2 \nu')f(\nu')}  \\
        =\label{eq:der2}\,& Q(\nu) - \lambda \sinh(\frac\pi 2 \nu) \int_{-\infty}^{+\infty}\dd{\nu'} \frac{\pm 2\pi i }{  2\sinh(\frac \pi 2 (\nu-\nu' \pm 2i))} \frac{Q(\nu')}{\sinh(\frac \pi 2 \nu')f(\nu')}.
        \end{align}
To go from \eqref{eq:der1} to \eqref{eq:der2} (and analogously for $Q(\nu-2i)$) we simply used the shifting properties of $\sinh$ together with the regularity of the kernel for $\nu = \nu'$. Instead, for the extra piece, more care is needed: we cannot naively use that  $\sinh(\frac{\pi}{2}(\nu - \nu' \pm 2i)) \equiv -\sinh(\frac{\pi}{2}(\nu - \nu'))$ as the kernel there would hit a singularity. Indeed, upon combining the two expressions we get a further boundary term:
\begin{equation}
\begin{split}
    &Q(\nu + 2i) +  Q(\nu - 2i) - 2\,Q(\nu) =\\ \label{eq:intplus} &=\lambda \sinh(\frac\pi 2 \nu) \int_{-\infty}^{+\infty}\dd{\nu'}  \left(\frac{\pi i}{\sinh(\frac\pi 2(\nu-\nu' - 2i))} - \frac{\pi i}{  \sinh(\frac\pi 2(\nu-\nu' + 2i))} \right) \frac{Q(\nu')}{\sinh(\frac \pi 2 \nu')f(\nu')}
\end{split}\end{equation}
To evaluate this, it is convenient to use the following representation:
\small
\begin{align}
\label{eq:lim}
    &\frac{1}{  \sinh\frac\pi 2(\nu-\nu' - 2i)} - \frac{1}{  \sinh\frac{\pi}{2}(\nu-\nu' + 2i)}=
   \lim_{\epsilon\to 0}\left[ \frac{1}{ \sinh(\frac\pi 2(\nu-\nu'  - i \epsilon))} - \frac{1}{  \sinh(\frac\pi 2(\nu-\nu' + i \epsilon))}\right],
\end{align}
\normalsize
and to split the integral  \eqref{eq:intplus}  as:
\be
    \int_{-\infty}^{+\infty} \dd{\nu'}(\cdots) = \int_{-\infty}^{\nu - \delta}\dd{\nu'}(\cdots) + \int_{\nu - \delta}^{\nu + \delta} \dd{\nu'}(\cdots) + \int_{\nu + \delta}^{\infty} \dd{\nu'}(\cdots), \qquad 0< \delta \ll 1
\ee
Now, observe that in the first and third integration domain where $(\nu-\nu') \not\sim 0$,  in the integrand \eqref{eq:intplus} we can safely take the limit $\epsilon \equiv 0$ as the kernel is regular and the two terms in the integrand identically cancel. Instead, for $\nu' \in [\nu-\delta, \nu + \delta]$ we make use of the standard  Sokhotski–Plemelj theorem:
\begin{equation}
\text{\eqref{eq:lim}} = \frac{2}{\pi} \lim_{\epsilon\to 0}\left[ \frac{1}{  \nu-\nu'  - i \epsilon} - \frac{1}{  \nu-\nu' + i \epsilon}\right] = 4 i \delta(\nu-\nu') ,
\end{equation}
to deduce:
\be 
Q(\nu + 2i) + Q(\nu- 2i ) - 2Q(\nu) =   \frac{- 4\pi\lambda}{ \nu\coth(\frac \pi 2 \nu )  + \frac{2\alpha}{\pi} } \, Q(\nu) \, \comma
\ee 
that is equivalent to \eqref{tqtilde}.

\pdfbookmark[1]{\refname}{references.bib}
\bibliographystyle{JHEP}
\bibliography{references}

	\end{document}